\documentclass[graybox]{svmult}

\usepackage{type1cm}        % activate if the above 3 fonts are
\usepackage{amsfonts,amssymb,amsbsy,tabulary,amsmath}
\smartqed  % flush right qed marks, e.g. at end of proof
\usepackage{graphicx}
\usepackage{ulem}
\usepackage{amsmath}
\usepackage{natbib}
\usepackage{makeidx}         % allows index generation
\usepackage{graphicx}        % standard LaTeX graphics tool
\usepackage{multicol}        % used for the two-column index
\usepackage[bottom]{footmisc}% places footnotes at page bottom

\usepackage{newtxtext}       % 
\usepackage[varvw]{newtxmath}       % selects Times Roman as basic font
\usepackage{makecell}

\usepackage{color}
\definecolor{Navy}		{RGB}{  0,   0, 128}
\definecolor{MidnightBlue}	{RGB}{ 25,  25, 112}
\definecolor{yellow}   	{RGB}{255, 215,   0}
\definecolor{darkorange}{RGB}{255, 140,   0}
\definecolor{dodgerblue}{RGB}{ 30, 144, 255}
\definecolor{black}     {RGB}{  0,   0,   0}
\definecolor{dimgray}   {RGB}{105, 105, 105}
\definecolor{gray}   {RGB}{128, 128, 128}

\usepackage{url,multirow,morefloats,floatflt,cancel,tfrupee}
\makeatletter

\usepackage{hyperref}
\hypersetup{colorlinks=true, linkcolor=Navy, citecolor=Navy, urlcolor=Navy}

\usepackage{sidecap, caption}

\@ifundefined{etal}{}{}

\newcommand{\ME}{$\text{M}_{\oplus}$} 
\newcommand{\RE}{$\text{R}_{\oplus}$} 
\newcommand{\ms}{m\,s$^{-1}$}
\newcommand{\cms}{cm\,s$^{-1}$}

\newcommand*\arcsec{\ensuremath{^{\prime\prime}}}

\def\modif{}
\def\modiff{}

\makeindex             % used for the subject index
\begin{document}

\title*{Recent development in high-precision high-fidelity spectrographs for exoplanet research and characterization}
% Use \titlerunning{Short Title} for an abbreviated version of
% your contribution title if the original one is too long

%\author{Name of First Author\orcidID{0000-1111-2222-3333} and\\ Name of Second Author\orcidID{1111-2222-3333-4444}}
\author{François Bouchy$^1$, Francesco Pepe$^1$, Xavier Dumusque$^1$, 
Tobias Schmidt$^1$, Christophe Lovis$^1$, and Stéphane Udry$^1$ 
}
\institute{1: Observatoire de Genève, Université de Genève, Chemin Pegasi 51, 1290 Versoix, Switzerland. \email{francois.bouchy@unige.ch}}
% Use \authorrunning{Short Title} for an abbreviated version of
% your contribution title if the original one is too long
%
\setcounter{chapter}{19}
\titlerunning{High-precision high-fidelity spectrographs for exoplanets}
\authorrunning{Bouchy et al.}
\maketitle

%
% Use the package "url.sty" to avoid
% problems with special characters
% used in your e-mail or web address
%

\abstract{
High-precision high-fidelity spectrographs are the most powerful instruments for exoplanets detection and characterization. The sub-{\ms} radial-velocity precision, required to detect Earth-mass exoplanets, necessitates tackling all the sources of instrumental and stellar instabilities. We present the new high-precision high-fidelity spectrographs ESPRESSO, NIRPS, ANDES and RISTRETTO designed, developed, and operated with support of $PlanetS$.  
}

\section{Introduction}
\label{sec:intro}

The pace of early history of exoplanet discoveries was dictated by the radial-velocity (RV) technique using high-precision spectrographs. The planet 51 Peg b \citep{Mayor1995} was detected through the RV variation it induced on its parent star, in what became one of the most efficient techniques for planets detection and characterization. High-precision spectrographs also offer the required performance to characterise the exoplanet atmosphere by transmission spectroscopy. The study of planetary atmospheres from the ground \citep[e.g.][]{Snellen2008,Redfield2008,Charbonneau2002} represents a breakthrough in our understanding of what exoplanets are made of (chemical composition), which is essential for determining planetary evolution and may be key to better understanding planet formation processes. 
A review of past and present spectrographs dedicated to the search for exoplanets is given by \citet{Pepe2018}. From CORAVEL \citep{Baranne1979} to HARPS \citep{Mayor2003}, there has been an impressive evolution of technology based on an approach of small steps, every time improvements had become necessary to overcome precision limitations. When aiming at sub-{\ms} precision, required to detect Earth-mass exoplanets, the motion of the spectral line in the focal plane of the spectrograph becomes so tiny that it is actually comparable or eventually much smaller than any physical instrumental drift which then requires a dedicated strategy for wavelength calibration. The RV semi-amplitude induced by the planet is also comparable or smaller than any RV variation due to the stellar activity and photosphere inhomogeneity \citep{Dumusque2014,Dumusque2016, Dumusque2018}. {\modif So far, only a handful of extremely precise radial-velocity (EPRV) instruments allow the detection of Keplerian signal with a semi-amplitude below 2 {\ms}: HIRES \citep{Vogt1994}, HARPS \citep{Mayor2003}, PFS \citep{Crane2010}, HARPS-N \citep{Cosentino2012}, SOPHIE \citep{Bouchy2013}, CARMENES \citep{Quirrenbach2014}, HPF \citep{Mahadevan2014}, NEID \citep{Schwab2016}, EXPRES \citep{Jurgenson2016}, ESPRESSO \citep{Pepe2021}, MAROON-X \citep{Seifahrt2020}, and KPF \citep{Gibson2024}.}   
In the following sections, we present new generations of high-fidelity high-precision spectrographs. They are all based on the original HARPS concept that definitively set new standards \citep{Mayor2003}, was continuously improved \citep{LoCurto2015, LoCurto2024}, and is still one of the state-of-the-art instruments almost entirely dedicated to exoplanet studies. We present concepts, designs and performances of these new instruments developed, maintained, or operated by researchers in $PlanetS$. Most of them are intensively used for the follow-up and the mass measurement of transiting super-Earths and sub-Neptunes (see the chapter 12 about ``Origin and characterization of super-Earths and sub-Neptunes''). We present in Section \ref{sec:espresso} the ESPRESSO spectrograph installed at the incoherent focus of the ESO 8.2m Very Large Telescopes (VLTs). Section \ref{sec:nirps} highlights the recent near-infrared spectrograph NIRPS on the ESO 3.6m telescope. We discuss in Section \ref{sec:andes} the project ANDES for the ESO 39m Extremely Large Telescope (ELT). The RISTRETTO project, which combines high-contrast imaging and high-resolution spectroscopy, is detailed in Section \ref{sec:ristretto}. We discuss in Section \ref{sec:wavelength} the strategies and challenges for deriving an accurate, precise, and reliable wavelength solution. Finally, stellar activity mitigation is addressed in Section \ref{sec:activity}.

\section{ESPRESSO@VLT}
\label{sec:espresso}

\subsection{The ESPRESSO project and the instrument}
\begin{figure}
 \includegraphics[width=0.487\linewidth]{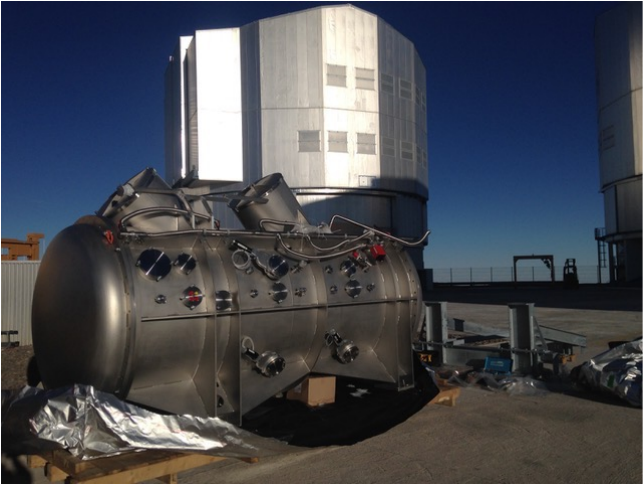}
 \includegraphics[width=0.513\linewidth]{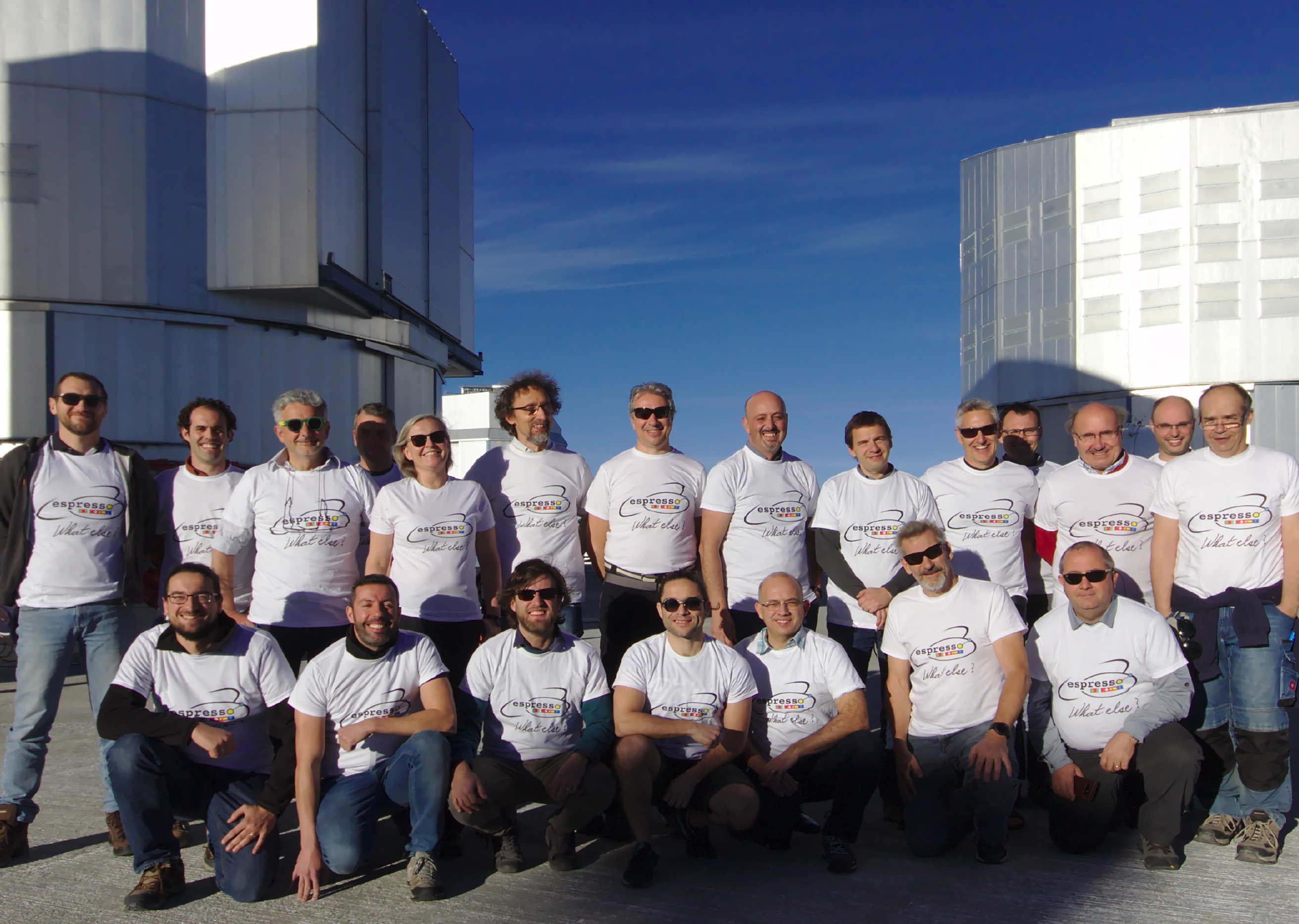}
 \caption{\small
        (Left) ESPRESSO vacuum chamber delivered on the VLT platform {\modiff(Credit: Francesco Pepe)}. (Right) ESPRESSO team during a commissioning phase {\modiff(Credit: ESO/Giorgio Calderone)}.}
 \label{Fig:ESPRESSO}
\end{figure}

ESPRESSO is a fiber-fed, cross-dispersed, high-resolution echelle spectrograph located in the Combined Coudé Laboratory (CCL) at the incoherent focus of the VLT, where it can be fed by the light of either one UT (1-UT configuration) or all four UTs (4-UT configuration). In each of its configurations, two fibers illuminate the spectrograph: 
One fiber carries the light from the science target, and the second carries either the light from the sky background (7\arcsec~away) or the light from a reference source for simultaneous drift measurements. A detailed description of the instrument and all its subsystems is given in \citet{Pepe2014}, \citet{Megevand2014}, \citet{Gonzalez2018}, \citet{Pepe2021} and references therein. The fiber-link is described more specifically in \citet{Genoni2020} and \citet{Aliverti2020}.
The control software and electronics are described more specifically in \citet{Calderone2018}, \citet{Baldini2016}, and \citet{Calderone2016}. The description of the scientific software, data flow and data products are presented in \citet{DiMarcantonio2018}. The integration of ESPRESSO started at Paranal Observatory in Oct 2017 (see Fig.\ref{Fig:ESPRESSO}) followed by several commissioning phases. Scientific operations started on 2 Sept 2018, a few days before the first TESS candidates release.  

\subsection{Observing with ESPRESSO}

The main instrument characteristics in the various observing configurations and instrument modes are summarized in Table\,\ref{tab:modes}. An ESPRESSO user can choose from two observing modes: the 1-UT configuration (one single telescope, any of the UTs) or the 4-UT configuration (all UTs simultaneously). In the 1-UT configuration, four different instrument modes exist. In HR mode, the light is collected using octagonal fibers with projected sky diameters of 1\arcsec. This mode is intended for spectroscopy and RV monitoring. Three different detector readout modes offer optimized solutions depending on the target magnitude. In case of bright targets (HR11, 1$\times$1 binning), the detector will be saturated quickly and the exposure time therefore has to be short. Overheads are reduced by a faster readout and the dynamic range is maximized by avoiding binning. Given the high signal level expected for a bright star, the higher readout noise (RON) is acceptable. For faint or very faint objects, the HR21 (2$\times$1 binning) or HR42 (4$\times$2 binning) should be used instead. The difference between these two modes is that the total RON per spectral bin is further reduced in the HR42 mode, and the limiting magnitude, at which the noise starts to be dominated by the detector RON, is improved. This gain comes at the cost of a slightly reduced resolving power since the resulting pixel sampling gets close to the Nyquist limit. 

Finally, the UHR11 mode delivers the highest resolving power, but the required use of a 0.5\arcsec diameter fiber leads to reduced spectral sampling and higher slit losses. Due to the latter issue, this mode is particularly suitable for ultra-high-resolution observations on relatively bright targets.

By design, the 4-UT observing configuration is intended for spectroscopy of very faint targets. The lower resolution is a price to pay for the usage of an equivalently wider (square) fiber in the spectrograph that can carry the light of the four telescopes. Collecting the light from the four telescopes reduces the contribution of detector readout and (partially) dark noise when adequate slit configurations and binning levels are adopted. The two instrument modes MR42 (4$\times$2 binning) and MR84 (8$\times$4 binning) therefore enable {\modif spectroscopy of very faint sources} and the study of fast transient events.

\begin{table*}
\caption{ESPRESSO's instrument modes and characteristics.}             
\label{tab:modes}      
\centering          
\begin{tabular}{l c c c c c c}    

\hline\hline
Observing configuration & \multicolumn{4}{c}{1-UT} & \multicolumn{2}{c}{4-UT}\\
 & \multicolumn{4}{c}{-----------------------------------} & \multicolumn{2}{c}{-----------------}\\
Instrument mode & UHR11 & HR11 & HR21 & HR42 & MR42 &  MR84 \\
\hline                    
%Simultaneous reference & \multicolumn{6}{c}{Available (no simultaneous sky background on fiber B)}\\
%Sky subtraction & \multicolumn{6}{c}{Available (no simultaneous reference on fiber B)}\\
Wavelength range  [nm]& \multicolumn{6}{c}{378.2--788.7}\\
Median spectral res. $R=\lambda/\Delta\lambda$ & $>$190k & 138k & 138k & 130k &  72.5k & 70k\\
Aperture on sky [\arcsec] & 0.5. & 1.0  & 1.0   & 1.0  &  4$\times$1.0 & 4$\times$1.0 \\
%Spectral sampling [pixels] & 2.5 & 4.5 & 4.5 & 4.5 & 10 & 10 \\
%Spatial sampling [slices$\times$physical pixels] & 2$\times$5  & 2$\times$9 & 2$\times$9 & 2$\times$9 & 2$\times$20 & 2$\times$20\\
Detector readout speed [kpix\,s$^{-1}$]& 300 & 300  & 100 & 100 & 100 & 100 \\
Detector binning [pixels]& 1$\times$1  & 1$\times$1  & 2$\times$1 & 4$\times$2  & 4$\times$2 & 8$\times$4\\
{\modif Effective} spectral sampling [pixels] & 2.5 & 4.5 & 4.5 & 2.25 & 5 & 2.5 \\
Detector RON red/blue [e$^-$]& 5/8  & 5/8  & 2/3  & 2/4  & 2/4 & 2/5\\
%Detector conversion factor red/blue [e$^-$/ADU] & 1.15/1.1  & 1.15/1.1  & 1.13/1.09 & 1.13/1.09  & 1.13/1.09 & 1.13/1.09\\
Total peak efficiency & 5\% & 10\% & 10\% & 10\% & 10\% & 10\% \\
\hline\hline
%\multicolumn{7}{l}{$^{(a)}$To be considered as a lower limit since many thorium lines used for the analysis are partially resolved at this resolution.}\\
%\multicolumn{7}{l}{$^{(b)}$The "FAST" readout modes (HR11 and UHR11) of the blue detector are still undergoing optimization.}
\end{tabular}
\end{table*}

\subsection{Exoplanets with ESPRESSO}
\label{sec:obs}

In exchange for funding and building the ESPRESSO instrument, the ESPRESSO Consortium was awarded 273\,nights of Guarantee Time of Observation (GTO) with the VLT, following a well-established and successful ESO scheme for funding second-generation instruments for the La Silla and Paranal observatories. 
%Although formally granted to the principal investigator representing the Consortium, this person has to present observing proposals to ESO prior to the start of operations. The GTO proposals have to be approved by ESO Observing Program Committee (OPC) before execution. All of the GTO proposals must follow the science program proposed by the Consortium to ESO at the time of the signature of the construction agreement. 
The ESPRESSO consortium dedicated 80\% of the granted GTO to the search for and characterization of exoplanets and 10\% to the study of the (possible) variability of fundamental physical constants. The remaining 10\% of the GTO was reserved for other science cases of interest. When officially proposed to ESO in 2007 (before the first results provided by the $Kepler$ mission and well before the discovery of Earth-mass planets around stars in the solar neighborhood), ESPRESSO had already been conceived to pursue the following scientific objectives in the exoplanetary field: the discovery and characterization of Earth-mass planets (possibly) within the habitable zone (HZ) of their parent stars, the characterization of the atmospheres of individual exoplanets, and the precise mass determination of low-mass transiting planets. In what follows, we describe these three sub-programs.

\subsubsection{Blind search for rocky planets in the habitable zone}

One of the main science drivers for ESPRESSO is the detection of Earth-mass planets in the HZ around quiet nearby GKM stars. It was this driver that set the most stringent constraints on spectral resolution, thermo-mechanical stability, and illumination stability, ultimately aiming for 10 {\cms} of RV precision. 
%This enables the search for Earth-mass planets inside the HZs of solar-type stars, provided that these stars are "quiet" and sufficiently bright, thus pointing towards late-G, K and M dwarfs of the solar neighborhood. 
%Their proximity make them particularly suitable for follow-up studies regarding, for example, direct detection and characterization of their atmospheres. Starting in 2010, we performed several preparatory studies to define our target sample. We investigated the possible contamination by faint, spatially unresolved companions \citep{Cunha2013} and the impact of micro-tellurics on precise RVs \citep{Cunha2014}, among other topics. 
%To avoid being biased by previous RV surveys, we built our blind search catalog from scratch, performing a spectroscopic characterization for the stars that had not yet been observed at high spectral resolution. 
%We obtained observing time on HARPS, HARPS-N, and UVES to complement HARPS observations retrieved from the ESO archive. 
%For all candidates, we performed a detailed spectroscopic characterization as well as calculated the rotational velocity and chromospheric activity indices, among several other studies. 
Initially our target list contained 77 close-by, quiet, non-rotating stars covering a wide range of right ascension values and being thus observable the entire year. For these stars, the photon noise attained in a {\modif 15-min} integration time (or {\modif 30\,min} for the faintest targets) allows us to detect the RV signal of a one Earth-mass planet orbiting inside the HZ. The details regarding target selection, spectral characterization, and further studies, along with reduced data access, are presented in \citet{Hojjatpanah2019}.

During the first year and a half of observations, we narrowed down our sample to $\sim45$ promising targets, of which 25 were followed intensively and on which we aimed to conduct an average of 60-70 visits by the end of the GTO. The remaining 30 targets were discarded due to the presence of stellar companions, or simply to avoid competition with other programs already following them. 

It is worth mentioning that for several stars of this sub-program, the RV scatter is less than 1\,{\ms}; we stress that this value is computed without any data post-processing and with the RVs provided by the publicly available pipeline. For $\tau$\,Cet, the RV scatter of the nightly binned RVs is of 50\,{\cms}. In \citet{Suarez-Mascareno2020}, 63 ESPRESSO RV-points are used to independently confirm and revisit the exoplanet Proxima Cen\,b ($m$\,sin$i$=1.173$\pm$0.086 \ME, P=11.184 days), get a clear measurement of the stellar rotation (87 $\pm$ 12 days), and find some evidence for the presence of a second short period signal at 5.15 days with K=40 {\cms}. Once the planetary and stellar activity models are removed, the residuals scatter is at the level set by photon-noise of $\sim$30\,{\cms}, demonstrating unprecedented precision. Additional 52 ESPRESSO measurements allowed us to confirm the 5.12 days signal corresponding to a Sub-Earth-mass planet \citep{Faria2022} with a minimum mass of about twice the mass of Mars (0.26 $\pm$ 0.05 \ME).   
The first ESPRESSO stand-alone planet discovery HD\,22496b ($m$\,sin$i$=5.6$\pm$0.7 \ME, P=5.09 days) was published by \citet{Lillo-Box2021}. Two temperate Earth-mass planets ($m$\,sin$i$=1.08$\pm$0.13 \ME, P=10.35 days and $m$\,sin$i$=1.36$\pm$0.17 \ME, P=20.20 days) were discovered around the nearby star GJ\,1002 \citep{Suarez-Mascareno2023}. The compact multi-planetary systems of super-Earths GJ\,9827 and HD\,20794 were revisited with ESPRESSO \citep{Passegger2024, Nari2025}. HD\,20794d is likely to be a rocky planet ($m$\,sin$i$=5.8$\pm$0.6 \ME) in the HZ (P=647 days) of its G-dwarf star. Recently a sub-Earth-mass planet ($m$\,sin$i$=0.37$\pm$0.05 \ME) was identified orbiting Barnard's star \citep{Gonzalez2024}. 

%- Revisiting the multi-planetary system of the nearby star HD 20794: Confirmation of a low-mass planet in the habitable zone of a nearby G-dwarf \citep{Nari2025}\\
%- A sub-Earth-mass planet orbiting Barnard's star \citep{Gonzalez2024}\\
%- The compact multi-planet system GJ 9827 revisited with ESPRESSO \citep{Passegger2024}\\
%- Two temperate Earth-mass planets orbiting the nearby star GJ 1002 \citep{Suarez-Mascareno2023}\\
%- A candidate short-period sub-Earth orbiting Proxima Centauri \citep{Faria2022}\\
%- HD 22496 b: The first ESPRESSO stand-alone planet discovery \citep{Lillo-Box2021}\\

\subsubsection{Characterization of planetary atmospheres}

One of the first ESPRESSO-GTO results in exoplanetary atmosphere was the detection of an asymmetric absorption signature due to neutral iron during the primary transit of the ultra-hot giant planet WASP-76b \citep{Ehrenreich2020}. This feature is blueshifted in the trailing limb (probably owing to the combination of planetary rotation and atmospheric wind), but no signature appears in the leading limb. We interpreted this as evidence of the condensation of gaseous iron, which is injected from the hot (day) side atmosphere onto the cool (night) side of the planet. 

%In coming papers, the ESPRESSO-GTO team will show that this scenario likely happens in more planets with similar temperatures as those of WASP-76b and will report robust detections of atomic elements in the atmospheres of warm planets, such as lithium, that were not solidly established before (Tabernero et al. in prep.; Borsa et al. in prep.).\\ 

ESPRESSO allowed us to detect the atmospheric Rossiter-McLaughlin effect {\modif as well as} the transmission spectroscopy of WASP-121b \citep{Borsa2021}. The high-resolution transmission spectroscopy of WASP-76b and MASCARA-1b are presented by \citet{Tabernero2021} and \citet{Casasayas-Barris2022}, respectively. \citet{AzevedoSilva2022} present the detection of barium in the atmospheres of the ultra-hot gas giants WASP-76b and WASP-121b as well as new detections of Cobalt and Strontium ion on WASP-121b. More recently, ESPRESSO revealed blueshifted neutral iron emission lines on the dayside of WASP-76 b \citep{CostaSilva2024}.    

%- ESPRESSO reveals blueshifted neutral iron emission lines on the dayside of WASP-76 b \citep{CostaSilva2024}\\
%- Detection of barium in the atmospheres of the ultra-hot gas giants WASP-76b and WASP-121b. Together with new detections of Co and Sr+ on WASP-121b \citep{AzevedoSilva2022}\\
%- Transmission spectroscopy of MASCARA-1b with ESPRESSO: Challenges of overlapping orbital and Doppler tracks \citep{Casasayas-Barris2022}\\
%- ESPRESSO high-resolution transmission spectroscopy of WASP-76 b \citep{Tabernero2021}\\
%- Atmospheric Rossiter-McLaughlin effect and transmission spectroscopy of WASP-121b with ESPRESSO \citep{Borsa2021}\\

\subsubsection{RV follow-up of \textit{K2} and \textit{TESS} transiting planets}

ESPRESSO represents a breakthrough in the mass measurements of transiting exoplanets: Its 2-magnitude gain and improved RV precision with respect {\modif to HARPS} enable a thorough exploration of the rocky planet population detected by \textit{K2} and \textit{TESS}. These rocky worlds have radii below $\sim$2\,R$_\oplus$ and their masses are typically below 10\,M$_\oplus$. ESPRESSO-GTO observations enable to constrain the internal composition of these objects, namely their iron/rock/water mass fraction, under the reasonable assumption that they do not have any H/He envelopes. %Moreover, ESPRESSO explored how the properties of this population vary with stellar irradiation, stellar mass, planetary architecture, and even stellar composition \citep{Santos2017}, shedding new light on the formation and evolution pathways of these systems. 
%Ultimately, ESPRESSO will be able to characterize transiting rocky planets within the HZ of their host star with unprecedented detail.

The ESPRESSO-GTO \textit{K2}/\textit{TESS} planet sample has been defined based on the following criteria: a) confirmed or validated planet candidate; b) planet radius $<2.0$\,R$_\oplus$; c) host star $V<14.5$\,mag (spectral-type dependent); and d) no precise mass measurement available. In addition, we intended to probe the rocky-to-gaseous transition at a given stellar irradiation level with the goal to constrain the evaporation processes that transform sub-Neptunes into naked rocky cores under the influence of extreme UV (XUV) irradiation from the host star. According to the results of \citet{Fulton2017}, a particularly interesting irradiation range to carry out this experiment is $\sim50-200$\,$\times$ Earth irradiation levels, where both gas-rich sub-Neptunes and gas-poor rocky planets appear to coexist. Mass and density measurements from ESPRESSO thus probe the mass and density thresholds, leading to significant evaporation of planetary envelopes. The target sample for this experiment is the same as the one defined above, complemented in the following ways: 1) adding the larger planets that are in the same systems as the rocky planets defined above, and 2) adding a few more sub-Neptunes ($<4$\,R$_\oplus$) in other systems, which are under irradiation levels of 50-200\,$\times$ Earth level. 
%The main objective of this sub-program was to measure precise masses and bulk densities for all these objects. 
%In total, we aim at characterizing 50-100 small planets by the end of the GTO period.\\

To illustrate the potential of ESPRESSO for this sub-program, we refer to the case of the planetary system $\pi$\,Men. $\pi$\,Men is a bright, ``naked-eye'' G0V-star ($V=5.6$\,mag) known to host a sub-stellar companion on a long-period and very eccentric orbit of $P_{\rm b}\sim$2\,000\,days. On September 6, 2018, NASA announced the discovery by TESS (\citealt{Ricker2015}) of a second planet, the transiting super-Earth $\pi$\,Men\,c ($P_{\rm c}=$\,6.27 days; $R_{\rm c}=2$\RE). Following the announcement, \cite{Huang2018} and \citet{Gandolfi2018} independently detected the spectroscopic orbit of $\pi$\,Men\,c by analyzing archival RVs of HARPS and UCLES, and confirmed its planetary nature (4.5 $\pm$ 0.8 \ME). The brightness of the star makes $\pi$\,Men a perfect target for testing the performance of ESPRESSO in measuring the masses and bulk densities of low-mass planets. From September 5, 2018, to March 25, 2019, we recorded therefore a total of 275 spectra and covered a time span of 201 days. The spectra were acquired with a typical exposure time of 120\,s, providing 
%median $S/N=243$ per extracted pixel at $\lambda=$\,500\,nm and
a median RV (internal) precision of 25 {\cms}. 
%During each observing night, we collected series of multiple spectra at a rate of two to 12 consecutive exposures, leading to a typical in-night RV precision of the binned data points of about 10 {\cms}. 
%Furthermore, \textit{TESS} reobserved the host star during cycle 1 (sectors 4, 8, and 11-13) from October 2018 to July 2019, collecting 19 additional transits of planet\,c in short-cadence mode. 
A detailed and complete analysis of the system architecture of $\pi$\,Men is presented by \cite{Damasso2020}. 
%The authors combined the exquisite RVs of ESPRESSO with the \textit{TESS} photometry and with the \textit{Gaia} and \textit{HIPPARCOS} astrometry. 
Besides refining the mass of the transiting super-Earth and reducing the error bars on its orbital parameters, we also constrained the mutual orbital inclination of the two planets, the real mass of the long-period planet\,b, and the system architecture in general.

As part of this sub-program, we also characterized the K2-38 planetary system \citep{Todelo2020}, including a dense iron-rich Mercury-like planet (1.54 $\pm$0.14 \RE, 7.3$\pm$1.0 \ME) and a sub-Neptune (2.29$\pm$0.26 \RE, 8.3$\pm$1.3 \ME). We unveiled a sub-Neptune and a non-transiting Neptune-mass companion around the bright late-F dwarf HD 5278 \citep{Sozzetti2021}. We measured the mass and bulk density of a warm terrestrial planet with half the mass of Venus transiting the nearby star L\,98-59 \citep{Demangeon2021} {\modif hosting} 5 planets system. We detected a multi-planetary system with three super-Earths and two potential super-Mercuries around HD\,23472 \citep{Barros2022}. 
We unveiled the multi-planetary system in a triple M-dwarf system LTT 1445 \citep{Lavie2023}. ESPRESSO measured the unusually low-density super-Earth transiting the bright early-type M-dwarf GJ 1018 \citep{Castro-Gonzalez2023}. ESPRESSO radial velocities lead to the detection of transits with low SNR in the compact multi-planet system HIP\,29442 \citep{Damasso2023}. More recently, several new super-Earths and sub-Neptunes were discovered by TESS and ESPRESSO  \citep{Suarez-Mascareno2024, Hobson2024}.

\section{NIRPS - The Near-InraRed Planet Searcher}
\label{sec:nirps}

\begin{figure}
 \includegraphics[width=0.542\linewidth]{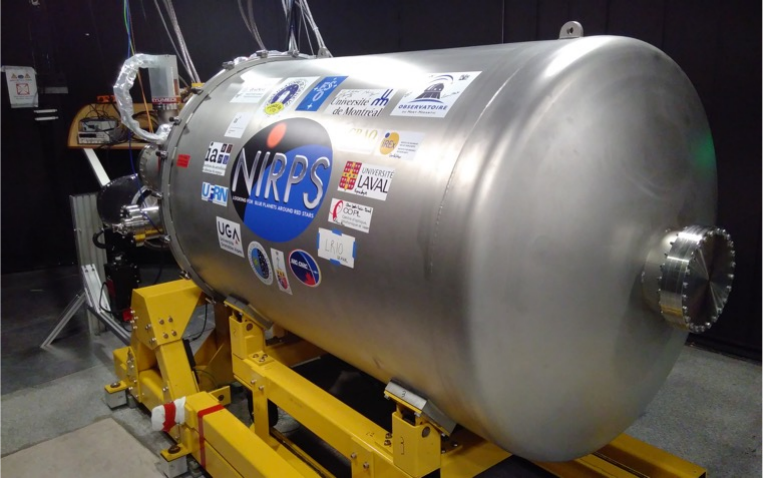}
 \includegraphics[width=0.458\linewidth]{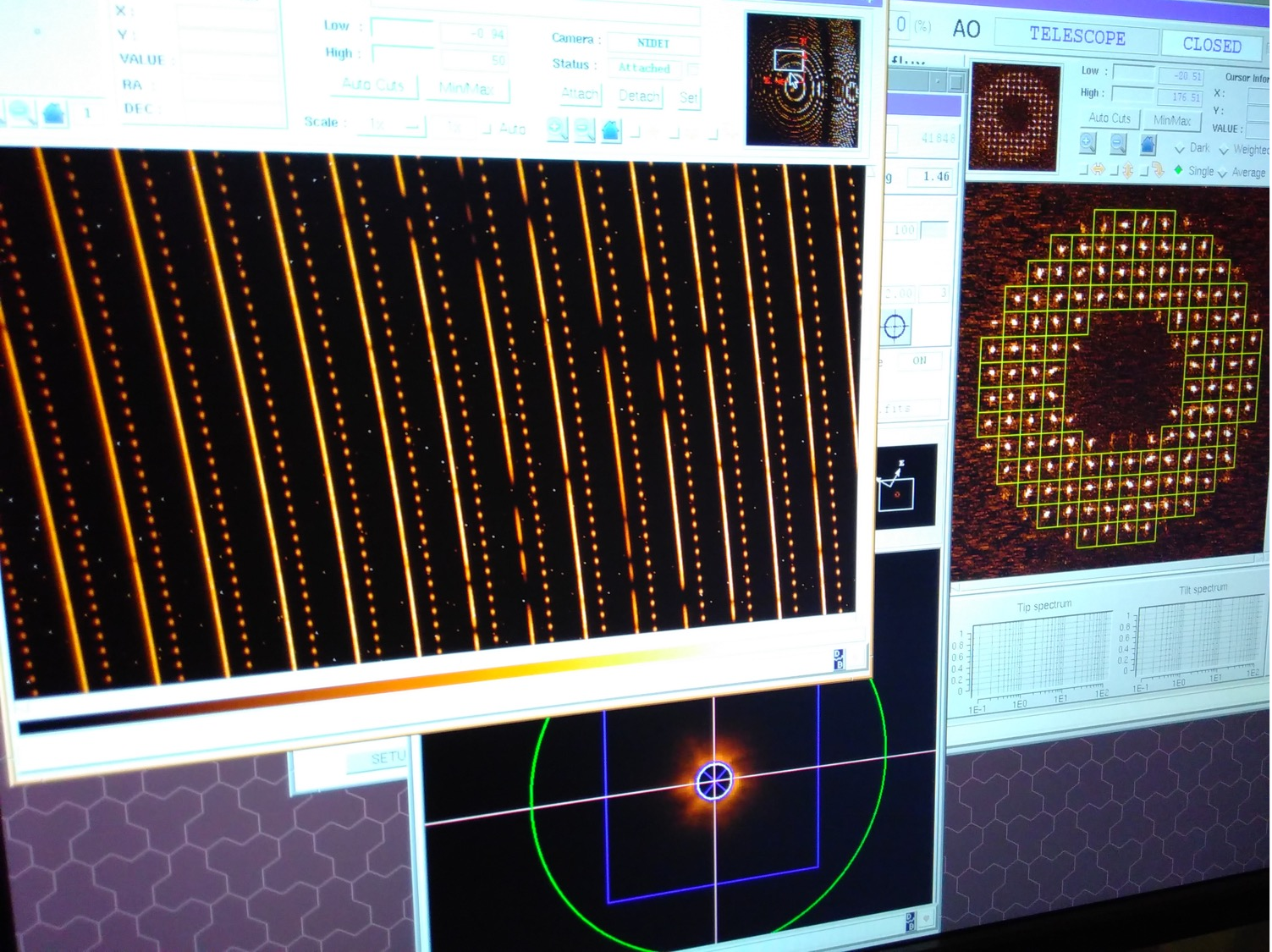}
 \caption{\small
        (Left) Vaccum vessel of NIRPS spectrograph in the Coudé Room of the 3.6m telescope. (Right) Screenshot of the NIRPS first light made on 17 May 2022 showing a portion of the raw spectra, the guiding camera and the AO wave-front sensor {\modiff{(Credit: François Bouchy)}}.}
 \label{Fig:NIRPS}
\end{figure}

In response to an ESO call for new instruments for La Silla Observatory in February 2015, the NIRPS consortium{\footnote{The NIRPS consortium is composed of: Université de Montréal (co-project lead, Canada), Observatoire Astronomique de l'Université de Genève (co-project lead, Switzerland), Instituto de Astrofisica e Ciencias do Espaco hosted by the Centro de Astrofisica da Universidade do Porto and the Faculdade de Ciências da Universidade de Lisboa (Portugal), Instituto de Astrofisica de Canarias (Spain), Université de Grenoble-Alpes (France) and Universidade Federal do Rio Grande do Norte (Brazil). ESO participated in the NIRPS project as an associated partner. }} proposed, designed and built a dedicated near-infrared (nIR) spectrograph to enable radial velocity (RV) measurements of low-mass exoplanets around M dwarfs and to characterise exoplanet atmospheres in the nIR \citep{Bouchy2017}. NIRPS is a high-resolution, high-stability near-infrared spectrograph equipped with an adaptive optics (AO) system. NIRPS was built upon the lessons learned from the first generation of nIR velocimeters GIANO \citep{Oliva2012}, CARMENES \citep{Quirrenbach2014} and SPIRou \citep{Donati2020}, and the success of exoplanet optical hunters HARPS \citep{Mayor2003} and ESPRESSO \citep{Pepe2021}. NIRPS installation at the ESO 3.6-m telescope started in 2019. Intensive on-sky testing phases were conducted between 2019 (with the front-end and adaptive optics system), and March 2023 (with the entire instrument), with the official first light on 17 May 2022 (See Fig.\ref{Fig:NIRPS}). The start of operations took place on April 1, 2023, the date on which NIRPS was offered by ESO both to the community for open-time observations and to the NIRPS consortium for Guaranteed Time Observation (GTO). For a more comprehensive description of NIRPS, readers are referred to \citet{Bouchy2025}.

\subsection{A dedicated design for high-fidelity nIR spectroscopy}

The NIRPS instrument includes a Front-End at the Cassegrain focus, linked via optical fibres to the cryogenic spectrograph in the Coudé room of the 3.6m telescope. The Front-End integrates an adaptive optics (AO) system which corrects for atmospheric turbulence to enhance coupling efficiency, minimise the instrument's size and offers a unique high-angular resolution capability with a fibre acceptance of only 0.4 arcsec. NIRPS operates in the Y, J, and H bands, covering a wavelength range from 972.4 nm to 1919.6 nm. It offers two fibre sizes (0.4 and 0.9 arcsec) yielding resolving powers of R = 90'000 for the high-accuracy (HA) mode and R = 75’000 for the high-efficiency (HE) mode. The fibre link transports light to the spectrograph and incorporates octagonal fibres, fibre stretchers and double scrambler to reduce the impact the modal noise. Modal noise is also mitigated through AO scanning by using the tip/tilt mirror to move the star within the fibre core randomly during an exposure \citep{Frensch2022}. Together, both stretching and AO-scanning yield a modal noise reduction by a factor of 5 for the HE mode, leaving a residual noise of 0.43\% (SNR~230) which translates into RV noise of 0.9 {\ms}. The calibration unit includes hollow-cathode (HC) Uranium-Neon lamps and a Fabry-Pérot (FP) étalon to generate absolute wavelength calibration. The Back End spectrograph is housed within a vacuum vessel (see Fig.\ref{Fig:NIRPS}), with its optical bench stabilized to 75 K. The instrument’s thermal control system maintains ~0.2 mK RMS stability over several months, leading to extreme intrinsic stability. The optical design features a reflective double-pass collimator, a 13-lines/mm R4 echelle grating, a carousel of five ZnSe prisms for cross-dispersion, and a refractive camera that feeds a 4k × 4k pixels Hawaii-4RG (H4RG) infrared detector with 15-$\mu$m pixels. The full wavelength range spans 71 orders, with the line spread function sampled by 3 pixels in HA mode. The overall throughput from the top of the atmosphere to the detector peaks at 13\% in the H band. The AO system significantly improves fibre coupling efficiency, achieving typical encircled energy of 55\% and 70\% for the HA and HE modes, respectively; this performance being constant up to I = 11. The HE mode, although providing a slightly lower spectral resolution, is recommended to reach the highest SNR and RV precision. The HA mode is recommended in cases high angular resolution is requested (e.g., visual binaries with angular separation less than 2 arcsec. The fact that the spectrograph is intrinsically ultra-stable (with a typical drift of less than 0.1 {\ms} per day) {\modif makes the use of the simultaneous tracking with Fabry-Pérot no longer required. Moreover, fibre B spectrum (especially with the HE mode) is affected by a higher level of modal noise, which renders it less useful.} Furthermore, the use of the OBJ-SKY mode is recommended for the sky OH lines correction. 

\subsection{NIRPS+HARPS: a unique dual optical-infrared precision velocimeter} 

A dichroic beam splitter simultaneously directs light to both HARPS and NIRPS, enabling simultaneous visible and nIR observations. This dual-wavelength approach offers the advantages 1) to cover simultaneously from 378 nm to 1920 nm (with a gap in between 691 nm and 972 nm), 2) to reduce significantly the radial-velocity photon-noise uncertainty on M dwarfs, and 3) to explore different approaches to disentangle stellar activity signals from exoplanet signals. This dual-wavelength configuration greatly enhances the potential for exoplanet atmospheric studies and lays the groundwork for future direct imaging efforts using extremely large telescopes (ELT). {\modif It is worth noting that the gap in wavelength coverage, used by the NIRPS AO system, includes the Z and I bands which are not negligible in terms of RV content for M dwarfs \citep[e.g.][]{Figueira2016, Reiners2020, Artigau2018}. These bands are included in MAROON-X \citep[500-920 nm;][]{Seifahrt2020}, CARMENES \citep[520-1710 nm;][]{Quirrenbach2014}, HPF \citep[810-1280 nm;][]{Mahadevan2014}, and NEID \citep[380-930 nm;][]{Schwab2016}.}
NIRPS and HARPS are also continuously monitoring during daytime the disk-integrated sunlight with HELIOS solar telescope.

\subsection{Two dedicated NIRPS data reduction software} 

The NIRPS data reduction software (NIRPS-DRS) is adapted from the ESPRESSO pipeline \citep{Pepe2021}, incorporating features inspired by the APERO pipeline \citep{Cook2022}, a versatile framework initially developed for SPIRou. While NIRPS-DRS and APERO share similarities, they differ in key aspects, such as their approaches to telluric correction \citep{Bouchy2025}. The two independent pipelines are particularly valuable for cross-validation and assessing the robustness of scientific results. As expected for nIR spectrographs, the telluric contamination is one of the main limitations on the RV precision. The NIRPS-DRS pipeline includes a telluric subtraction module developed for ESPRESSO \citep{Allart2022}, while APERO is using an ensemble of telluric standards, fast-rotating early-type stars, and a principal component analysis (PCA)-based method for modeling and removing telluric lines. Radial velocity extraction is performed using two complementary methods: the standard cross-correlation function (CCF) technique and the line-by-line (LBL) method \citep{Artigau2022}. The CCF provides immediate RV measurements for each observation, while the LBL method, which constructs a template spectrum from a time series, generally delivers more precise RVs with less sensitivity to outliers. The RV computed online and in real-time at the telescope should be considered as indicative and not as the nominal one. 

\subsection{NIRPS On-sky performance and first results} 

During commissioning, NIRPS demonstrated stable RV precision at the level of 1 {\ms} over several weeks on several RV standards with known planetary systems \citep{Wildi2022, Artigau2024, Doyon2025, Bouchy2025}. NIRPS is the first nIR velocimeter to demonstrate sub-metre-per-second performance, partly due to the excellent sub-Kelvin thermal stability of the spectrograph yielding typical drifts of 3-4 cm/s per day and wavelength uncertainties at the level of 50–70 {\cms}. This level of stability enables for reliable, long-term RV monitoring of low-mass stars. The instrument’s high throughput, particularly in the H band, offers a notable improvement over previous spectrographs, enhancing the ability to detect small exoplanets. 

The M5.5V dwarf Proxima Centauri was intensively observed with NIRPS. The 149 nightly-binned RVs are presented by \cite{Suarez-Mascareno2025} confirming the exoplanet Proxima b and finding evidence of the presence of the sub-Earth Proxima d. The standard deviation of the residuals of NIRPS RVs after the modelling is $\sim$0.80 {\ms}, showcasing the potential of NIRPS to measure precise radial velocities in the near-infrared. 

The relatively quiet M3V star TOI-406 (V = 13.8 and J = 9.7) was observed to measure the mass of the transiting sub-Neptune (2.1 {modif R$_\oplus$}) revealed by TESS with a period of 13.2 days. An additional transit signal of a super-Earth with a period of 3.3 days was detected later as a Community-TOI (CTOI). This target was also observed during the same season (P111) with ESPRESSO (PI: E. Pallé). We combined our data sets and published the masses of TOI-406 b and TOI-406 c \citep{Lacedelli2024}. Scaled to the same exposure time of 1800 s, the median RV uncertainty is 3.1 and 3.7 {\ms} for NIRPS and HARPS, respectively, illustrating the gain in photon-noise uncertainty between the two instruments. The dispersion of the residuals is 3.4 {\ms} and 5.1 {\ms} for NIRPS and HARPS, respectively. 

NIRPS also led to the discovery of : 1) a transiting sub-Neptune and a cold eccentric giant orbiting the M-dwarf TOI-756 \citep{Parc2025}; 2) the updated characterization of the multi-planetary system including a transiting sub-Neptune and orbiting the nearby M-dwarf GJ3090 \citep{Lamontagne2026}; 3) the two giant planets transiting TOI-3832 and TOI-4666 \citep{Frensch2026}; and 4) the ultra-short period super-Earth TOI-4552b citep{Srivastava2026}. 

The NIRPS capability for probing exoplanet atmosphere and measuring the spin-orbit angle through the Rossiter-MCLaughlin (RM) effect was demonstrated by observing 3 transit events of the warm Saturn WASP-69b \citep{Allart2025}. NIRPS successfully detected the helium triplet near 1083 nm in the planet’s atmosphere, with evidence of variability indicative of cometary-like tail mass loss. The RM measurements suggest a slightly misaligned orbit with a true obliquity of 28.7 $\pm$ 6 deg.

The detailed analysis of the optical (with HARPS) and near-infrared (with NIRPS) transmission spectrum of the ultra-hot gas giant WASP-189b is presented by \cite{Vaulato2025}. It reveals that hydride ion continuum hides absorption signatures in the near-infrared transmission spectrum of this exoplanet atmosphere. 
The atmospheric analysis of the daysides of the ultra-hot Jupiter WASP-121b from emission observations using NIRPS is presented by \cite{Bazinet2025}. It reveals evidence of thermal dissociation and redshifted water detection in this exoplanet atmosphere. 

Initial results confirm that NIRPS, in combination with HARPS, opens new possibilities for RV measurements, stellar characterisation, and exoplanet atmosphere studies by extending observations into the NIR with high precision and high spectral fidelity. NIRPS offers the astronomical community high-resolution and high-stability near-infrared spectroscopy. The instrument’s unique features open a new parameter space in ground-based observations that can address a large diversity of science cases.

%Development, tests, performance\\
%Key results \\
%Lessons learned \\
%- Bouchy et al., 2017, The Messenger, 169, 21, Near-InfraRed Planet Searcher to Join HARPS on the ESO 3.6-metre Telescope\\
%- Bouchy et al. 2025, A\&A, submitted: NIRPS joining HARPS at the ESO 3.6m /On-sky performance and science objectives\\
%- Artigau, É., et al., 2024, NIRPS first light and early science: breaking the 1 m/s RV precision barrier at infrared wavelengths, SPIE, Ground-based and Airborne Instrumentation for Astronomy X, 13096, 130960C\\
%- Doyon et al. 2025, NIRPS Joins HARPS: Setting New Standards at Infrared Wavelengths, Messenger\\
%- Frensch, Y. G. C., et al., 2022, NIRPS fiber-link design, performances and modal noise mitigation performances tested on sky, SPIE, Ground-based and Airborne Instrumentation for Astronomy IX, 12184, 1218451\\
%- Wildi, F., et al., 2022, First light of NIRPS, the near-infrared adaptive-optics assisted high resolution spectrograph for the ESO 3.6m, SPIE, Ground-based and Airborne Instrumentation for Astronomy IX, 12184, 121841H\\

\section{ANDES@ELT}
\label{sec:andes}

\subsection{The European Extremely Large Telescope}
\begin{figure}
 \includegraphics[width=0.505\linewidth]{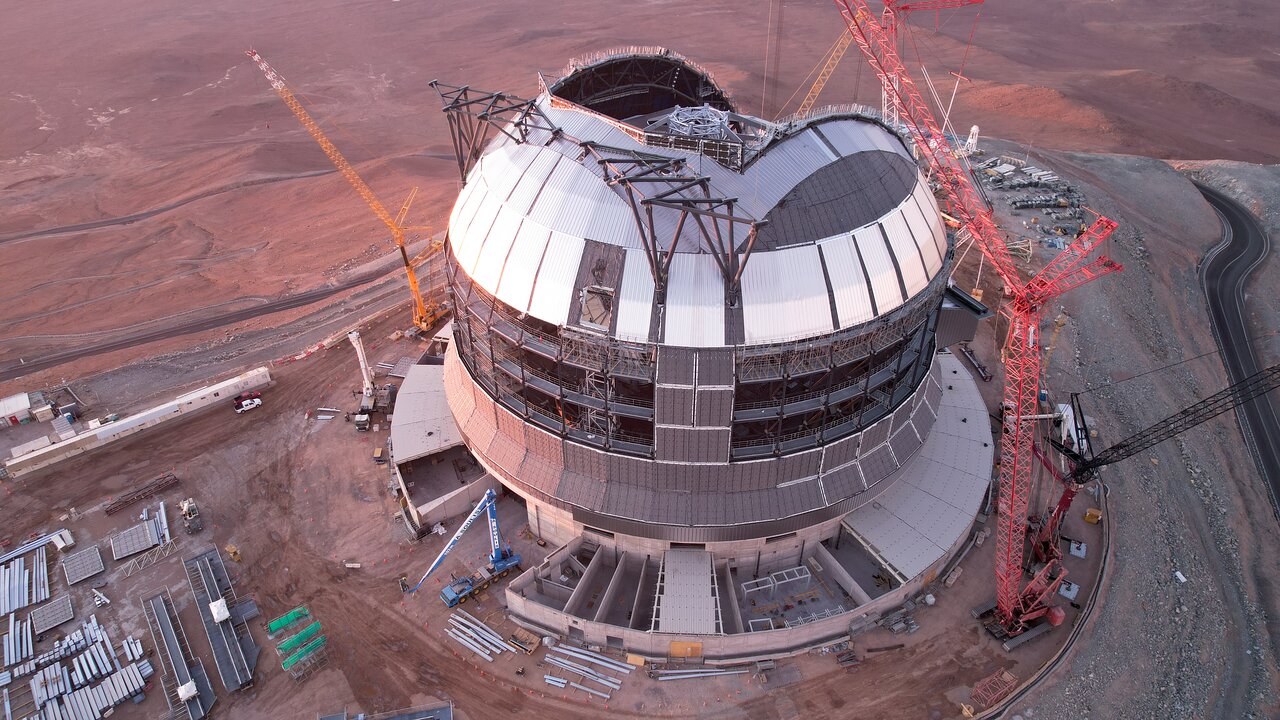}
 \includegraphics[width=0.495\linewidth]{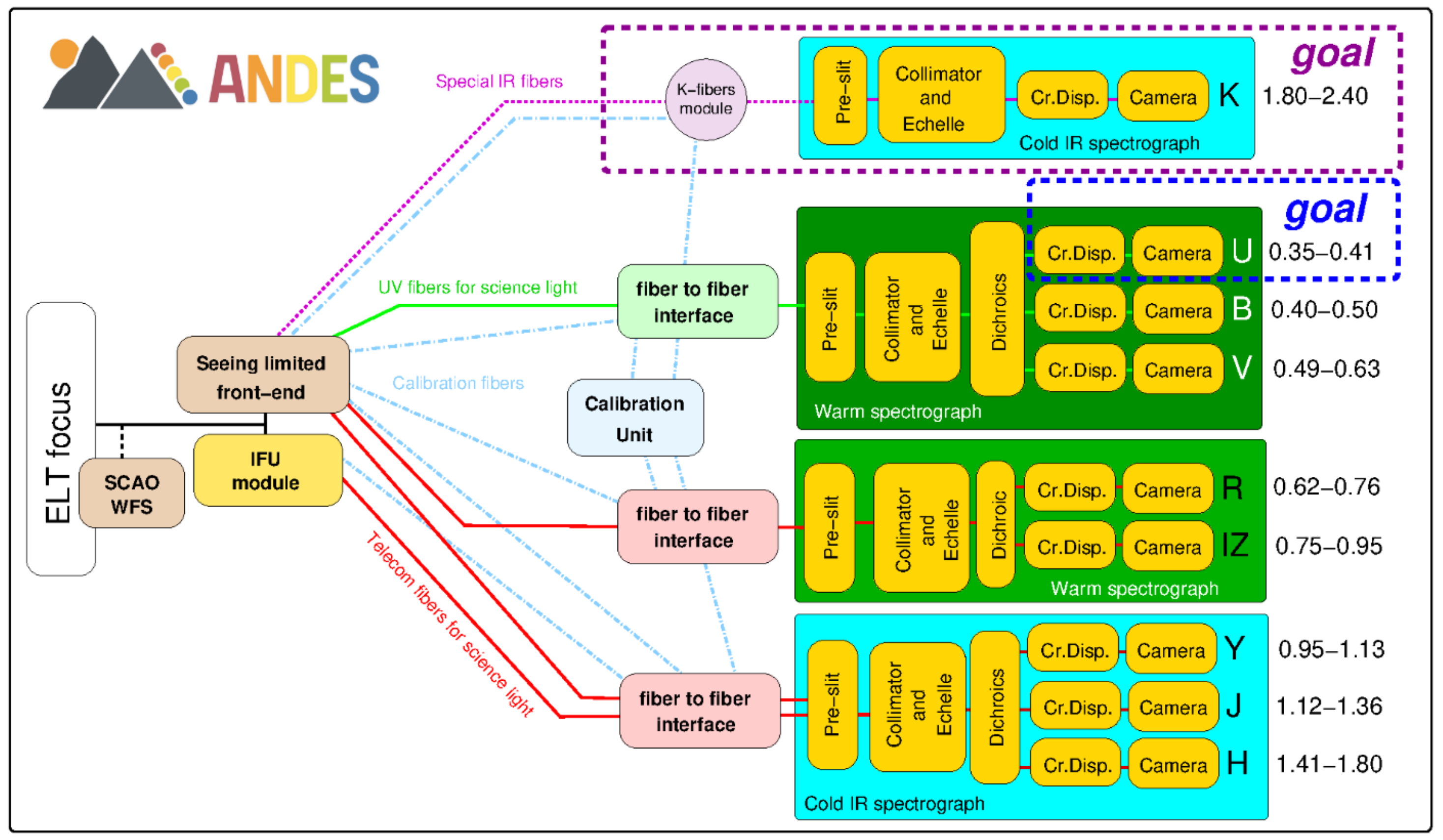}
 \caption{\small
        (Left) Picture of the construction of ELT@ESO (Credit: ESO). (Right) ANDES architectural design {\modiff \citep[from][]{Marconi2024}}.}
 \label{Fig:ANDES}
\end{figure}

The European Southern Observatory (ESO) is currently building a giant 39-m telescope on Cerro Armazones, Chile (see Fig.\ref{Fig:ANDES}), with a {\modif telescope first light planned for 2029}. It will be the largest optical telescope ever built and is likely to remain so for decades to come. The European Extremely Large Telescope (ELT\footnote{\url{https://https://elt.eso.org/}}) will be almost half the length of a soccer pitch in diameter and will gather 15 times more light {\modif per unit of time} than the largest optical telescopes operating today. The telescope has an innovative five-mirror design that includes advanced adaptive optics (AO) to correct for the turbulent atmosphere, giving exceptional image quality. The main mirror will be made up from almost 800 individual hexagonal segments. Thanks to its unprecedented light-collecting power and angular resolution, the ELT {\modif and its suite of planned instruments} will address and revolutionize all major science cases of contemporary astrophysics, including: extrasolar planets, star and planet formation, stellar evolution, compact objects, resolved stellar populations in other galaxies, the physics of high-redshift galaxies, cosmology, and fundamental physics.

\subsection{The ELT-ANDES project}
ANDES (ArmazoNes high-Dispersion Echelle Spectrograph), previously known as HIRES, is the optical/near-IR high-resolution spectrograph for the ELT \citep{Marconi2022, Marconi2024}. It is made of three fiber-fed spectrographs spanning the BV, RIZ and YJH wavelength domains, providing simultaneous coverage between 0.4-1.8 $\mu$m at a spectral resolution R=100,000. As a goal the wavelength coverage could be extended to 0.35-2.4 $\mu$m with the addition of a U-band arm and a K-band spectrograph. ANDES includes both a seeing-limited and diffraction-limited (single-conjugate adaptive optics or SCAO) mode. The seeing-limited mode enables high-efficiency and high-stability spectroscopic observations without the need of an adaptive optics system. The SCAO mode exploits the unprecedented angular resolution of the ELT by using an integral-field unit feeding the YJH (and possibly RIZ) spectrograph. Fig.\ref{Fig:ANDES} shows a the architecture of the instrument. The ANDES project is carried out by the ANDES Consortium, which is composed of 35 institutes from 13 countries and gathers most of the scientific and technical expertise in high-resolution spectroscopy available across Europe.

%- Marconi, A., et al., 2024, ANDES, the high resolution spectrograph for the ELT: science goals, project overview, and future developments, SPIE,  Ground-based and Airborne Instrumentation for Astronomy X, 13096, 1309613\\
%- Marconi, A., et al., 2022, ANDES, the high resolution spectrograph for the ELT: science case, baseline design and path to construction, Ground-based and Airborne Instrumentation for Astronomy IX, 12184, 1218424\\

The main sub-systems of the ANDES instrument are:
\begin{itemize}
\item Seeing-limited front-end (lead: Portugal)
\item SCAO-IFU front-end (lead: Italy)
\item Fiber link (lead: Italy)
\item Calibration unit (lead: Germany)
\item UBV spectrograph (lead: Germany)
\item RIZ spectrograph (lead: Switzerland)
\item YJH spectrograph (lead: United Kingdom)
\item K spectrograph (goal, lead: Germany)
\end{itemize}

$PlanetS$ is a major partner within the ANDES Consortium and is responsible for building the RIZ spectrograph, which is led by the University of Geneva (UNIGE). This is a recognition of the world-leading technical expertise developed by $PlanetS$ in building high-resolution spectrographs, such as HARPS, HARPS-N, ESPRESSO and NIRPS. $PlanetS$ also has key roles in other ANDES sub-systems, namely the calibration unit, the SCAO-IFU front-end, and the data reduction software. The RIZ spectrograph preliminary design is described by \citet{Chazelas2024a} and \citet{Lanotte2024}. This echelle spectrograph will be a warm, vacuum-operated, thermally controlled and fiber-fed instrument. Following the phase A design, the large etendue of the telescope in seeing-limited mode will be reformatted into a long slit made of smaller fibers. Covering wavelengths from 620 to 960 nm, the RIZ spectrograph deals with a recomposed 40-mm-long entrance slit and a pupil anamorphic magnification to overcome the limited (although record-large) size of a mosaic 1.6-m R4 echelle grating. It requires two fast cameras with F/D ratio close to unity. The vacuum vessel and thermal enclosure are described in \citet{SantosDiaz2024}. The detector unit is presented in \citet{Genolet2024}.

%- Chazelas, B., et al., 2024, ANDES, the high-resolution spectrograph for the ELT: RIZ spectrograph preliminary design, Ground-based and Airborne Instrumentation for Astronomy X, 13096, 130964J\\
%- Lanotte, A. A., et al., 2024, ANDES, the high-resolution spectrograph for the ELT: RIZ spectrograph preliminary optical design, SPIE, Ground-based and Airborne Instrumentation for Astronomy X, 13096, 130964F\\
%- Genolet, L., et al., 2024, ANDES, the high-resolution spectrograph for the ELT: RIZ and UBV spectrographs' preliminary design of the detector unit, SPIE, X-Ray, Optical, and Infrared Detectors for Astronomy XI, 13103, 131031V\\
%- Santos Diaz, P., et al., 2024, ANDES, the high-resolution spectrograph for the ELT: RIZ \& UBV spectrographs' preliminary design, analysis, and integration of the vacuum vessel and thermal enclosure, SPIE, Ground-based and Airborne Instrumentation for Astronomy X, 13096, 130964K\\

\subsection{The ELT-ANDES science cases}

During the previous phases of the project, the ANDES Consortium has developed a prioritized list of science cases from which top-level requirements for the instrument have been derived. These are:
\begin{itemize}
\item Priority 1: Exoplanet atmospheres in transmission
\item Priority 2: Variation of the fundamental constants of physics
\item Priority 3: Detection of exoplanet atmospheres in reflected light
\item Priority 4: Sandage test (redshift drift of distant extragalactic sources)
\end{itemize}

As can be seen, exoplanets feature prominently among the top science cases of ANDES. Building on their strong expertise in the field, $PlanetS$ scientists have been involved in the project since the beginning and played a key role in developing these priorities. In the past two years, the ANDES science team has conducted a large exploration of the various science cases for ANDES, culminating in the publication of four white papers covering all fields of astrophysics. These are:
\begin{itemize}
\item “Ground-breaking Exoplanet Science with the ANDES spectrograph at the ELT” \citep{Palle2025} 
\item “The discovery space of ELT-ANDES. Stars and stellar populations” \citep{Roederer2024} 
\item “Galaxy Formation and Symbiotic Evolution with the Inter-Galactic Medium in the Age of ELT-ANDES” \citep{DOdorico2024} 
\item “Cosmology and fundamental physics with the ELT-ANDES spectrograph” \citep{Martins2024} 
\end{itemize}
These white papers demonstrate the potential for ANDES to become a high-demand workhorse instrument at the ELT. We detail below one of the scientific highlights from the exoplanet white paper.

%\subsubsection{Exoplanets and planet formation with ELT-ANDES}
\subsection{Characterization of habitable exoplanets with ELT-ANDES}

The discovery 30 years ago of the first giant planet outside of the solar system \citep{Mayor1995} triggered a revolution in astronomy and led to the award of the 2019 Nobel Prize in physics to Michel Mayor and Didier Queloz (UNIGE). Since the first discoveries we came to know more than 5000 exoplanets today, detected primarily through radial velocity measurements and transit photometry. We now know that exoplanets come with an extreme variety of mass, size, internal structure, temperature, atmospheric composition, and dynamical properties. Although enormous progress has been made in the past three decades, we are still lacking a global picture of how exoplanetary systems form and evolve, and which mechanisms determine exoplanet composition and surface conditions. One of the primary objectives in the next decade will be a thorough investigation of exoplanet atmospheres for a wide range of objects, from gas giants to rocky planets, and from hot to temperate planets.

In the ELT era, one focus of exoplanet science will certainly be on “habitable” terrestrial planets, with the ultimate goal of detecting possible biosignatures. ANDES will be the only approved instrument on the ELT capable of achieving this breakthrough. Although the concept of habitability is far from being well explored, in practice it generally involves the study of solid, rocky objects whose surface temperature enables the presence of liquid water. The role of ANDES on the ELT will be to characterize these worlds well beyond their basic parameters of mass and radius, including their atmospheric structure and composition, as well as surface conditions. This is the most compelling science case for ANDES. In particular, the detection of molecules such as H$_2$O, O$_2$ or CH$_4$ in Earth- or super-Earth sized planets is considered to be truly transformational, as they may be regarded as signatures of habitability (although evidence for the presence of life will require a combination of various spectroscopic signatures, many of which however are included within the ANDES spectral range). This science case defines the minimum top-level requirements (TLRs) for ANDES exoplanet science and, in fact, for the instrument as a whole.

Two aspects make the ELT unique in its ability to study exoplanets: its unprecedented light collecting power and its angular resolution. These two properties lead to two separate applicable techniques: transmission spectroscopy of transiting planets, and the direct detection of the planet reflected light. The TRLs for ANDES enable both, but a major difference between the two is that only the latter relies on the need of an adaptive optics (AO) system. ANDES will perform in-depth observations of the nearest transiting rocky planets (e.g. the TRAPPIST-1 system) as well as the nearest non-transiting planets (e.g. Proxima b), constrain their atmospheric and surface properties, and search for potential biosignatures on these worlds.

\section{RISTRETTO: high-resolution spectroscopy at the diffraction limit of the VLT}
\label{sec:ristretto}

\subsection{The RISTRETTO concept}

In 2016, the discovery of Proxima b clearly marked a milestone in the exoplanet field: the star closest to the Sun, at just 1.3 pc, hosts a temperate rocky planet in its habitable zone \citep{Anglada2016}. Given its proximity, angular separation and favorable contrast with respect to the star, there will never be a “better” potentially habitable planet than Proxima b in terms of observability. This triggered the original idea to develop RISTRETTO, a pioneering experiment for reflected-light exoplanet spectroscopy at the VLT \citep{Lovis2017, Lovis2022, Lovis2024}. Combining an XAO system to a high-resolution spectrograph in the visible, we could show that Proxima b would be amenable to direct detection with an 8-m telescope. RISTRETTO would thus become a pathfinder for this science case, which requires an ensemble of state-of-the-art but existing technologies to be combined together for the first time. This will pave the way towards the development of similar instrumentation at the European ELT, namely ANDES and PCS. Technical specifications for RISTRETTO were derived from the requirement to characterize Proxima b, which serves as the sizing science case for the instrument. However they broadly apply to reflected-light exoplanet spectroscopy in general. They can be summarized as follows:

\begin{itemize}
\item A raw contrast of $<10^{-4}$ at 37 mas from the star
\item A total system throughput $>5\%$ for planets located at 37 mas from the star
\item A spectral resolution $R>100,000$
\item A spectral range $\Delta\lambda/\lambda > 25\%$ over the visible $I$-band
\end{itemize}

\begin{figure}
 \includegraphics[width=\linewidth]{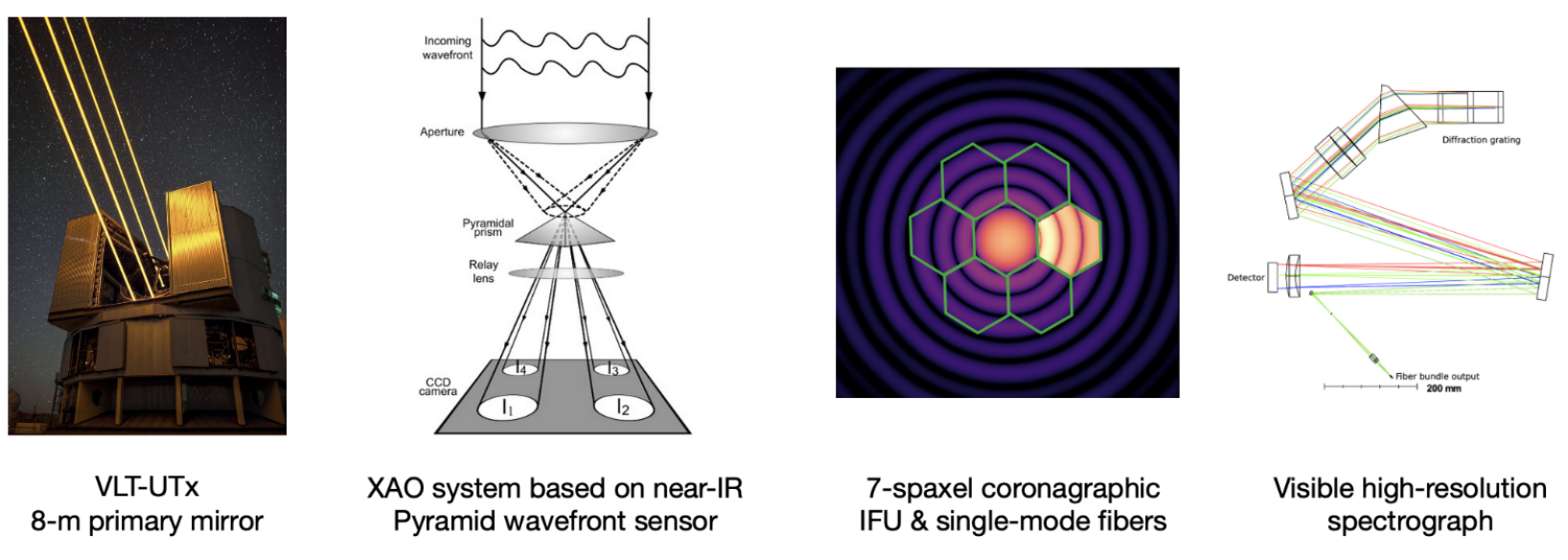}
 \caption{\small
        Schematic of the RISTRETTO instrument \citep[from][]{Lovis2022}}
 \label{Fig:ristretto}
\end{figure}

We thus started to design an instrument that would be made of three essential parts: a high-Strehl XAO system \citep{Blind2024,Shinde2024}, a coronagraphic integral-field unit working at the diffraction limit \citep{Restori2024, Blind2025}, and a visible high-resolution spectrograph covering the 620-840 nm range \citep{Chazelas2024b}. Fig.\ref{Fig:ristretto} shows a schematic view of RISTRETTO. We also developed a RISTRETTO simulator \citep{Bugatti2024} that allows us to explore the feasibility of various science cases, which are briefly described below.

\subsection{Exoplanets in reflected light}

Over the past three decades, RV surveys have discovered many exoplanets orbiting very nearby stars. Some of them reach angular separations at maximum elongation that are large enough to make them resolvable by 8-m-class telescopes in the visible. We used the NASA Exoplanet Archive and the RISTRETTO simulator to derive a target list of exoplanets that RISTRETTO will be able to characterize, opening a new window into the exoplanet population. A dozen of known objects are accessible to RISTRETTO, from cold Jupiters to warm Neptunes to the temperate rocky planet Proxima b. The cold gas giants GJ876b and GJ876c can be detected in just 1-2 hours of observing time. The warm Saturn HD3651b can be detected in about 1 night of integration. The Neptune-mass objects HD192310b, HD102365b and 61 Vir d can be studied in a few nights. The warm super-Earth GJ887c can be reached in less than a night. And Proxima b can be detected in about 5 nights if it has an Earth-like atmosphere. For all these objects, the level of characterization will depend on a priori unknown planet properties such as albedo and atmospheric composition. Broadband and chromatic albedos may be derived from the RISTRETTO detection of the reflected stellar spectral lines. Phase curves may be obtained for planets that can be observed at different phase angles. Molecular absorption by water vapor, methane and oxygen may be detected if present in sufficient quantities. Thus RISTRETTO is by no means a “single-target instrument” but truly has the capability to open a new field in exoplanet studies.

\subsection{Other science cases}

Besides exoplanets in reflected light, RISTRETTO will address a variety of other science cases that will be of interest to the broader planetary science community and beyond. We briefly summarize them below.

\begin{itemize}
\item Accreting protoplanets in H-alpha: the RISTRETTO wavelength range covers the H-alpha line, which is a well-known tracer of accretion in young stars and protoplanets. Thus we will be able to study the spectrally-resolved H-alpha line profile in objects such as PDS70b/c and constrain the geometry and kinematics of the accretion flows onto the planet. Moreover, RISTRETTO will be able to search for new protoplanets at 30-50 mas from young stars, which corresponds to $\sim$3-5 AU at the distance of nearby star-forming regions. This is the typical orbital distance at which most giant planets are thought to form, and thus RISTRETTO will be able to access this crucial parameter space for the first time.
\item Kinematics of protoplanetary disks: RISTRETTO will be able to derive spatially-resolved Doppler velocity maps of protoplanetary disks in scattered light. This capability would usefully complement the sub-mm velocity maps obtained by ALMA. Resolved disk kinematics including localized velocity anomalies linked to forming planets could be probed with RISTRETTO.
\item Spatially-resolved stellar surfaces: with an angular resolution of 19 mas, RISTRETTO will be able to resolve the surfaces of the largest stars in the sky, e.g. Betelgeuse (45 mas in diameter) or Antares (41 mas). It will thus be able to study how spectral lines vary in shape and radial velocity as a function of disk position, and how they evolve with time.
\item Solar System science: RISTRETTO will be able to study local structures at the surfaces of the icy moons of Jupiter and Saturn (e.g. Io, Europa, Titan) with a spatial resolution of 120 km at Jupiter and 240 km at Saturn. It may also measure local wind velocities and chemical abundances in the atmospheres of Uranus and Neptune.
\end{itemize}

\subsection{Status of the RISTRETTO spectrograph}

The RISTRETTO high-resolution spectrograph has been developed as the first part of the instrument, building on $PlanetS$ extensive experience in high-resolution spectroscopy (e.g. HARPS, HARPS-N, ESPRESSO, NIRPS). The spectrograph is a single-mode, fiber-fed spectrograph with 7 available spaxels \citep{Chazelas2024}. It is a classical echelle design, equipped with a deep-depletion 4Kx4K CCD detector. It covers the 620-840 nm wavelength range at a spectral resolution of 140,000. The requirement for this spectrograph is to be stable at the level of 10 {\ms} over 24 hours. We chose to keep the design concept of HARPS, having the spectrometer under vacuum and thermally controlled. Optical fibers are single-mode and the spectrograph is working at the diffraction limit. The spectrograph is self-contained and can be easily moved. It is designed to live on the Nasmyth platform of a VLT and thus to bring little to no disturbance to the environment. The instrument is currently under construction, with the vacuum tank, thermal enclosure and full detector head already completed. All optical components, mounts and the optical bench are close to completion. The whole spectrograph will be ready for laboratory commissioning towards the end of 2025. Once fully tested in the laboratory, it will be possible to validate it on sky using existing adaptive optics systems, e.g. on the Euler 1.2-m telescope in La Silla equipped with the KalAO system, or on the OHP 1.52-m telescope equipped with PAPYRUS.

\subsection{Status of the RISTRETTO XAO and high-contrast front-end}

The RISTRETTO front-end will feed the spectrograph with stellar and planetary light corrected {\modif for} atmospheric effects (turbulence, refraction) and telescope effects (vibrations, low-wind effect). Since no existing solutions have demonstrated appropriate performances for RISTRETTO, new approaches are developed on both the XAO and coronagraphic sides \citep{Blind2024,Blind2025}. The front-end is designed as a stand-alone Nasmyth XAO system, so that it can be installed on any of the 4 VLT UTs. It is based on a 2000-actuator deformable mirror driven by a near-IR Pyramid wavefront sensor. A novel coronagraphic integral-field unit working at the diffraction limit of the telescope has been developed, which meets the performance requirements for RISTRETTO. The combined XAO and high-contrast front-end is planned to be ready for installation at the VLT by 2029.

%- Lovis et al., 2022, RISTRETTO: high-resolution spectroscopy at the diffraction limit of the VLT, SPIE, Ground-based and Airborne Instrumentation for Astronomy IX, 12184, 121841Q\\
%- Blind, N., et al., 2022, RISTRETTO: coronagraph and AO designs enabling High Dispersion Coronagraphy at 2 $\lambda$/D, Adaptive Optics Systems VIII, 12185, 1218573\\
%- Blind, N., et al., 2024, RISTRETTO: a VLT XAO design to reach Proxima Cen b in the visible, arXiv e-prints, arXiv:2409.08052 \\
%- Bugatti, M., et al., 2024, The RISTRETTO simulator: Exoplanet reflected spectra, arXiv e-prints, arXiv:2412.20879 \\
%- Lovis, C., et al., 2024, RISTRETTO: reflected-light exoplanet spectroscopy at the diffraction limit of the VLT, SPIE, Ground-based and Airborne Instrumentation for Astronomy X, 13096, 130961I \\
%- Lovis, C., et al., 2024, RISTRETTO: reflected-light exoplanet spectroscopy at the diffraction limit of the VLT, arXiv e-prints, arXiv:2409.02875 \\

\section{Strategies and challenges for wavelength calibration of high-resolution spectrographs}
\label{sec:wavelength}

The hunt for exoplanets using the radial-velocity method (RV) as well as the search for a possible variation of fundamental physical constants and the redshift drift experiment are key science cases which crucially rely on high quality wavelength calibration of the spectrograph and therefore require sophisticated algorithms and capable calibration light sources.

The method of choice for reliable and stable wavelength calibration is to use hollow-cathode lamps, typically thorium-argon in the visible regime or uranium-neon in the infrared (ThAr/UNe HCL), combined with a passively stabilized white-light Fabry-P\'erot etalon \citep[FP,][]{Wildi2010, Wildi2011, Wildi2012, Cersullo2017}. Here, the atomic transitions of the HCLs provide the absolute wavelength information, while the dense, equally spaced, and equally bright lines of the FP facilitate a clever interpolation in-between the sparse atomic lines of the HCL and allow to derive high-fidelity wavelength solutions for echelle spectrographs \citep[e.g][]{Cersullo2019}. In addition, the FP is also used as simultaneous reference during science exposures to track and correct the intra-night drift of the spectroraph.

\begin{figure}
 \includegraphics[width=\linewidth]{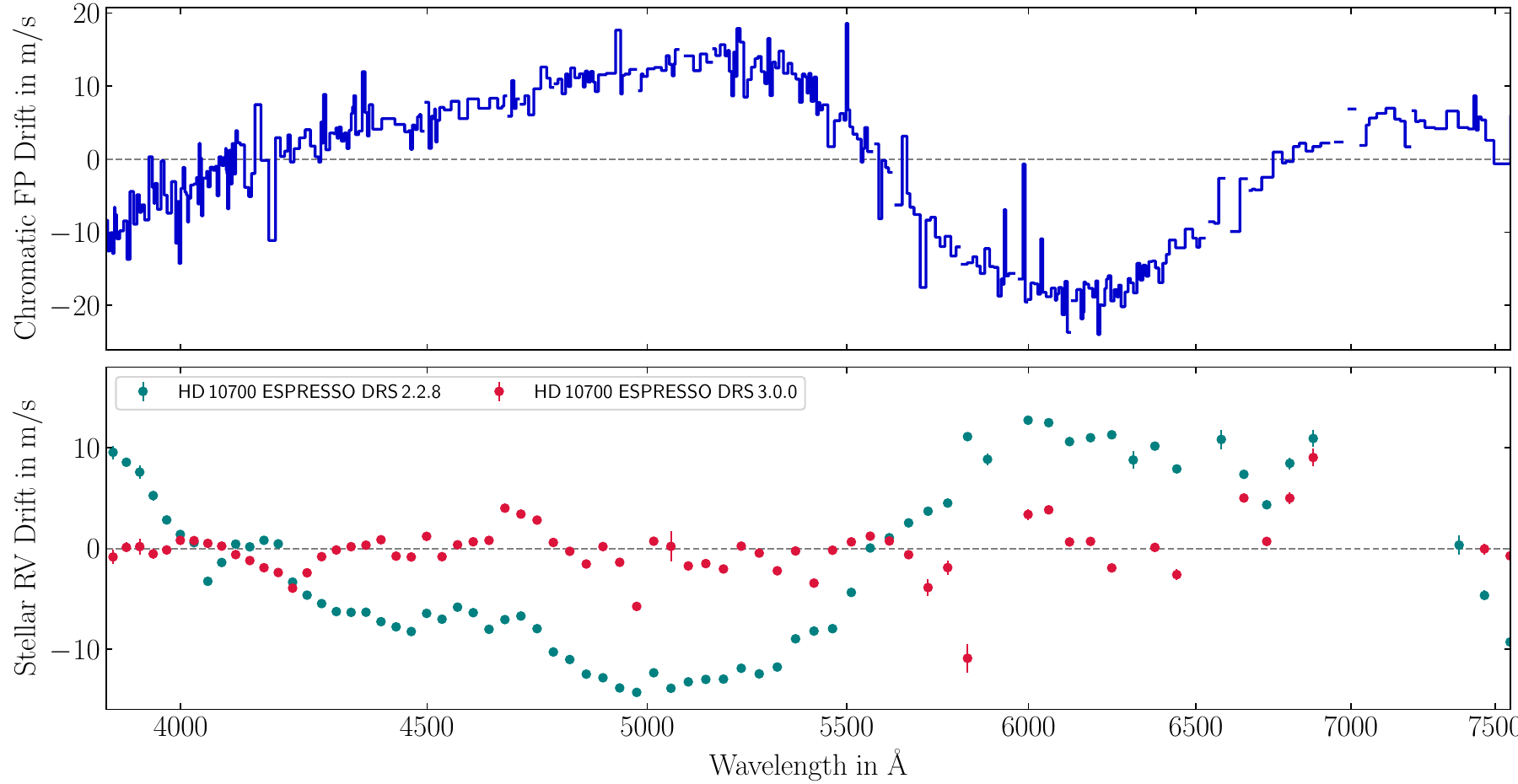}
 \caption{\small
        Chromatic drift of the ESPRESSO FP etalon, as inferred from the ThAr calibration, over a period of 900 days (top).
        Difference of inferred stellar RVs from two nights separated by 1078 days (bottom).
        Using the old data reduction system, which did not yet take this effect into account, the measurements exhibit a chromatic pattern that clearly resembles the inverse of the FP drift. In the new DRS, the FP is fully re-calibrated every day, which completely eliminated this issue.}
 \label{Fig:FP_ChromaticDrift}
\end{figure}

$PlanetS$ has extensive experience in the construction of FP devices, e.g. for SOPHIE, HARPS, SPIROU, NIRPS, and ESPRESSO, where they provide the backbone for high-quality RV observations. To achieve optimal performance, it was however crucial to understand that FPs do not only have a single degree of freedom, the spacing of the two mirrors, but also exhibit a chromatic drift, which can lead to severe inaccuracies after several years if not measured and properly taken into account (see Figure~\ref{Fig:FP_ChromaticDrift}).
As demonstrated in \citet{Schmidt2022}, the ESPRESSO FP exhibits achromatic as well as chromatic drifts at the level of a few {\cms} per day, which both can be calibrated to full satisfaction with just the daily ThAr exposures. The excellent stability of the combined ThAr/FP wavelength calibration has enabled e.g. the discovery of Proxima~Cen~d, the exoplanet with the smallest RV amplitude (39\,\cms) discovered by the RV method so far \citep{Faria2022}.

A completely different type of calibration source are laser frequency combs (LFCs). Derived from ultra-short laser pulses, they provide a dense comb of laser lines with a fixed frequency relation of the form $f_n = f_\mathrm{CEO} + n \cdot f_\mathrm{rep}$. Here, the carrier-envelope offset, $f_\mathrm{CEO}$, and the repetition rate, $f_\mathrm{rep}$, both radio frequencies in the GHz range, can be measured and actively stabilized against an atomic clock or GNSS signal, defining the frequency of each comb line to extreme accuracy ($\lesssim10^{-12}$). LFCs therefore bear the potential to provide the \textit{ideal} spectrograph calibration. However, in practice there are numerous technical challenges. Mode-locked lasers typically operate with pulse repetition rates and therefore line spacings of a few hundred MHz, which is insufficient to be resolved by astronomical high-resolution spectrographs. For these, a line spacing in the range of 15 to 35\,GHz is needed, requiring spectral filtering of the fundamental comb by high-finesse FP cavities. Furthermore, there are no suitable laser-gain media in the optical regime. All combs are therefore set-up over a narrow spectral range in the infrared and brought to the desired wavelength range via frequency conversion and spectral broadening in highly non-linear media at extreme optical power densities ($\gtrsim\mathrm{W{}/{}\mu{}m}^2$). These inherent non-linearities make them prone to instabilities and broadening into the green or even UV range becomes increasingly difficult. LFCs are thus highly complex and expensive devices, which so far often suffer from poor reliability, frequent breakdowns, and have\,--\,despite over a decade long attempts\,--\,seen up to now only limited use for routine calibration of astronomical observations. 
%and not led to particularly noteworthy astrophysical discoveries.

$PlanetS$ has been involved in numerous projects aiming at developing novel technologies for LFCs, in particular electro-optic modulation (EOM) combs \citep{Obrzud2018}. Here, a continuous-wave laser is modulated in phase and amplitude to inscribe a pulse train, which allows to directly produce LFCs with a line spacing up to 35\,GHz and thus eliminates the need for complex mode filtering.
Other efforts were directed towards more advanced spectral broadening techniques, e.g. the use of silicon-nitride optical chips, which provide high non-linearity \citep{Obrzud2019}. Using a combination of spectral broadening and harmonic generation in nano-photonic optical waveguides etched into periodically-poled lithium-niobate wavers, the BLUVES project demonstrated at OHP/SOPHIE for the first time LFC-based calibration of astronomical spectrographs in the near-UV, down to 390\,nm, although with quite significant spectral gaps in between the harmonics \citep{Ludwig2024}. They also highlighted the importance of minimizing phase noise in the optical pulse, particularly for EOM combs, as well as its influence on the intrinsic line with and the diffuse background, which presents a significant nuisance in the analysis of LFC spectra.

\begin{figure}[htb]
  \includegraphics[width=\linewidth]{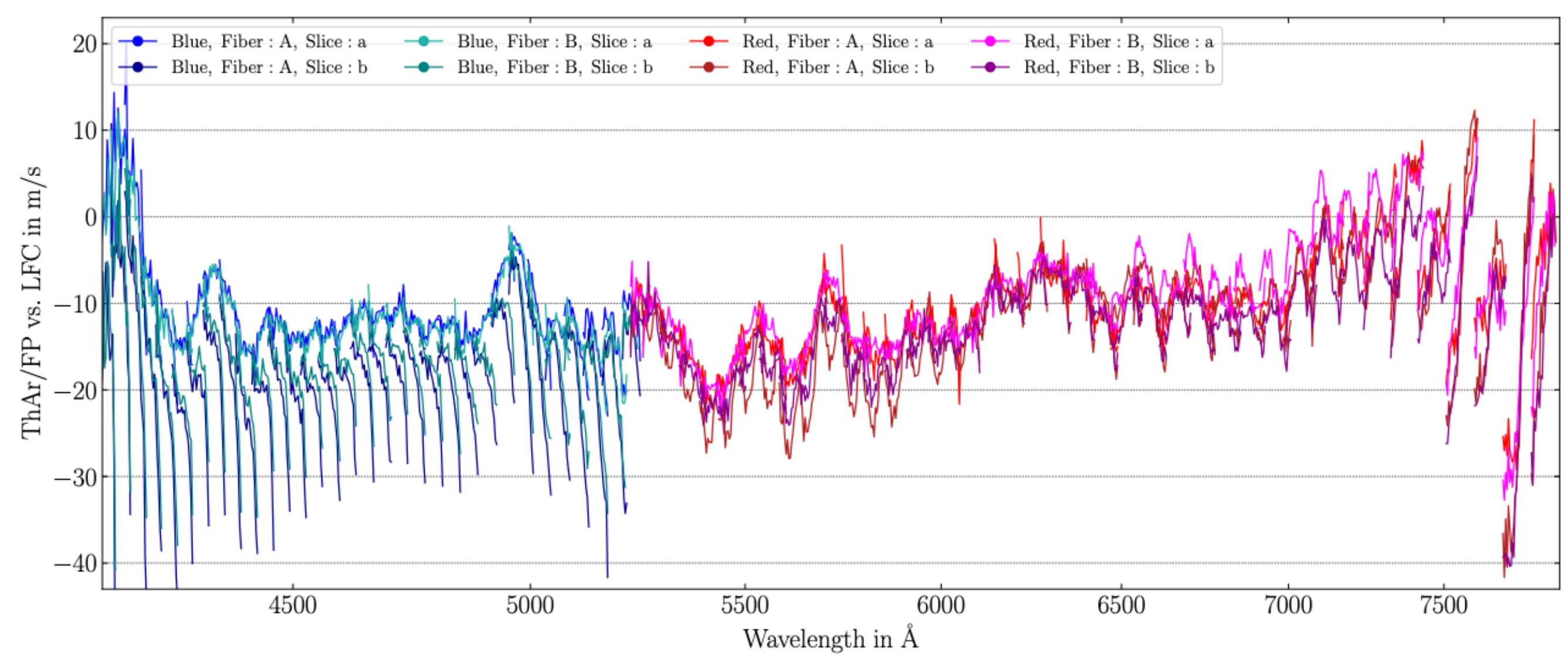}
 %\captionsetup{width=1.2\textwidth}
  \caption{\small
   Comparison between ESPRESSO ThAr/FP and LFC wavelength solutions from {\modiff \citet{Schmidt-Bouchy2024}}.
   Assuming a Gaussian LSF (light colors in the background) leads to significant intra-order modulations and discrepancies between fibers and slices up to $30\,\mathrm{m/s}$, in particular for the blue arm.
   Using the accurate, non-parametric LSF model (strong colors) one achieves an excellent agreement between fibers and slices with instrument-related discrepancies drastically reduced to below a few~m/s. The remaining offset and modulations are systematic and predominantly related to the ThAr lines.}
 \label{Fig:ESPRESSO_ComparisonWavelengthsolutions}
\end{figure}

Given that laser lines are narrow and unresolved, proper treatment of the instrumental line-spread function (LSF) {\modif is mandatory} to achieve accurate and stable wavelength calibration.
A comparison between the ESPRESSO ThAr/FP and LFC wavelength solutions by \citet{Schmidt2021} revealed significant, instrument-related systematics and inconsistencies, many of them correlated with the echelle order structure, up to 40 {\ms}.
Employing a careful, non-parametric model of the instrumental LSF, informed by the LFC which provides a dense forest of truly unresolved lines and therefore reveals the instrumental LSF, it was possible to reduce instrument-related inconsistencies to just a few {\ms} \citep[Figure~\ref{Fig:ESPRESSO_ComparisonWavelengthsolutions},][]{Schmidt2024}.
The complication is though, that an incorporation of the LSF is only possible for analyses that follow a forward-modeling approach.

Another issue is that non-identical fiber-injection geometries between light fed from the telescope and the calibration unit can lead to different LSFs on the detector and therefore systematics in the wavelength calibration \citep{Chazelas2012}.
%
%For ESPRESSO, \citet{Schmidt2025} tested this by temporarily installing an iodine absorption cell in the beam between telescope and spectrograph.
For ESPRESSO, this was tested by temporarily installing an iodine absorption cell in the beam between telescope and spectrograph \citep{Schmidt2025}. 
Observing featureless stars through the cell and comparing to an analytic model of the iodine transitions allowed to validate the existing ESPRESSO wavelength calibration and demonstrate excellent accuracy, better than 5\,\ms.
However, it was also shown that varying illumination geometries can cause significant systematics up to 10\,\ms.
With the same {\modif iodine cell}, RV stability of ESPRESSO observations at a level $\lesssim$20\,\cms could be demonstrated, as well in an end-to-end fashion, from sky to data product.

Another concept to tackle non-common path effects, is {\modif proposed by the $\nu$ANCESTOR concept \citep{Pepe2024}. It consists of embarking an optical frequency comb on-board a satellite equipped with an actively pointing telescope and precision orbitography. This calibration satellite shall be available and serve extremely precise radial velocity spectrographs in all major observatories around the world.}

%\section{Wavelength Calibration challenges and Laser Frequency Comb [TSc]}
%Wavelength calibration challenges \\
%Standard approach combining Hollow-Cathod + Fabry-Pérot \\
%Limitations and challenges of LFC \\
%- Obrzud, E. et al.., 2018, OExpr, 26, 34830: Broadband near-infrared astronomical spectrometer calibration and on-sky validation with an electro-optic laser frequency comb\\
%- Obrzud, E., et al.., 2019, NaPho, 13, 31: A microphotonic astrocomb \\
%- Schmidt et al., 2021, A\&A, 646, 144: Fundamental physics with ESPRESSO: Towards an accurate wavelength calibration for a precision test of the fine-structure constant \\
%- Schmidt, T. M. et al., 2022, Astronomy and Astrophysics, 664, A191, Chromatic drift of the Espresso Fabry-Pérot etalon \\
%- Schmidt. T. \& Bouchy, F., 2024, MNRAS, 530, 1252 : Characterization of the ESPRESSO line-spread function and improvement of the wavelength calibration accuracy \\
%- Ludwig, et al., 2024, Nature Communication, 15, 7614 : Ultraviolet astronomical spectrograph calibration with laser frequency combs from nanophotonic lithium niobate waveguides \\
%- Pepe, F., et al., 2024, $\nu$Ancestor: an artificial satellite-borne star for accurate frequency calibration of ground-based EPRV spectrographs, SPIE, Advances in Optical and Mechanical Technologies for Telescopes and Instrumentation VI, 13100, 131005W\\
%- Schmidt, T. M., 2024, New techniques for accurate and stable wavelength calibration, SPIE, Advances in Optical and Mechanical Technologies for Telescopes and Instrumentation VI, 13100, 131004P\\

\section{Solar telescope and stellar activity mitigation}
\label{sec:activity}

\begin{figure}
 \includegraphics[width=\linewidth]{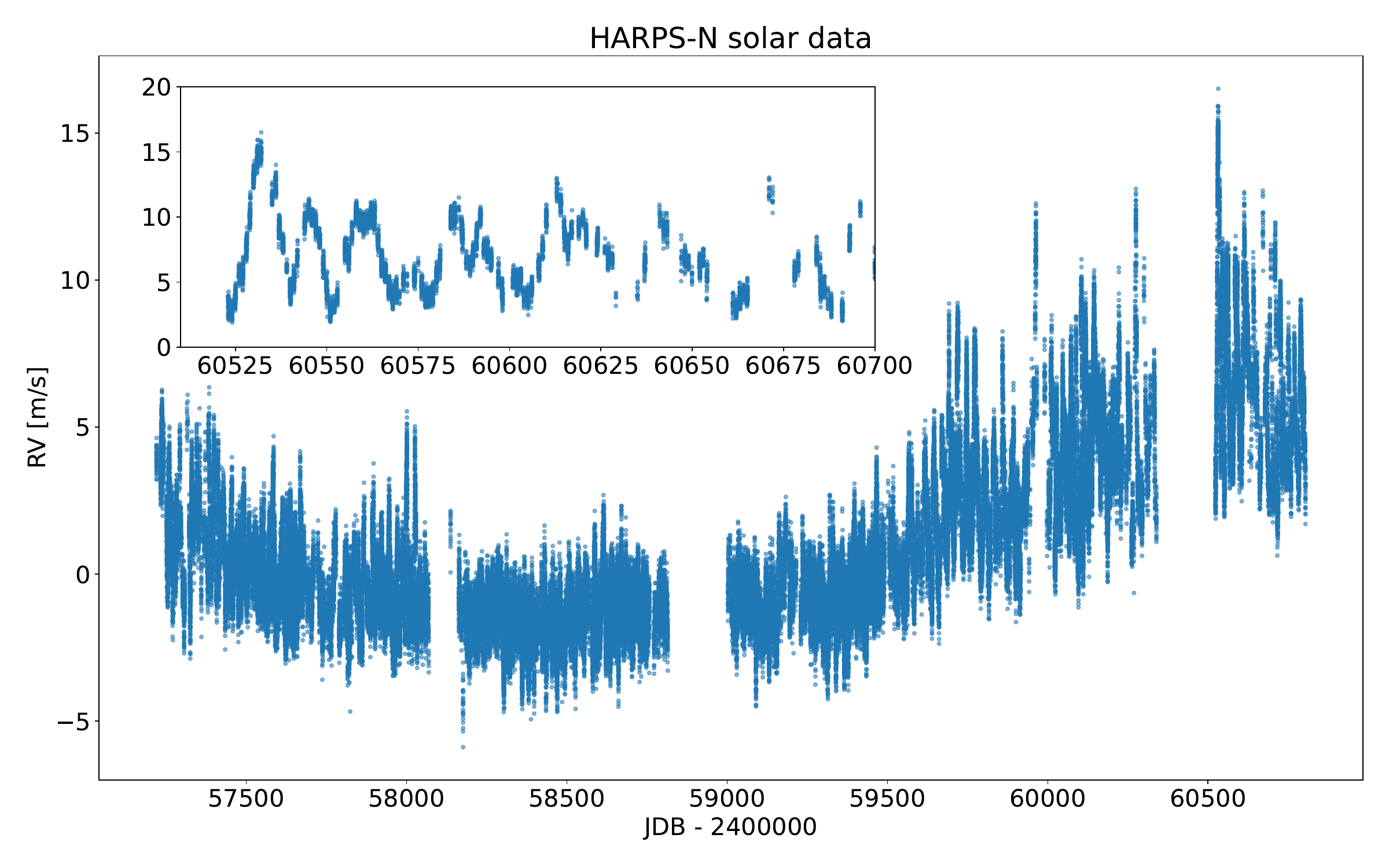}
 \caption{\small
        10-years of HARPS-N solar extremely precise RVs in the heliocentric rest-frames. The long-term variation is induced by the solar 11-year magnetic cycle.}
 \label{Fig:solar_RVs}
\end{figure}

Due to stellar signal timescales ranging from minutes for
stellar oscillations to years for magnetic cycles, standard radial
velocity observations obtained either from transit follow-up
observations or blind search surveys very often lack the sampling and time baselines necessary to characterize the diﬀerent types of stellar signals. The lack of good data makes it extremely
challenging to understand stellar signals in enough detail and, therefore, to find efficient techniques to mitigate them. The ideal
solution to make progress is to continuously observe a target
for which astrophysical parameters and activity behavior as
a function of time are perfectly known, with the same instrument employed for exoplanet characterization. This reflection
led to the development of the HARPS-N low-cost solar telescope \citep[][]{Dumusque-2015b, Phillips:2016aa}. Its main feature is the use of an integrating sphere, which allows to strongly mitigate guiding issues and inject an uniform integrated light of the solar disc into a fiber that is then routed to the calibration unit of HARPS-N. As seen in Fig.~\ref{Fig:solar_RVs} this solar telescope started operating in July 2015 and has now obtained 10 years of nearly continuous data, with several hours of observations per day \citep[][Dumusque et al. 2025 in prep.]{Collier-Cameron:2019aa, Dumusque:2021aa,AlMoulla:2023aa}. Attempts to observe the Sun with a lens directly focusing the solar disk into an optical fiber showed clearly guiding limitations at the {\ms} level \citep[e.g.][]{Lemke-2016}. Since this first experiment, most of the extreme-precision spectrographs around the world have been or will be equipped with a similar solar feed, all using the solution of an integrating sphere. Here is a non-exhaustive list of existing solar telescopes: HELIOS feeding sunlight into HARPS and more recently NIRPS \citep{Bouchy2025}, the NEID solar telescope \citep[][]{Lin:2022aa}, the LOST telescope feeding sunlight into EXPRES \citep[][]{Llama:2022aa}, SoCal the KPF solar telescope \citep[][]{Rubenzahl:2023aa} and in the near future the POET solar telescope feeding integrated-disc but also disc-resolved images of the Sun into ESPRESSO \citep[][]{Santos:2023aa} and the ABORAS solar telescope capable of observing the Sun in spectro-polarimetry using HARPS-3 \citep[][]{Farret-Jentink:2022aa}. We note that in \citet{Zhao:2023ab}, the authors performed a detailed comparison of the solar data obtained over one month from HARPS-N, HARPS, NEID and EXPRES and found an excellent agreement between all the datasets, {\modif with residual intraday scatter of only 15-30 \cms}.

The obtained HARPS-N solar high-fidelity spectra and high-precision radial-velocities were already used in many different studies to understand better stellar signals and to develop new efficient mitigation techniques. On the front of a better characterisation of stellar magnetic activity, several studies looked at different types of activity indicators and how they correlate with the RVs \citep[][]{Maldonado:2019aa, Haywood:2020aa, Thompson:2020aa, Lienhard:2023aa,Cretignier:2024aa}. Comparison between magnetic activity simulations {\modif of the Sun}, informed by the known type of active regions and position on the solar disc, and HARPS-N spectral and RV data also bring a much deeper understanding of the origin and link between activity and observed RV signal \citep[][]{Milbourne:2019aa, Meunier:2024aa,Klein:2024aa,Zhao:2025aa}. On the side of granulation characterisation,  \citet{AlMoulla:2023aa}, \citet{Lakeland:2024aa} and {\modif \citet{Sullivan2025}} made also significant progress.

The HARPS-N solar telescope data were also used to develop new mitigation techniques for stellar signals. From a pure data-driven approach, the SCALPEL \citep[][]{Collier-Cameron:2021aa} and YARARA \citep[][]{Cretignier:2021aa} frameworks were developed and optimized thanks to the HARPS-N solar data. Those same data, due to the significant number of measurements, was also used to develop neural-network based approaches, as published in \citet{Beurs:2022aa} and \citet{Zhao:2024cc}, where the authors demonstrate that planets as small as 20 cm/s could be detected.

Finally, not expected at the beginning, the data from a solar telescope can be used to probe instrumental systematics through time. If spectrograph calibrations are not tracking perfectly the instrument instabilities, any stellar data will be affected. Those effects might be difficult to track on stars due to the uneven and sparse sampling, however, can be easily spotted on nearly continuous solar data. A solar telescope can therefor be seen as a calibrator for high-resolution spectrographs aiming at sub-m/s precision. Although this is a path of research that barely started, results presented in \citet{Dumusque:2021aa}, \citet{Cretignier:2021aa}, \citet{Ford:2024aa} and Dumusque et al. (2025, in prep) show the potential of solar data for characterising better the spectrographs that are used in {\modif EPRV} instruments.

\section{Conclusion and future prospects}

ESPRESSO@VLT and HARPS+NIRPS@3.6m offer the European astronomical community high-resolution, high-fidelity, and high-precision wide-range spectroscopy. Their performance opened up a new parameter space in ground-based observations that can address a large variety of scientific cases, especially in exoplanet studies. ANDES@ELT will give access to unprecedented high-efficiency spectroscopic observations.  

The discovery of temperate Earth-like planets and the search for biomarkers in their atmospheres are some of the main objectives of XXIst century’s astronomy. The primary goal of the PLATO (PLAnetary Transits and Oscillations of stars) mission \citep{Rauer2025}, which is set to be launched in end 2026, is to open a new avenue in exoplanetary science by detecting terrestrial exoplanets up to the temperate zone and characterizing their bulk properties. PLATO will provide the key information (radii, bulk densities, insolation, architecture as well as the age) needed to determine the habitability of these unexpectedly diverse new worlds. The first PLATO field (LOPS2) was recently selected in the southern hemisphere with a significant overlap with the TESS South Continuous Viewing Zone. The simulated PLATO planet yield \citep{Matuszewski2023} predicts that PLATO will detect several hundreds of temperate Earths, super-Earths, and sub-Neptunes, with one-third of them transiting stars brighter than V=11. One important challenge will be to reach and secure the RV precision requested to measure the mass of temperate exoplanets revealed by PLATO. For this purpose, HARPS+NIRPS and ESPRESSO will be key players, providing reliable and accurate long-term wavelength calibration and optimal correction for stellar activity. ANDES and RISTRETTO will give access to unique detailed characterization of temperate exoplanet atmospheres.

%\begin{acknowledgement}
%\end{acknowledgement}
%

%Custom format ADS
%    - %2i, %Y, %T, %J, %V, %p \\\

\bibliographystyle{spbasic}
\footnotesize
\bibliography{bib.bib}

@INCOLLECTION{Pepe2018,
       author = {{Pepe}, Francesco and {Bouchy}, Fran{\c{c}}ois and {Mayor}, Michel and {Udry}, St{\'e}phane},
        title = "{High-Precision Spectrographs for Exoplanet Research: CORAVEL, ELODIE, CORALIE, SOPHIE and HARPS}",
    booktitle = {Handbook of Exoplanets},
         year = 2018,
       editor = {{Deeg}, Hans J. and {Belmonte}, Juan Antonio},
          eid = {190},
        pages = {190},
          doi = {10.1007/978-3-319-55333-7_190},
       adsurl = {https://ui.adsabs.harvard.edu/abs/2018haex.bookE.190P}
}

@ARTICLE{LoCurto2015,
       author = {{Lo Curto}, G. and {Pepe}, F. and {Avila}, G. and {Boffin}, H. and {Bovay}, S. and {Chazelas}, B. and {Coffinet}, A. and {Fleury}, M. and {Hughes}, I. and {Lovis}, C. and {Maire}, C. and {Manescau}, A. and {Pasquini}, L. and {Rihs}, S. and {Sinclaire}, P. and {Udry}, S.},
        title = "{HARPS Gets New Fibres After 12 Years of Operations}",
      journal = {The Messenger},
         year = 2015,
        month = dec,
       volume = {162},
        pages = {9-15},
       adsurl = {https://ui.adsabs.harvard.edu/abs/2015Msngr.162....9L}
}

@ARTICLE{LoCurto2024,
       author = {{Lo Curto}, G. and {Pepe}, F. and {Fleury}, M. and {Hughes}, I. and {Schnell}, R. and {Wimmer}, V. and {Fuerte Rodriguez}, P.~A. and {Silber}, A. and {Saviane}, I. and {Lovis}, C. and {Sosnowska}, D. and {Dumusque}, X. and {Jakob}, G. and {Accardo}, M.},
        title = "{HARPS at 20: Evolving Through Continuous Improvements}",
      journal = {The Messenger},
         year = 2024,
        month = mar,
       volume = {192},
        pages = {38-40},
          doi = {10.18727/0722-6691/5354},
       adsurl = {https://ui.adsabs.harvard.edu/abs/2024Msngr.192...38C}
}

@ARTICLE{Pepe2014,
       author = {{Pepe}, F. and {Molaro}, P. and {Cristiani}, S. and {Rebolo}, R. and {Santos}, N.~C. and {Dekker}, H. and {M{\'e}gevand}, D. and {Zerbi}, F.~M. and {Cabral}, A. and {Di Marcantonio}, P. and {Abreu}, M. and {Affolter}, M. and {Aliverti}, M. and {Allende Prieto}, C. and {Amate}, M. and {Avila}, G. and {Baldini}, V. and {Bristow}, P. and {Broeg}, C. and {Cirami}, R. and {Coelho}, J. and {Conconi}, P. and {Coretti}, I. and {Cupani}, G. and {D'Odorico}, V. and {De Caprio}, V. and {Delabre}, B. and {Dorn}, R. and {Figueira}, P. and {Fragoso}, A. and {Galeotta}, S. and {Genolet}, L. and {Gomes}, R. and {Gonz{\'a}lez Hern{\'a}ndez}, J.~I. and {Hughes}, I. and {Iwert}, O. and {Kerber}, F. and {Landoni}, M. and {Lizon}, J. -L. and {Lovis}, C. and {Maire}, C. and {Mannetta}, M. and {Martins}, C. and {Monteiro}, M. and {Oliveira}, A. and {Poretti}, E. and {Rasilla}, J.~L. and {Riva}, M. and {Santana Tschudi}, S. and {Santos}, P. and {Sosnowska}, D. and {Sousa}, S. and {Span{\'o}}, P. and {Tenegi}, F. and {Toso}, G. and {Vanzella}, E. and {Viel}, M. and {Zapatero Osorio}, M.~R.},
        title = "{ESPRESSO: The next European exoplanet hunter}",
      journal = {Astronomische Nachrichten},
         year = 2014,
        month = jan,
       volume = {335},
       number = {1},
        pages = {8},
          doi = {12.1002/asna.201312004},
       adsurl = {https://ui.adsabs.harvard.edu/abs/2014AN....335....8P}
}

@INPROCEEDINGS{Megevand2014,
       author = {{M{\'e}gevand}, Denis and {Zerbi}, Filippo M. and {Di Marcantonio}, Paolo and {Cabral}, Alexandre and {Riva}, Marco and {Abreu}, Manuel and {Pepe}, Francesco and {Cristiani}, Stefano and {Rebolo Lopez}, Rafael and {Santos}, Nuno C. and {Dekker}, Hans and {Aliverti}, Matteo and {Allende Prieto}, Carlos and {Amate}, Manuel and {Avila}, Gerardo and {Baldini}, Veronica and {Bandy}, Timothy and {Bristow}, Paul and {Broeg}, Christopher and {Cirami}, Roberto and {Coelho}, Jo{\~a}o. and {Conconi}, Paolo and {Coretti}, Igor and {Cupani}, Guido and {D'Odorico}, Valentina and {De Caprio}, Vincenzo and {Delabre}, Bernard and {Dorn}, Reinhold and {Figueira}, Pedro and {Fragoso}, Ana and {Galeotta}, Samuele and {Genolet}, Ludovic and {Gomes}, Ricardo and {Gonz{\'a}lez Hern{\'a}ndez}, Jonay and {Hughes}, Ian and {Iwert}, Olaf and {Kerber}, Florian and {Landoni}, Marco and {Lizon}, Jean-Louis and {Lovis}, Christophe and {Maire}, Charles and {Mannetta}, Marco and {Martins}, Carlos C.~J.~A.~P. and {Molaro}, Paolo and {Monteiro}, Manuel A.~S. and {Moschetti}, Manuele and {Oliveira}, Antonio and {Zapatero Osorio}, Maria Rosa and {Poretti}, Ennio and {Rasilla}, Jos{\'e} L. and {Santana Tschudi}, Samuel and {Santos}, Pedro and {Sosnowska}, Danuta and {Sousa}, S{\'e}rgio and {Tenegi}, Fabio and {Toso}, Giorgio and {Vanzella}, Eros and {Viel}, Matteo},
        title = "{ESPRESSO: the radial velocity machine for the VLT}",
    booktitle = {Ground-based and Airborne Instrumentation for Astronomy V},
         year = 2014,
       editor = {{Ramsay}, Suzanne K. and {McLean}, Ian S. and {Takami}, Hideki},
       series = {Society of Photo-Optical Instrumentation Engineers (SPIE) Conference Series},
       volume = {9147},
        month = jul,
          eid = {91471H},
        pages = {91471H},
          doi = {10.1117/12.2055816},
       adsurl = {https://ui.adsabs.harvard.edu/abs/2014SPIE.9147E..1HM}
}

@INCOLLECTION{Gonzalez2018,
       author = {{Gonz{\'a}lez Hern{\'a}ndez}, Jonay I. and {Pepe}, Francesco and {Molaro}, Paolo and {Santos}, Nuno C.},
        title = "{ESPRESSO on VLT: An Instrument for Exoplanet Research}",
    booktitle = {Handbook of Exoplanets},
         year = 2018,
       editor = {{Deeg}, Hans J. and {Belmonte}, Juan Antonio},
          eid = {157},
        pages = {157},
          doi = {10.1007/978-3-319-55333-7_15710.1007/978-3-319-30648-3-157-1},
       adsurl = {https://ui.adsabs.harvard.edu/abs/2018haex.bookE.157G}
}

@INPROCEEDINGS{Calderone2018,
       author = {{Calderone}, Giorgio and {Baldini}, Veronica and {Cirami}, Roberto and {Coretti}, Igor and {Cristiani}, Stefano and {Di Marcantonio}, Paolo and {M{\'e}gevand}, Denis},
        title = "{ESPRESSO instrument control software and electronics: commissioning in Paranal}",
    booktitle = {Software and Cyberinfrastructure for Astronomy V},
         year = 2018,
       editor = {{Guzman}, Juan C. and {Ibsen}, Jorge},
       series = {Society of Photo-Optical Instrumentation Engineers (SPIE) Conference Series},
       volume = {10707},
        month = jul,
          eid = {107072G},
        pages = {107072G},
          doi = {10.1117/12.2312736},
       adsurl = {https://ui.adsabs.harvard.edu/abs/2018SPIE10707E..2GC}
}

@INPROCEEDINGS{Baldini2016,
       author = {{Baldini}, V. and {Calderone}, G. and {Cirami}, R. and {Coretti}, I. and {Cristiani}, S. and {Di Marcantonio}, P. and {M{\'e}gevand}, D. and {Riva}, M. and {Santin}, P.},
        title = "{Integration of the instrument control electronics for the ESPRESSO spectrograph at ESO-VLT}",
    booktitle = {Software and Cyberinfrastructure for Astronomy IV},
         year = 2016,
       editor = {{Chiozzi}, Gianluca and {Guzman}, Juan C.},
       series = {Society of Photo-Optical Instrumentation Engineers (SPIE) Conference Series},
       volume = {9913},
        month = jul,
          eid = {99132H},
        pages = {99132H},
          doi = {10.1117/12.2231588},
       adsurl = {https://ui.adsabs.harvard.edu/abs/2016SPIE.9913E..2HB}
}

@INPROCEEDINGS{Calderone2016,
       author = {{Calderone}, G. and {Baldini}, V. and {Cirami}, R. and {Coretti}, I. and {Cristiani}, S. and {Di Marcantonio}, P. and {Landoni}, M. and {M{\'e}gevand}, D. and {Riva}, M. and {Santin}, P.},
        title = "{The technical CCDs in ESPRESSO: usage, performances, and network requirements}",
    booktitle = {Software and Cyberinfrastructure for Astronomy IV},
         year = 2016,
       editor = {{Chiozzi}, Gianluca and {Guzman}, Juan C.},
       series = {Society of Photo-Optical Instrumentation Engineers (SPIE) Conference Series},
       volume = {9913},
        month = jul,
          eid = {99132K},
        pages = {99132K},
          doi = {10.1117/12.2231650},
       adsurl = {https://ui.adsabs.harvard.edu/abs/2016SPIE.9913E..2KC}
}

@ARTICLE{Mayor1995,
       author = {{Mayor}, Michel and {Queloz}, Didier},
        title = "{A Jupiter-mass companion to a solar-type star}",
      journal = {\nat},
         year = 1995,
        month = nov,
       volume = {378},
       number = {6555},
        pages = {355-359},
          doi = {10.1038/378355a0},
       adsurl = {https://ui.adsabs.harvard.edu/abs/1995Natur.378..355M}
}

@ARTICLE{Anglada2016,
       author = {{Anglada-Escud{\'e}}, Guillem and {Amado}, Pedro J. and {Barnes}, John and {Berdi{\~n}as}, Zaira M. and {Butler}, R. Paul and {Coleman}, Gavin A.~L. and {de La Cueva}, Ignacio and {Dreizler}, Stefan and {Endl}, Michael and {Giesers}, Benjamin and {Jeffers}, Sandra V. and {Jenkins}, James S. and {Jones}, Hugh R.~A. and {Kiraga}, Marcin and {K{\"u}rster}, Martin and {L{\'o}pez-Gonz{\'a}lez}, Mar{\'\i}a J. and {Marvin}, Christopher J. and {Morales}, Nicol{\'a}s and {Morin}, Julien and {Nelson}, Richard P. and {Ortiz}, Jos{\'e} L. and {Ofir}, Aviv and {Paardekooper}, Sijme-Jan and {Reiners}, Ansgar and {Rodr{\'\i}guez}, Eloy and {Rodr{\'\i}guez-L{\'o}pez}, Cristina and {Sarmiento}, Luis F. and {Strachan}, John P. and {Tsapras}, Yiannis and {Tuomi}, Mikko and {Zechmeister}, Mathias},
        title = "{A terrestrial planet candidate in a temperate orbit around Proxima Centauri}",
      journal = {\nat},
         year = 2016,
        month = aug,
       volume = {536},
       number = {7617},
        pages = {437-440},
          doi = {10.1038/nature19106},
archivePrefix = {arXiv},
       eprint = {1609.03449},
 primaryClass = {astro-ph.EP},
       adsurl = {https://ui.adsabs.harvard.edu/abs/2016Natur.536..437A}
}

@ARTICLE{Hojjatpanah2019,
       author = {{Hojjatpanah}, S. and {Figueira}, P. and {Santos}, N.~C. and {Adibekyan}, V. and {Sousa}, S.~G. and {Delgado-Mena}, E. and {Alibert}, Y. and {Cristiani}, S. and {Gonz{\'a}lez Hern{\'a}ndez}, J.~I. and {Lanza}, A.~F. and {Di Marcantonio}, P. and {Martins}, J.~H.~C. and {Micela}, G. and {Molaro}, P. and {Neves}, V. and {Oshagh}, M. and {Pepe}, F. and {Poretti}, E. and {Rojas-Ayala}, B. and {Rebolo}, R. and {Su{\'a}rez Mascare{\~n}o}, A. and {Zapatero Osorio}, M.~R.},
        title = "{Catalog for the ESPRESSO blind radial velocity exoplanet survey}",
      journal = {\aap},
         year = 2019,
        month = sep,
       volume = {629},
          eid = {A80},
        pages = {A80},
          doi = {10.1051/0004-6361/201834729},
archivePrefix = {arXiv},
       eprint = {1908.04627},
 primaryClass = {astro-ph.EP},
       adsurl = {https://ui.adsabs.harvard.edu/abs/2019A&A...629A..80H}
}

@ARTICLE{Snellen2008,
       author = {{Snellen}, I.~A.~G. and {Albrecht}, S. and {de Mooij}, E.~J.~W. and {Le Poole}, R.~S.},
        title = "{Ground-based detection of sodium in the transmission spectrum of exoplanet HD 209458b}",
      journal = {\aap},
         year = 2008,
        month = aug,
       volume = {487},
       number = {1},
        pages = {357-362},
          doi = {10.1051/0004-6361:200809762},
archivePrefix = {arXiv},
       eprint = {0805.0789},
 primaryClass = {astro-ph},
       adsurl = {https://ui.adsabs.harvard.edu/abs/2008A&A...487..357S}
}

@ARTICLE{Redfield2008,
       author = {{Redfield}, Seth and {Endl}, Michael and {Cochran}, William D. and {Koesterke}, Lars},
        title = "{Sodium Absorption from the Exoplanetary Atmosphere of HD 189733b Detected in the Optical Transmission Spectrum}",
      journal = {\apjl},
         year = 2008,
        month = jan,
       volume = {673},
       number = {1},
        pages = {L87},
          doi = {10.1086/527475},
archivePrefix = {arXiv},
       eprint = {0712.0761},
 primaryClass = {astro-ph},
       adsurl = {https://ui.adsabs.harvard.edu/abs/2008ApJ...673L..87R}
}

@ARTICLE{Charbonneau2002,
       author = {{Charbonneau}, David and {Brown}, Timothy M. and {Noyes}, Robert W. and {Gilliland}, Ronald L.},
        title = "{Detection of an Extrasolar Planet Atmosphere}",
      journal = {\apj},
         year = 2002,
        month = mar,
       volume = {568},
       number = {1},
        pages = {377-384},
          doi = {10.1086/338770},
archivePrefix = {arXiv},
       eprint = {astro-ph/0111544},
 primaryClass = {astro-ph},
       adsurl = {https://ui.adsabs.harvard.edu/abs/2002ApJ...568..377C}
}

@ARTICLE{Ehrenreich2020,
       author = {{Ehrenreich}, David and {Lovis}, Christophe and {Allart}, Romain and {Zapatero Osorio}, Mar{\'\i}a Rosa and {Pepe}, Francesco and {Cristiani}, Stefano and {Rebolo}, Rafael and {Santos}, Nuno C. and {Borsa}, Francesco and {Demangeon}, Olivier and {Dumusque}, Xavier and {Gonz{\'a}lez Hern{\'a}ndez}, Jonay I. and {Casasayas-Barris}, N{\'u}ria and {S{\'e}gransan}, Damien and {Sousa}, S{\'e}rgio and {Abreu}, Manuel and {Adibekyan}, Vardan and {Affolter}, Michael and {Allende Prieto}, Carlos and {Alibert}, Yann and {Aliverti}, Matteo and {Alves}, David and {Amate}, Manuel and {Avila}, Gerardo and {Baldini}, Veronica and {Bandy}, Timothy and {Benz}, Willy and {Bianco}, Andrea and {Bolmont}, {\'E}meline and {Bouchy}, Fran{\c{c}}ois and {Bourrier}, Vincent and {Broeg}, Christopher and {Cabral}, Alexandre and {Calderone}, Giorgio and {Pall{\'e}}, Enric and {Cegla}, H.~M. and {Cirami}, Roberto and {Coelho}, Jo{\~a}o M.~P. and {Conconi}, Paolo and {Coretti}, Igor and {Cumani}, Claudio and {Cupani}, Guido and {Dekker}, Hans and {Delabre}, Bernard and {Deiries}, Sebastian and {D'Odorico}, Valentina and {Di Marcantonio}, Paolo and {Figueira}, Pedro and {Fragoso}, Ana and {Genolet}, Ludovic and {Genoni}, Matteo and {G{\'e}nova Santos}, Ricardo and {Hara}, Nathan and {Hughes}, Ian and {Iwert}, Olaf and {Kerber}, Florian and {Knudstrup}, Jens and {Landoni}, Marco and {Lavie}, Baptiste and {Lizon}, Jean-Louis and {Lendl}, Monika and {Lo Curto}, Gaspare and {Maire}, Charles and {Manescau}, Antonio and {Martins}, C.~J.~A.~P. and {M{\'e}gevand}, Denis and {Mehner}, Andrea and {Micela}, Giusi and {Modigliani}, Andrea and {Molaro}, Paolo and {Monteiro}, Manuel and {Monteiro}, Mario and {Moschetti}, Manuele and {M{\"u}ller}, Eric and {Nunes}, Nelson and {Oggioni}, Luca and {Oliveira}, Ant{\'o}nio and {Pariani}, Giorgio and {Pasquini}, Luca and {Poretti}, Ennio and {Rasilla}, Jos{\'e} Luis and {Redaelli}, Edoardo and {Riva}, Marco and {Santana Tschudi}, Samuel and {Santin}, Paolo and {Santos}, Pedro and {Segovia Milla}, Alex and {Seidel}, Julia V. and {Sosnowska}, Danuta and {Sozzetti}, Alessandro and {Span{\`o}}, Paolo and {Su{\'a}rez Mascare{\~n}o}, Alejandro and {Tabernero}, Hugo and {Tenegi}, Fabio and {Udry}, St{\'e}phane and {Zanutta}, Alessio and {Zerbi}, Filippo},
        title = "{Nightside condensation of iron in an ultrahot giant exoplanet}",
      journal = {\nat},
         year = 2020,
        month = apr,
       volume = {580},
       number = {7805},
        pages = {597-601},
          doi = {10.1038/s41586-020-2107-1},
archivePrefix = {arXiv},
       eprint = {2003.05528},
 primaryClass = {astro-ph.EP},
       adsurl = {https://ui.adsabs.harvard.edu/abs/2020Natur.580..597E}
}

@ARTICLE{Fulton2017,
       author = {{Fulton}, Benjamin J. and {Petigura}, Erik A. and {Howard}, Andrew W. and {Isaacson}, Howard and {Marcy}, Geoffrey W. and {Cargile}, Phillip A. and {Hebb}, Leslie and {Weiss}, Lauren M. and {Johnson}, John Asher and {Morton}, Timothy D. and {Sinukoff}, Evan and {Crossfield}, Ian J.~M. and {Hirsch}, Lea A.},
        title = "{The California-Kepler Survey. III. A Gap in the Radius Distribution of Small Planets}",
      journal = {\aj},
         year = 2017,
        month = sep,
       volume = {154},
       number = {3},
          eid = {109},
        pages = {109},
          doi = {10.3847/1538-3881/aa80eb},
archivePrefix = {arXiv},
       eprint = {1703.10375},
 primaryClass = {astro-ph.EP},
       adsurl = {https://ui.adsabs.harvard.edu/abs/2017AJ....154..109F}
}

@ARTICLE{Ricker2015,
       author = {{Ricker}, George R. and {Winn}, Joshua N. and {Vanderspek}, Roland and {Latham}, David W. and {Bakos}, G{\'a}sp{\'a}r {\'A}. and {Bean}, Jacob L. and {Berta-Thompson}, Zachory K. and {Brown}, Timothy M. and {Buchhave}, Lars and {Butler}, Nathaniel R. and {Butler}, R. Paul and {Chaplin}, William J. and {Charbonneau}, David and {Christensen-Dalsgaard}, J{\o}rgen and {Clampin}, Mark and {Deming}, Drake and {Doty}, John and {De Lee}, Nathan and {Dressing}, Courtney and {Dunham}, Edward W. and {Endl}, Michael and {Fressin}, Francois and {Ge}, Jian and {Henning}, Thomas and {Holman}, Matthew J. and {Howard}, Andrew W. and {Ida}, Shigeru and {Jenkins}, Jon M. and {Jernigan}, Garrett and {Johnson}, John Asher and {Kaltenegger}, Lisa and {Kawai}, Nobuyuki and {Kjeldsen}, Hans and {Laughlin}, Gregory and {Levine}, Alan M. and {Lin}, Douglas and {Lissauer}, Jack J. and {MacQueen}, Phillip and {Marcy}, Geoffrey and {McCullough}, Peter R. and {Morton}, Timothy D. and {Narita}, Norio and {Paegert}, Martin and {Palle}, Enric and {Pepe}, Francesco and {Pepper}, Joshua and {Quirrenbach}, Andreas and {Rinehart}, Stephen A. and {Sasselov}, Dimitar and {Sato}, Bun'ei and {Seager}, Sara and {Sozzetti}, Alessandro and {Stassun}, Keivan G. and {Sullivan}, Peter and {Szentgyorgyi}, Andrew and {Torres}, Guillermo and {Udry}, Stephane and {Villasenor}, Joel},
        title = "{Transiting Exoplanet Survey Satellite (TESS)}",
      journal = {Journal of Astronomical Telescopes, Instruments, and Systems},
         year = 2015,
        month = jan,
       volume = {1},
          eid = {014003},
        pages = {014003},
          doi = {10.1117/1.JATIS.1.1.014003},
       adsurl = {https://ui.adsabs.harvard.edu/abs/2015JATIS...1a4003R}
}

@ARTICLE{Huang2018,
       author = {{Huang}, Chelsea X. and {Burt}, Jennifer and {Vanderburg}, Andrew and {G{\"u}nther}, Maximilian N. and {Shporer}, Avi and {Dittmann}, Jason A. and {Winn}, Joshua N. and {Wittenmyer}, Rob and {Sha}, Lizhou and {Kane}, Stephen R. and {Ricker}, George R. and {Vanderspek}, Roland K. and {Latham}, David W. and {Seager}, Sara and {Jenkins}, Jon M. and {Caldwell}, Douglas A. and {Collins}, Karen A. and {Guerrero}, Natalia and {Smith}, Jeffrey C. and {Quinn}, Samuel N. and {Udry}, St{\'e}phane and {Pepe}, Francesco and {Bouchy}, Fran{\c{c}}ois and {S{\'e}gransan}, Damien and {Lovis}, Christophe and {Ehrenreich}, David and {Marmier}, Maxime and {Mayor}, Michel and {Wohler}, Bill and {Haworth}, Kari and {Morgan}, Edward H. and {Fausnaugh}, Michael and {Ciardi}, David R. and {Christiansen}, Jessie and {Charbonneau}, David and {Dragomir}, Diana and {Deming}, Drake and {Glidden}, Ana and {Levine}, Alan M. and {McCullough}, P.~R. and {Yu}, Liang and {Narita}, Norio and {Nguyen}, Tam and {Morton}, Tim and {Pepper}, Joshua and {P{\'a}l}, Andr{\'a}s and {Rodriguez}, Joseph E. and {Stassun}, Keivan G. and {Torres}, Guillermo and {Sozzetti}, Alessandro and {Doty}, John P. and {Christensen-Dalsgaard}, J{\o}rgen and {Laughlin}, Gregory and {Clampin}, Mark and {Bean}, Jacob L. and {Buchhave}, Lars A. and {Bakos}, G. {\'A}. and {Sato}, Bun'ei and {Ida}, Shigeru and {Kaltenegger}, Lisa and {Palle}, Enric and {Sasselov}, Dimitar and {Butler}, R.~P. and {Lissauer}, Jack and {Ge}, Jian and {Rinehart}, S.~A.},
        title = "{TESS Discovery of a Transiting Super-Earth in the pi Mensae System}",
      journal = {\apjl},
         year = 2018,
        month = dec,
       volume = {868},
       number = {2},
          eid = {L39},
        pages = {L39},
          doi = {10.3847/2041-8213/aaef91},
archivePrefix = {arXiv},
       eprint = {1809.05967},
 primaryClass = {astro-ph.EP},
       adsurl = {https://ui.adsabs.harvard.edu/abs/2018ApJ...868L..39H}
}

@ARTICLE{Gandolfi2018,
       author = {{Gandolfi}, D. and {Barrag{\'a}n}, O. and {Livingston}, J.~H. and {Fridlund}, M. and {Justesen}, A.~B. and {Redfield}, S. and {Fossati}, L. and {Mathur}, S. and {Grziwa}, S. and {Cabrera}, J. and {Garc{\'\i}a}, R.~A. and {Persson}, C.~M. and {Van Eylen}, V. and {Hatzes}, A.~P. and {Hidalgo}, D. and {Albrecht}, S. and {Bugnet}, L. and {Cochran}, W.~D. and {Csizmadia}, Sz. and {Deeg}, H. and {Eigm{\"u}ller}, Ph. and {Endl}, M. and {Erikson}, A. and {Esposito}, M. and {Guenther}, E. and {Korth}, J. and {Luque}, R. and {Monta{\~n}es Rodr{\'\i}guez}, P. and {Nespral}, D. and {Nowak}, G. and {P{\"a}tzold}, M. and {Prieto-Arranz}, J.},
        title = "{TESS's first planet. A super-Earth transiting the naked-eye star {\ensuremath{\pi}} Mensae}",
      journal = {\aap},
         year = 2018,
        month = nov,
       volume = {619},
          eid = {L10},
        pages = {L10},
          doi = {10.1051/0004-6361/201834289},
archivePrefix = {arXiv},
       eprint = {1809.07573},
 primaryClass = {astro-ph.EP},
       adsurl = {https://ui.adsabs.harvard.edu/abs/2018A&A...619L..10G}
}

@ARTICLE{Damasso2020,
       author = {{Damasso}, M. and {Sozzetti}, A. and {Lovis}, C. and {Barros}, S.~C.~C. and {Sousa}, S.~G. and {Demangeon}, O.~D.~S. and {Faria}, J.~P. and {Lillo-Box}, J. and {Cristiani}, S. and {Pepe}, F. and {Rebolo}, R. and {Santos}, N.~C. and {Zapatero Osorio}, M.~R. and {Gonz{\'a}lez Hern{\'a}ndez}, J.~I. and {Amate}, M. and {Pasquini}, L. and {Zerbi}, F.~M. and {Adibekyan}, V. and {Abreu}, M. and {Affolter}, M. and {Alibert}, Y. and {Aliverti}, M. and {Allart}, R. and {Allende Prieto}, C. and {{\'A}lvarez}, D. and {Alves}, D. and {Avila}, G. and {Baldini}, V. and {Bandy}, T. and {Benz}, W. and {Bianco}, A. and {Borsa}, F. and {Bossini}, D. and {Bourrier}, V. and {Bouchy}, F. and {Broeg}, C. and {Cabral}, A. and {Calderone}, G. and {Cirami}, R. and {Coelho}, J. and {Conconi}, P. and {Coretti}, I. and {Cumani}, C. and {Cupani}, G. and {D'Odorico}, V. and {Deiries}, S. and {Dekker}, H. and {Delabre}, B. and {Di Marcantonio}, P. and {Dumusque}, X. and {Ehrenreich}, D. and {Figueira}, P. and {Fragoso}, A. and {Genolet}, L. and {Genoni}, M. and {G{\'e}nova Santos}, R. and {Hughes}, I. and {Iwert}, O. and {Kerber}, F. and {Knudstrup}, J. and {Landoni}, M. and {Lavie}, B. and {Lizon}, J. -L. and {Lo Curto}, G. and {Maire}, C. and {Martins}, C.~J.~A.~P. and {M{\'e}gevand}, D. and {Mehner}, A. and {Micela}, G. and {Modigliani}, A. and {Molaro}, P. and {Monteiro}, M.~A. and {Monteiro}, M.~J.~P.~F.~G. and {Moschetti}, M. and {Mueller}, E. and {Murphy}, M.~T. and {Nunes}, N. and {Oggioni}, L. and {Oliveira}, A. and {Oshagh}, M. and {Pall{\'e}}, E. and {Pariani}, G. and {Poretti}, E. and {Rasilla}, J.~L. and {Rebord{\~a}o}, J. and {Redaelli}, E.~M. and {Riva}, M. and {Santana Tschudi}, S. and {Santin}, P. and {Santos}, P. and {S{\'e}gransan}, D. and {Schmidt}, T.~M. and {Segovia}, A. and {Sosnowska}, D. and {Span{\`o}}, P. and {Su{\'a}rez Mascare{\~n}o}, A. and {Tabernero}, H. and {Tenegi}, F. and {Udry}, S. and {Zanutta}, A.},
        title = "{A precise architecture characterization of the {\ensuremath{\pi}} Mensae planetary system}",
      journal = {\aap},
         year = 2020,
        month = oct,
       volume = {642},
          eid = {A31},
        pages = {A31},
          doi = {10.1051/0004-6361/202038416},
archivePrefix = {arXiv},
       eprint = {2007.06410},
 primaryClass = {astro-ph.EP},
       adsurl = {https://ui.adsabs.harvard.edu/abs/2020A&A...642A..31D}
}

@ARTICLE{Pepe2021,
       author = {{Pepe}, F. and {Cristiani}, S. and {Rebolo}, R. and {Santos}, N.~C. and {Dekker}, H. and {Cabral}, A. and {Di Marcantonio}, P. and {Figueira}, P. and {Lo Curto}, G. and {Lovis}, C. and {Mayor}, M. and {M{\'e}gevand}, D. and {Molaro}, P. and {Riva}, M. and {Zapatero Osorio}, M.~R. and {Amate}, M. and {Manescau}, A. and {Pasquini}, L. and {Zerbi}, F.~M. and {Adibekyan}, V. and {Abreu}, M. and {Affolter}, M. and {Alibert}, Y. and {Aliverti}, M. and {Allart}, R. and {Allende Prieto}, C. and {{\'A}lvarez}, D. and {Alves}, D. and {Avila}, G. and {Baldini}, V. and {Bandy}, T. and {Barros}, S.~C.~C. and {Benz}, W. and {Bianco}, A. and {Borsa}, F. and {Bourrier}, V. and {Bouchy}, F. and {Broeg}, C. and {Calderone}, G. and {Cirami}, R. and {Coelho}, J. and {Conconi}, P. and {Coretti}, I. and {Cumani}, C. and {Cupani}, G. and {D'Odorico}, V. and {Damasso}, M. and {Deiries}, S. and {Delabre}, B. and {Demangeon}, O.~D.~S. and {Dumusque}, X. and {Ehrenreich}, D. and {Faria}, J.~P. and {Fragoso}, A. and {Genolet}, L. and {Genoni}, M. and {G{\'e}nova Santos}, R. and {Gonz{\'a}lez Hern{\'a}ndez}, J.~I. and {Hughes}, I. and {Iwert}, O. and {Kerber}, F. and {Knudstrup}, J. and {Landoni}, M. and {Lavie}, B. and {Lillo-Box}, J. and {Lizon}, J. -L. and {Maire}, C. and {Martins}, C.~J.~A.~P. and {Mehner}, A. and {Micela}, G. and {Modigliani}, A. and {Monteiro}, M.~A. and {Monteiro}, M.~J.~P.~F.~G. and {Moschetti}, M. and {Murphy}, M.~T. and {Nunes}, N. and {Oggioni}, L. and {Oliveira}, A. and {Oshagh}, M. and {Pall{\'e}}, E. and {Pariani}, G. and {Poretti}, E. and {Rasilla}, J.~L. and {Rebord{\~a}o}, J. and {Redaelli}, E.~M. and {Santana Tschudi}, S. and {Santin}, P. and {Santos}, P. and {S{\'e}gransan}, D. and {Schmidt}, T.~M. and {Segovia}, A. and {Sosnowska}, D. and {Sozzetti}, A. and {Sousa}, S.~G. and {Span{\`o}}, P. and {Su{\'a}rez Mascare{\~n}o}, A. and {Tabernero}, H. and {Tenegi}, F. and {Udry}, S. and {Zanutta}, A.},
        title = "{ESPRESSO at VLT. On-sky performance and first results}",
      journal = {\aap},
         year = 2021,
        month = jan,
       volume = {645},
          eid = {A96},
        pages = {A96},
          doi = {10.1051/0004-6361/202038306},
archivePrefix = {arXiv},
       eprint = {2010.00316},
 primaryClass = {astro-ph.IM},
       adsurl = {https://ui.adsabs.harvard.edu/abs/2021A&A...645A..96P}
}

@INPROCEEDINGS{Genoni2020,
       author = {{Genoni}, M. and {Pariani}, G. and {Avila}, G. and {Aliverti}, M. and {Oggioni}, L. and {Redaelli}, E.~M.~A. and {Riva}, M. and {Pepe}, F. and {Hughes}, I. and {Meg{\`e}vand}, D. and {Chazelas}, B. and {Temich}, F.~G. and {Rasilla}, J.~L.},
        title = "{ESPRESSO Fiber-Link upgrade. I: project overview and performances}",
    booktitle = {Ground-based and Airborne Instrumentation for Astronomy VIII},
         year = 2020,
       editor = {{Evans}, Christopher J. and {Bryant}, Julia J. and {Motohara}, Kentaro},
       series = {Society of Photo-Optical Instrumentation Engineers (SPIE) Conference Series},
       volume = {11447},
        month = dec,
          eid = {114473G},
        pages = {114473G},
          doi = {10.1117/12.2561464},
       adsurl = {https://ui.adsabs.harvard.edu/abs/2020SPIE11447E..3GG}
}

@INPROCEEDINGS{Aliverti2020,
       author = {{Aliverti}, Matteo and {Pariani}, Giorgio and {Avila}, Gerardo and {Oggioni}, Luca and {Redaelli}, Edoardo Maria Alberto and {Riva}, Marco and {Pepe}, Francesco and {Huges}, Ian and {Meg{\`e}vand}, Denis and {Chazelas}, Bruno and {Temich}, Felix Garcia and {Rasilla}, Jose Luis},
        title = "{ESPRESSO fiber-Link upgrade: III - alignment and integration activities}",
    booktitle = {Ground-based and Airborne Instrumentation for Astronomy VIII},
         year = 2020,
       editor = {{Evans}, Christopher J. and {Bryant}, Julia J. and {Motohara}, Kentaro},
       series = {Society of Photo-Optical Instrumentation Engineers (SPIE) Conference Series},
       volume = {11447},
        month = dec,
          eid = {114474I},
        pages = {114474I},
          doi = {10.1117/12.2562129},
       adsurl = {https://ui.adsabs.harvard.edu/abs/2020SPIE11447E..4IA}
}

@INPROCEEDINGS{DiMarcantonio2018,
       author = {{Di Marcantonio}, P. and {Sosnowska}, D. and {Cupani}, G. and {D'Odorico}, V. and {Lovis}, C. and {Segovia}, A. and {Sousa}, S. and {Gonz{\'a}lez Hern{\'a}ndez}, J.~I. and {Calderone}, G. and {Cirami}, R. and {Modigliani}, A. and {Lo Curto}, G. and {Cristiani}, S. and {Molaro}, P. and {Pepe}, F. and {M{\'e}gevand}, D.},
        title = "{ESPRESSO data flow in operations: results of commissioning activities}",
    booktitle = {Observatory Operations: Strategies, Processes, and Systems VII},
         year = 2018,
       series = {Society of Photo-Optical Instrumentation Engineers (SPIE) Conference Series},
       volume = {10704},
        month = jul,
          eid = {107040F},
        pages = {107040F},
          doi = {10.1117/12.2311285},
       adsurl = {https://ui.adsabs.harvard.edu/abs/2018SPIE10704E..0FD}
}

@INPROCEEDINGS{Marconi2022,
       author = {{Marconi}, A. and {Abreu}, M. and {Adibekyan}, V. and {Alberti}, V. and {Albrecht}, S. and {Alcaniz}, J. and {Aliverti}, M. and {Allende Prieto}, C. and {Alvarado G{\'o}mez}, J.~D. and {Amado}, P.~J. and {Amate}, M. and {Andersen}, M.~I. and {Artigau}, E. and {Baker}, C. and {Baldini}, V. and {Balestra}, A. and {Barnes}, S.~A. and {Baron}, F. and {Barros}, S.~C.~C. and {Bauer}, S.~M. and {Beaulieu}, M. and {Bellido-Tirado}, O. and {Benneke}, B. and {Bensby}, T. and {Bergin}, E.~A. and {Biazzo}, K. and {Bik}, A. and {Birkby}, J.~L. and {Blind}, N. and {Boisse}, I. and {Bolmont}, E. and {Bonaglia}, M. and {Bonfils}, X. and {Borsa}, F. and {Brandeker}, A. and {Brandner}, W. and {Broeg}, C.~H. and {Brogi}, M. and {Brousseau}, D. and {Brucalassi}, A. and {Brynnel}, J. and {Buchhave}, L.~A. and {Buscher}, D.~F. and {Cabral}, A. and {Calderone}, G. and {Calvo-Ortega}, R. and {Canto Martins}, B.~L. and {Cantalloube}, F. and {Carbonaro}, L. and {Chauvin}, G. and {Chazelas}, B. and {Cheffot}, A. -L. and {Cheng}, Y.~S. and {Chiavassa}, A. and {Christensen}, L. and {Cirami}, R. and {Cook}, N.~J. and {Cooke}, R.~J. and {Coretti}, I. and {Covino}, S. and {Cowan}, N. and {Cresci}, G. and {Cristiani}, S. and {Cunha Parro}, V. and {Cupani}, G. and {D'Odorico}, V. and {de Castro Le{\~a}o}, I. and {De Cia}, A. and {De Medeiros}, J.~R. and {Debras}, F. and {Debus}, M. and {Demangeon}, O. and {Dessauges-Zavadsky}, M. and {Di Marcantonio}, P. and {Dionies}, F. and {Doyon}, R. and {Dunn}, J. and {Ehrenreich}, D. and {Faria}, J.~P. and {Feruglio}, C. and {Fisher}, M. and {Fontana}, A. and {Fumagalli}, M. and {Fusco}, T. and {Fynbo}, J. and {Gabella}, O. and {Gaessler}, W. and {Gallo}, E. and {Gao}, X. and {Genolet}, L. and {Genoni}, M. and {Giacobbe}, P. and {Giro}, E. and {Gon{\c{c}}alves}, R.~S. and {Gonzalez}, O. and {Gonz{\'a}lez Hern{\'a}ndez}, J.~I. and {Gracia T{\'e}mich}, F. and {Haehnelt}, M.~G. and {Haniff}, C. and {Hatzes}, A. and {Helled}, R. and {Hoeijmakers}, H.~J. and {Huke}, P. and {J{\"a}rvinen}, S. and {J{\"a}rvinen}, A. and {Kaminski}, A. and {Korn}, A. and {Kouach}, D. and {Kowzan}, G. and {Kreidberg}, L. and {Landoni}, M. and {Lanotte}, A. and {Lavail}, A. and {Li}, J. and {Liske}, J. and {Lovis}, C. and {Lucatello}, S. and {Lunney}, D. and {MacIntosh}, M. and {Madhusudhan}, N. and {Magrini}, L. and {Maiolino}, R. and {Malo}, L. and {Man}, A. and {Marquart}, T. and {Marques}, E.~L. and {Martins}, A.~M. and {Martins}, C.~J.~A.~P. and {Maslowski}, P. and {Mason}, C. and {Mason}, E. and {McCracken}, R.~A. and {Mergo}, P. and {Micela}, G. and {Mitchell}, T. and {Molli{\`e}re}, P. and {Monteiro}, M. and {Montgomery}, D. and {Mordasini}, C. and {Morin}, J. and {Mucciarelli}, A. and {Murphy}, M.~T. and {N'Diaye}, M. and {Neichel}, B. and {Niedzielski}, A.~T. and {Niemczura}, E. and {Nortmann}, L. and {Noterdaeme}, P. and {Nunes}, N. and {Oggioni}, L. and {Oliva}, E. and {{\"O}nel}, H. and {Origlia}, L. and {{\"O}stlin}, G. and {Palle}, E. and {Papaderos}, P. and {Pariani}, G. and {Pe{\~n}ate Castro}, J. and {Pepe}, F. and {Perreault Levasseur}, L. and {Petit}, P. and {Pino}, L. and {Piqueras}, J. and {Pollo}, A. and {Poppenhaeger}, K. and {Quirrenbach}, A. and {Rauscher}, E. and {Rebolo}, R. and {Redaelli}, E.~M.~A. and {Reffert}, S. and {Reid}, D.~T. and {Reiners}, A. and {Richter}, P. and {Riva}, M. and {Rivoire}, S. and {Rodr{\'\i}guez-L{\'o}pez}, C. and {Roederer}, I.~U. and {Romano}, D. and {Rousseau}, S. and {Rowe}, J. and {Salvadori}, S. and {Santos}, N. and {Santos Diaz}, P. and {Sanz-Forcada}, J. and {Sarajlic}, M. and {Sauvage}, J. -F. and {Sch{\"a}fer}, S. and {Schiavon}, R.~P. and {Schmidt}, T.~M. and {Selmi}, C. and {Sivanandam}, S. and {Sordet}, M. and {Sordo}, R. and {Sortino}, F. and {Sosnowska}, D. and {Sousa}, S.~G. and {Stempels}, E. and {Strassmeier}, K.~G. and {Su{\'a}rez Mascare{\~n}o}, A. and {Sulich}, A.},
        title = "{ANDES, the high resolution spectrograph for the ELT: science case, baseline design and path to construction}",
    booktitle = {Ground-based and Airborne Instrumentation for Astronomy IX},
         year = 2022,
       editor = {{Evans}, Christopher J. and {Bryant}, Julia J. and {Motohara}, Kentaro},
       series = {Society of Photo-Optical Instrumentation Engineers (SPIE) Conference Series},
       volume = {12184},
        month = aug,
          eid = {1218424},
        pages = {1218424},
          doi = {10.1117/12.2628689},
       adsurl = {https://ui.adsabs.harvard.edu/abs/2022SPIE12184E..24M}
}

@INPROCEEDINGS{Marconi2024,
       author = {{Marconi}, A. and {Abreu}, M. and {Adibekyan}, V. and {Alberti}, V. and {Albrecht}, S. and {Alcaniz}, J. and {Aliverti}, M. and {Allende Prieto}, C. and {Alvarado-Gomez}, J.~D. and {Alves}, C.~S. and {Amado}, P.~J. and {Amate}, M. and {Andersen}, M.~I. and {Antoniucci}, S. and {Artigau}, E. and {Bailet}, C. and {Baker}, C. and {Baldini}, V. and {Balestra}, A. and {Barnes}, S.~A. and {Baron}, F. and {Barros}, S.~C.~C. and {Bauer}, S.~M. and {Beaulieu}, M. and {Bellido-Tirado}, O. and {Benneke}, B. and {Bensby}, T. and {Bergin}, E.~A. and {Berio}, P. and {Biazzo}, K. and {Bigot}, L. and {Bik}, A. and {Birkby}, J.~L. and {Blind}, N. and {Boebion}, O. and {Boisse}, I. and {Bolmont}, E. and {Bolton}, J.~S. and {Bonaglia}, M. and {Bonfils}, X. and {Bonhomme}, L. and {Borsa}, F. and {Bouret}, J. -C. and {Brandeker}, A. and {Brandner}, W. and {Broeg}, C.~H. and {Brogi}, M. and {Brousseau}, D. and {Brucalassi}, A. and {Brynnel}, J. and {Buchhave}, L.~A. and {Buscher}, D.~F. and {Cabona}, L. and {Cabral}, A. and {Calderone}, G. and {Calvo-Ortega}, R. and {Cantalloube}, F. and {Canto Martins}, B.~L. and {Carbonaro}, L. and {Caujolle}, Y. and {Chauvin}, G. and {Chazelas}, B. and {Cheffot}, A. -L. and {Cheng}, Y.~S. and {Chiavassa}, A. and {Christensen}, L. and {Cirami}, R. and {Cirasuolo}, M. and {Cook}, N.~J. and {Cooke}, R.~J. and {Coretti}, I. and {Covino}, S. and {Cowan}, N. and {Cresci}, G. and {Cristiani}, S. and {Cunha Parro}, V. and {Cupani}, G. and {D'Odorico}, V. and {Dadi}, K. and {de Castro Le{\~a}o}, I. and {De Cia}, A. and {De Medeiros}, J.~R. and {Debras}, F. and {Debus}, M. and {Delorme}, A. and {Demangeon}, O. and {Derie}, F. and {Dessauges-Zavadsky}, M. and {Di Marcantonio}, P. and {Di Stefano}, S. and {Dionies}, F. and {Domiciano de Souza}, A. and {Doyon}, R. and {Dunn}, J. and {Egner}, S. and {Ehrenreich}, D. and {Faria}, J.~P. and {Ferruzzi}, D. and {Feruglio}, C. and {Fisher}, M. and {Fontana}, A. and {Frank}, B.~S. and {Fuesslein}, C. and {Fumagalli}, M. and {Fusco}, T. and {Fynbo}, J. and {Gabella}, O. and {Gaessler}, W. and {Gallo}, E. and {Gao}, X. and {Genolet}, L. and {Genoni}, M. and {Giacobbe}, P. and {Giro}, E. and {Gon{\c{c}}alves}, R.~S. and {Gonzalez}, O.~A. and {Gonz{\'a}lez-Hern{\'a}ndez}, J.~I. and {Gouvret}, C. and {Gracia T{\'e}mich}, F. and {Haehnelt}, M.~G. and {Haniff}, C. and {Hatzes}, A. and {Helled}, R. and {Hoeijmakers}, H.~J. and {Hughes}, I. and {Huke}, P. and {Ivanisenko}, Y. and {J{\"a}rvinen}, A.~S. and {J{\"a}rvinen}, S.~P. and {Kaminski}, A. and {Kern}, J. and {Knoche}, J. and {Kordt}, A. and {Korhonen}, H. and {Korn}, A.~J. and {Kouach}, D. and {Kowzan}, G. and {Kreidberg}, L. and {Landoni}, M. and {Lanotte}, A.~A. and {Lavail}, A. and {Lavie}, B. and {Lee}, D. and {Lehmitz}, M. and {Li}, J. and {Li}, W. and {Liske}, J. and {Lovis}, C. and {Lucatello}, S. and {Lunney}, D. and {MacIntosh}, M.~J. and {Madhusudhan}, N. and {Magrini}, L. and {Maiolino}, R. and {Maldonado}, J. and {Malo}, L. and {Man}, A.~W.~S. and {Marquart}, T. and {Marques}, C.~M.~J. and {Marques}, E.~L. and {Martinez}, P. and {Martins}, A. and {Martins}, C.~J.~A.~P. and {Martins}, J.~H.~C. and {Maslowski}, P. and {Mason}, C. and {Mason}, E. and {McCracken}, R.~A. and {Melo e Sousa}, M.~A.~F. and {Mergo}, P. and {Micela}, G. and {Milakovi{\'c}}, D. and {Molli{\`e}re}, P. and {Monteiro}, M.~A. and {Montgomery}, D. and {Mordasini}, C. and {Morin}, J. and {Mucciarelli}, A. and {Murphy}, M.~T. and {N'Diaye}, M. and {Nardetto}, N. and {Neichel}, B. and {Neri}, N. and {Niedzielski}, A.~T. and {Niemczura}, E. and {Nisini}, B. and {Nortmann}, L. and {Noterdaeme}, P. and {Nunes}, N.~J. and {Oggioni}, L. and {Olchewsky}, F. and {Oliva}, E. and {{\"O}nel}, H. and {Origlia}, L. and {{\"O}stlin}, G. and {Ouellette}, N.~N. -Q. and {Pall{\'e}}, E. and {Papaderos}, P. and {Pariani}, G. and {Pasquini}, L.},
        title = "{ANDES, the high resolution spectrograph for the ELT: science goals, project overview, and future developments}",
    booktitle = {Ground-based and Airborne Instrumentation for Astronomy X},
         year = 2024,
       editor = {{Bryant}, Julia J. and {Motohara}, Kentaro and {Vernet}, Jo{\"e}l. R.~D.},
       series = {Society of Photo-Optical Instrumentation Engineers (SPIE) Conference Series},
       volume = {13096},
        month = jul,
          eid = {1309613},
        pages = {1309613},
          doi = {10.1117/12.3017966},
archivePrefix = {arXiv},
       eprint = {2407.14601},
 primaryClass = {astro-ph.IM},
       adsurl = {https://ui.adsabs.harvard.edu/abs/2024SPIE13096E..13M}
}

@INPROCEEDINGS{Chazelas2024a,
       author = {{Chazelas}, Bruno and {Ivanisenko}, Yevgeniy and {Lanotte}, Audrey and {Santos Diaz}, Pablo and {Genolet}, Ludovic and {Sordet}, Michael and {Hughes}, Ian and {Lovis}, Christophe and {Schmidt}, Tobias M. and {Amate Plasencia}, Manuel and {Pe{\~n}ate Castro}, Jose and {Vega-Moreno}, Afrodisio and {Tenegi}, Fabio and {Simoes}, Roberto and {Gonz{\'a}lez-Hern{\'a}ndez}, Jonay I. and {Zapatero Osorio}, Maria-Rosa and {Piqueras}, Javier and {Belenguer D{\'a}vila}, Tom{\'a}s. and {Calvo-Ortega}, Roc{\'\i}o. and {Varas Gonz{\'a}lez}, Roberto and {Gonz{\'a}lez Fern{\'a}ndez}, Luis Miguel and {Amado}, Pedro J. and {Kern}, Jonathan and {Dionies}, Frank and {Bauer}, Svend-Marian and {{\"O}nel}, Hakan and {J{\"a}rvinen}, Arto and {Brynnel}, Joar and {F{\"u}{\ss}lein}, Christine and {Bellido}, Olga and {Weingrill}, J{\"o}rg and {Giannone}, Domenico and {Gaessler}, Wolfgang and {Lehmitz}, Michael and {Kaminski}, Adrian and {Stilz}, Ingo and {Sigwarth}, Michael and {Marconi}, Alessandro and {Di Marcantonio}, Paolo and {Oliva}, Ernesto and {Coretti}, Igor and {Aliverti}, Matteo and {Pariani}, Giorgio and {Cabona}, Lorenzo and {Radaelli}, Edouardo Maria Alberto and {Scalera}, Marcello Agostino and {Balestera}, Andrea},
        title = "{ANDES, the high-resolution spectrograph for the ELT: RIZ spectrograph preliminary design}",
    booktitle = {Ground-based and Airborne Instrumentation for Astronomy X},
         year = 2024,
       editor = {{Bryant}, Julia J. and {Motohara}, Kentaro and {Vernet}, Jo{\"e}l. R.~D.},
       series = {Society of Photo-Optical Instrumentation Engineers (SPIE) Conference Series},
       volume = {13096},
        month = jul,
          eid = {130964J},
        pages = {130964J},
          doi = {10.1117/12.3019996},
       adsurl = {https://ui.adsabs.harvard.edu/abs/2024SPIE13096E..4JC}
}

@INPROCEEDINGS{Lanotte2024,
       author = {{Lanotte}, Audrey A. and {Chazelas}, Bruno and {Lovis}, Christophe and {Weber}, Michael and {Woche}, Manfred and {Oliva}, Ernesto and {Delabre}, Bernard and {Genolet}, Ludovic and {Santos Diaz}, Pablo and {Marconi}, Alessandro and {Di Martantonio}, Paolo and {Zanutta}, Alessio and {Aliverti}, Matteo and {Balestra}, Andrea and {Cabona}, Lorenzo and {Coretti}, Igor and {Pariani}, Giorgio and {Redaelli}, Edoardo and {Scalera}, Marcello and {Scaudo}, Andrea},
        title = "{ANDES, the high-resolution spectrograph for the ELT: RIZ spectrograph preliminary optical design}",
    booktitle = {Ground-based and Airborne Instrumentation for Astronomy X},
         year = 2024,
       editor = {{Bryant}, Julia J. and {Motohara}, Kentaro and {Vernet}, Jo{\"e}l. R.~D.},
       series = {Society of Photo-Optical Instrumentation Engineers (SPIE) Conference Series},
       volume = {13096},
        month = jul,
          eid = {130964F},
        pages = {130964F},
          doi = {10.1117/12.3017976},
       adsurl = {https://ui.adsabs.harvard.edu/abs/2024SPIE13096E..4FL}
}

@INPROCEEDINGS{SantosDiaz2024,
       author = {{Santos Diaz}, P. and {Kern}, J. and {Genolet}, L. and {Chazelas}, B. and {Hughes}, I. and {Lanotte}, A.~A. and {Weber}, M. and {F{\"u}{\ss}lein}, C. and {Lovis}, C. and {Dionies}, F. and {Vega-Moreno}, A. and {Pe{\~n}ate Castro}, J. and {Calvo Ortega}, Rocio and {Varas Gonzalez}, Roberto and {Simoes}, Roberto and {Tenegi}, F. and {Ivanisenko}, Y. and {Brynnel}, J. and {Weingrill}, J. and {Sigwarth}, M. and {Sordet}, M. and {Bellido}, O. and {J{\"a}rvinen}, A. and {Aliverti}, M. and {Cabona}, L. and {Coretti}, I. and {Di Marcantonio}, P. and {Marconi}, A. and {Oliva}, E. and {Pariani}, G. and {Redaelli}, E.~M.~A. and {Riva}, M. and {Zanutta}, A.},
        title = "{ANDES, the high-resolution spectrograph for the ELT: RIZ \& UBV spectrographs' preliminary design, analysis, and integration of the vacuum vessel and thermal enclosure}",
    booktitle = {Ground-based and Airborne Instrumentation for Astronomy X},
         year = 2024,
       editor = {{Bryant}, Julia J. and {Motohara}, Kentaro and {Vernet}, Jo{\"e}l. R.~D.},
       series = {Society of Photo-Optical Instrumentation Engineers (SPIE) Conference Series},
       volume = {13096},
        month = jul,
          eid = {130964K},
        pages = {130964K},
          doi = {10.1117/12.3020150},
       adsurl = {https://ui.adsabs.harvard.edu/abs/2024SPIE13096E..4KS}
}

@INPROCEEDINGS{Genolet2024,
       author = {{Genolet}, L. and {Aliverti}, M. and {Balestra}, A. and {Bellido}, O. and {Brynnel}, J. and {Cabona}, L. and {Chazelas}, B. and {Coretti}, I. and {Di Marcantonio}, P. and {Fuesslein}, C. and {Ivanisenko}, Y. and {J{\"a}rvinen}, A. and {Kaminski}, A. and {Kern}, J. and {Lanotte}, A. and {Lovis}, C. and {Marconi}, A. and {Oliva}, E. and {Pariani}, G. and {Poppenh{\"a}ger}, K. and {Redaelli}, E. and {Riva}, M. and {Santos Diaz}, P. and {Scalera}, M. and {Sordet}, M. and {Strassmeier}, K. and {Weber}, M. and {Weingrill}, J. and {Zanutta}, A.},
        title = "{ANDES, the high-resolution spectrograph for the ELT: RIZ and UBV spectrographs' preliminary design of the detector unit}",
    booktitle = {X-Ray, Optical, and Infrared Detectors for Astronomy XI},
         year = 2024,
       editor = {{Holland}, Andrew D. and {Minoglou}, Kyriaki},
       series = {Society of Photo-Optical Instrumentation Engineers (SPIE) Conference Series},
       volume = {13103},
        month = aug,
          eid = {131031V},
        pages = {131031V},
          doi = {10.1117/12.3019925},
       adsurl = {https://ui.adsabs.harvard.edu/abs/2024SPIE13103E..1VG}
}

@ARTICLE{Roederer2024,
       author = {{Roederer}, Ian U. and {Alvarado-G{\'o}mez}, Juli{\'a}n D. and {Allende Prieto}, Carlos and {Adibekyan}, Vardan and {Aguado}, David S. and {Amado}, Pedro J. and {Amazo-G{\'o}mez}, Eliana M. and {Baratella}, Martina and {Barnes}, Sydney A. and {Bensby}, Thomas and {Bigot}, Lionel and {Chiavassa}, Andrea and {Domiciano de Souza}, Armando and {Gonz{\'a}lez Hern{\'a}ndez}, J.~I. and {Hansen}, Camilla Juul and {J{\"a}rvinen}, Silva P. and {Korn}, Andreas J. and {Lucatello}, Sara and {Magrini}, Laura and {Maiolino}, Roberto and {Di Marcantonio}, Paolo and {Marconi}, Alessandro and {De Medeiros}, Jos{\'e} R. and {Mucciarelli}, Alessio and {Nardetto}, Nicolas and {Origlia}, Livia and {Peroux}, Celine and {Poppenh{\"a}ger}, Katja and {Reiners}, Ansgar and {Rodr{\'\i}guez-L{\'o}pez}, Cristina and {Romano}, Donatella and {Salvadori}, Stefania and {Tisserand}, Patrick and {Venn}, Kim and {Wade}, Gregg A. and {Zanutta}, Alessio},
        title = "{The discovery space of ELT-ANDES. Stars and stellar populations}",
      journal = {Experimental Astronomy},
         year = 2024,
        month = apr,
       volume = {57},
       number = {2},
          eid = {17},
        pages = {17},
          doi = {10.1007/s10686-024-09938-8},
archivePrefix = {arXiv},
       eprint = {2311.16320},
 primaryClass = {astro-ph.IM},
       adsurl = {https://ui.adsabs.harvard.edu/abs/2024ExA....57...17R}
}

@ARTICLE{DOdorico2024,
       author = {{D'Odorico}, Valentina and {Bolton}, James S. and {Christensen}, Lise and {De Cia}, Annalisa and {Zackrisson}, Erik and {Kordt}, Aron and {Izzo}, Luca and {Li}, Jiangtao and {Maiolino}, Roberto and {Marconi}, Alessandro and {Richter}, Philipp and {Saccardi}, Andrea and {Salvadori}, Stefania and {Vanni}, Irene and {Feruglio}, Chiara and {Fumagalli}, Michele and {Fynbo}, Johan P.~U. and {Noterdaeme}, Pasquier and {Papaderos}, Polychronis and {P{\'e}roux}, C{\'e}line and {Verma}, Aprajita and {Di Marcantonio}, Paolo and {Origlia}, Livia and {Zanutta}, Alessio},
        title = "{Galaxy formation and symbiotic evolution with the inter-galactic medium in the age of ELT-ANDES}",
      journal = {Experimental Astronomy},
         year = 2024,
        month = dec,
       volume = {58},
       number = {3},
          eid = {21},
        pages = {21},
          doi = {10.1007/s10686-024-09967-3},
archivePrefix = {arXiv},
       eprint = {2311.16803},
 primaryClass = {astro-ph.GA},
       adsurl = {https://ui.adsabs.harvard.edu/abs/2024ExA....58...21D}
}

@ARTICLE{Martins2024,
       author = {{Martins}, C.~J.~A.~P. and {Cooke}, R. and {Liske}, J. and {Murphy}, M.~T. and {Noterdaeme}, P. and {Schmidt}, T.~M. and {Alcaniz}, J.~S. and {Alves}, C.~S. and {Balashev}, S. and {Cristiani}, S. and {Di Marcantonio}, P. and {G{\'e}nova Santos}, R. and {Gon{\c{c}}alves}, R.~S. and {Gonz{\'a}lez Hern{\'a}ndez}, J.~I. and {Maiolino}, R. and {Marconi}, A. and {Marques}, C.~M.~J. and {Melo e Sousa}, M.~A.~F. and {Nunes}, N.~J. and {Origlia}, L. and {P{\'e}roux}, C. and {Vinzl}, S. and {Zanutta}, A.},
        title = "{Cosmology and fundamental physics with the ELT-ANDES spectrograph}",
      journal = {Experimental Astronomy},
         year = 2024,
        month = feb,
       volume = {57},
       number = {1},
          eid = {5},
        pages = {5},
          doi = {10.1007/s10686-024-09928-w},
archivePrefix = {arXiv},
       eprint = {2311.16274},
 primaryClass = {astro-ph.CO},
       adsurl = {https://ui.adsabs.harvard.edu/abs/2024ExA....57....5M}
}

@ARTICLE{Palle2025,
       author = {{Palle}, Enric and {Biazzo}, Katia and {Bolmont}, Emeline and {Molli{\`e}re}, Paul and {Poppenhaeger}, Katja and {Birkby}, Jayne and {Brogi}, Matteo and {Chauvin}, Gael and {Chiavassa}, Andrea and {Hoeijmakers}, Jens and {Lellouch}, Emmanuel and {Lovis}, Christophe and {Maiolino}, Roberto and {Nortmann}, Lisa and {Parviainen}, Hannu and {Pino}, Lorenzo and {Turbet}, Martin and {Weder}, Jesse and {Albrecht}, Simon and {Antoniucci}, Simone and {Barros}, Susana C. and {Beaudoin}, Andre and {Benneke}, Bjorn and {Boisse}, Isabelle and {Bonomo}, Aldo S. and {Borsa}, Francesco and {Brandeker}, Alexis and {Brandner}, Wolfgang and {Buchhave}, Lars A. and {Cheffot}, Anne-Laure and {Deborde}, Robin and {Debras}, Florian and {Doyon}, Rene and {Di Marcantonio}, Paolo and {Giacobbe}, Paolo and {Gonz{\'a}lez Hern{\'a}ndez}, Jonay I. and {Helled}, Ravit and {Kreidberg}, Laura and {Machado}, Pedro and {Maldonado}, Jesus and {Marconi}, Alessandro and {Martins}, B.~L. Canto and {Miceli}, Adriano and {Mordasini}, Christoph and {N'Diaye}, Mamadou and {Niedzielski}, Andrzej and {Nisini}, Brunella and {Origlia}, Livia and {Peroux}, Celine and {Pietrow}, Alexander G.~M. and {Pinna}, Enrico and {Rauscher}, Emily and {Reffert}, Sabine and {Rodr{\'\i}guez-L{\'o}pez}, Cristina and {Rousselot}, Philippe and {Sanna}, Nicoletta and {Santos}, Nuno C. and {Simonnin}, Adrien and {Su{\'a}rez Mascare{\~n}o}, Alejandro and {Zanutta}, Alessio and {Zapatero-Osorio}, Maria Rosa and {Zechmeister}, Mathias},
        title = "{Ground-breaking exoplanet science with the ANDES spectrograph at the ELT}",
      journal = {Experimental Astronomy},
         year = 2025,
        month = jun,
       volume = {59},
       number = {3},
          eid = {29},
        pages = {29},
          doi = {10.1007/s10686-025-10000-4},
archivePrefix = {arXiv},
       eprint = {2311.17075},
 primaryClass = {astro-ph.IM},
       adsurl = {https://ui.adsabs.harvard.edu/abs/2025ExA....59...29P}
}

@INPROCEEDINGS{Wildi2010,
       author = {{Wildi}, Fran{\c{c}}ois and {Pepe}, Francesco and {Chazelas}, Bruno and {Lo Curto}, Gaspare and {Lovis}, Ch.},
        title = "{A Fabry-Perot calibrator of the HARPS radial velocity spectrograph: performance report}",
    booktitle = {Ground-based and Airborne Instrumentation for Astronomy III},
         year = 2010,
       editor = {{McLean}, Ian S. and {Ramsay}, Suzanne K. and {Takami}, Hideki},
       series = {Society of Photo-Optical Instrumentation Engineers (SPIE) Conference Series},
       volume = {7735},
        month = jul,
          eid = {77354X},
        pages = {77354X},
          doi = {10.1117/12.857951},
       adsurl = {https://ui.adsabs.harvard.edu/abs/2010SPIE.7735E..4XW}
}

@INPROCEEDINGS{Wildi2011,
       author = {{Wildi}, Francois and {Pepe}, Francesco and {Chazelas}, Bruno and {Lo Curto}, Gaspare and {Lovis}, Christophe},
        title = "{The performance of the new Fabry-Perot calibration system of the radial velocity spectrograph HARPS}",
    booktitle = {Techniques and Instrumentation for Detection of Exoplanets V},
         year = 2011,
       editor = {{Shaklan}, Stuart},
       series = {Society of Photo-Optical Instrumentation Engineers (SPIE) Conference Series},
       volume = {8151},
        month = oct,
          eid = {81511F},
        pages = {81511F},
          doi = {10.1117/12.901550},
       adsurl = {https://ui.adsabs.harvard.edu/abs/2011SPIE.8151E..1FW}
}

@INPROCEEDINGS{Wildi2012,
       author = {{Wildi}, Fran{\c{c}}cis and {Chazelas}, Bruno and {Pepe}, Francesco},
        title = "{A passive cost-effective solution for the high accuracy wavelength calibration of radial velocity spectrographs}",
    booktitle = {Ground-based and Airborne Instrumentation for Astronomy IV},
         year = 2012,
       editor = {{McLean}, Ian S. and {Ramsay}, Suzanne K. and {Takami}, Hideki},
       series = {Society of Photo-Optical Instrumentation Engineers (SPIE) Conference Series},
       volume = {8446},
        month = sep,
          eid = {84468E},
        pages = {84468E},
          doi = {10.1117/12.926841},
       adsurl = {https://ui.adsabs.harvard.edu/abs/2012SPIE.8446E..8EW}
}

@ARTICLE{Cersullo2017,
       author = {{Cersullo}, Federica and {Wildi}, Fran{\c{c}}ois and {Chazelas}, Bruno and {Pepe}, Francesco},
        title = "{A new infrared Fabry-P{\'e}rot-based radial-velocity-reference module for the SPIRou radial-velocity spectrograph}",
      journal = {\aap},
         year = 2017,
        month = may,
       volume = {601},
          eid = {A102},
        pages = {A102},
          doi = {10.1051/0004-6361/201629972},
archivePrefix = {arXiv},
       eprint = {1702.05067},
 primaryClass = {astro-ph.IM},
       adsurl = {https://ui.adsabs.harvard.edu/abs/2017A&A...601A.102C}
}

@ARTICLE{Cersullo2019,
       author = {{Cersullo}, F. and {Coffinet}, A. and {Chazelas}, B. and {Lovis}, C. and {Pepe}, F.},
        title = "{New wavelength calibration for echelle spectrographs using Fabry-P{\'e}rot etalons}",
      journal = {\aap},
         year = 2019,
        month = apr,
       volume = {624},
          eid = {A122},
        pages = {A122},
          doi = {10.1051/0004-6361/201833852},
archivePrefix = {arXiv},
       eprint = {1903.01942},
 primaryClass = {astro-ph.IM},
       adsurl = {https://ui.adsabs.harvard.edu/abs/2019A&A...624A.122C}
}

@ARTICLE{Schmidt2022,
       author = {{Schmidt}, Tobias M. and {Chazelas}, Bruno and {Lovis}, Christophe and {Dumusque}, Xavier and {Bouchy}, Fran{\c{c}}ois and {Pepe}, Francesco and {Figueira}, Pedro and {Sosnowska}, Danuta},
        title = "{Chromatic drift of the Espresso Fabry-P{\'e}rot etalon}",
      journal = {\aap},
         year = 2022,
        month = aug,
       volume = {664},
          eid = {A191},
        pages = {A191},
          doi = {10.1051/0004-6361/202243270},
       adsurl = {https://ui.adsabs.harvard.edu/abs/2022A&A...664A.191S}
}

@ARTICLE{Obrzud2018,
       author = {{Obrzud}, E. and {Rainer}, M. and {Harutyunyan}, A. and {Chazelas}, B. and {Cecconi}, M. and {Ghedina}, A. and {Molinari}, E. and {Kundermann}, S. and {Lecomte}, S. and {Pepe}, F. and {Wildi}, F. and {Bouchy}, F. and {Herr}, T.},
        title = "{Broadband near-infrared astronomical spectrometer calibration and on-sky validation with an electro-optic laser frequency comb}",
      journal = {Optics Express},
         year = 2018,
        month = dec,
       volume = {26},
       number = {26},
        pages = {34830},
          doi = {10.1364/OE.26.034830},
archivePrefix = {arXiv},
       eprint = {1808.00860},
 primaryClass = {physics.optics},
       adsurl = {https://ui.adsabs.harvard.edu/abs/2018OExpr..2634830O}
}

@ARTICLE{Obrzud2019,
       author = {{Obrzud}, Ewelina and {Rainer}, Monica and {Harutyunyan}, Avet and {Anderson}, Miles H. and {Liu}, Junqiu and {Geiselmann}, Michael and {Chazelas}, Bruno and {Kundermann}, Stefan and {Lecomte}, Steve and {Cecconi}, Massimo and {Ghedina}, Adriano and {Molinari}, Emilio and {Pepe}, Francesco and {Wildi}, Fran{\c{c}}ois and {Bouchy}, Fran{\c{c}}ois and {Kippenberg}, Tobias J. and {Herr}, Tobias},
        title = "{A microphotonic astrocomb}",
      journal = {Nature Photonics},
         year = 2019,
        month = jan,
       volume = {13},
       number = {1},
        pages = {31-35},
          doi = {10.1038/s41566-018-0309-y},
       adsurl = {https://ui.adsabs.harvard.edu/abs/2019NaPho..13...31O}
}

@ARTICLE{Ludwig2024,
       author = {{Ludwig}, Markus and {Ayhan}, Furkan and {Schmidt}, Tobias M. and {Wildi}, Thibault and {Voumard}, Thibault and {Blum}, Roman and {Ye}, Zhichao and {Lei}, Fuchuan and {Wildi}, Fran{\c{c}}ois and {Pepe}, Francesco and {Gaafar}, Mahmoud A. and {Obrzud}, Ewelina and {Grassani}, Davide and {Hefti}, Olivia and {Karlen}, Sylvain and {Lecomte}, Steve and {Moreau}, Fran{\c{c}}ois and {Chazelas}, Bruno and {Sottile}, Rico and {Torres-Company}, Victor and {Brasch}, Victor and {Villanueva}, Luis G. and {Bouchy}, Fran{\c{c}}ois and {Herr}, Tobias},
        title = "{Ultraviolet astronomical spectrograph calibration with laser frequency combs from nanophotonic lithium niobate waveguides}",
      journal = {Nature Communications},
         year = 2024,
        month = dec,
       volume = {15},
       number = {1},
          eid = {7614},
        pages = {7614},
          doi = {10.1038/s41467-024-51560-x},
archivePrefix = {arXiv},
       eprint = {2306.13609},
 primaryClass = {physics.optics},
       adsurl = {https://ui.adsabs.harvard.edu/abs/2024NatCo..15.7614L}
}

@INPROCEEDINGS{Schmidt2024,
       author = {{Schmidt}, Tobias M.},
        title = "{New techniques for accurate and stable wavelength calibration}",
    booktitle = {Advances in Optical and Mechanical Technologies for Telescopes and Instrumentation VI},
         year = 2024,
       editor = {{Navarro}, Ram{\'o}n and {Jedamzik}, Ralf},
       series = {Society of Photo-Optical Instrumentation Engineers (SPIE) Conference Series},
       volume = {13100},
        month = aug,
          eid = {131004P},
        pages = {131004P},
          doi = {10.1117/12.3020497},
       adsurl = {https://ui.adsabs.harvard.edu/abs/2024SPIE13100E..4PS}
}

@ARTICLE{Schmidt2021,
       author = {{Schmidt}, Tobias M. and {Molaro}, Paolo and {Murphy}, Michael T. and {Lovis}, Christophe and {Cupani}, Guido and {Cristiani}, Stefano and {Pepe}, Francesco A. and {Rebolo}, Rafael and {Santos}, Nuno C. and {Abreu}, Manuel and {Adibekyan}, Vardan and {Alibert}, Yann and {Aliverti}, Matteo and {Allart}, Romain and {Allende Prieto}, Carlos and {Alves}, David and {Baldini}, Veronica and {Broeg}, Christopher and {Cabral}, Alexandre and {Calderone}, Giorgio and {Cirami}, Roberto and {Coelho}, Jo{\~a}o and {Coretti}, Igor and {D'Odorico}, Valentina and {Di Marcantonio}, Paolo and {Ehrenreich}, David and {Figueira}, Pedro and {Genoni}, Matteo and {G{\'e}nova Santos}, Ricardo and {Gonz{\'a}lez Hern{\'a}ndez}, Jonay I. and {Kerber}, Florian and {Landoni}, Marco and {Leite}, Ana C.~O. and {Lizon}, Jean-Louis and {Lo Curto}, Gaspare and {Manescau}, Antonio and {Martins}, Carlos J.~A.~P. and {Meg{\'e}vand}, Denis and {Mehner}, Andrea and {Micela}, Giuseppina and {Modigliani}, Andrea and {Monteiro}, Manuel and {Monteiro}, Mario J.~P.~F.~G. and {Mueller}, Eric and {Nunes}, Nelson J. and {Oggioni}, Luca and {Oliveira}, Ant{\'o}nio and {Pariani}, Giorgio and {Pasquini}, Luca and {Redaelli}, Edoardo and {Riva}, Marco and {Santos}, Pedro and {Sosnowska}, Danuta and {Sousa}, S{\'e}rgio G. and {Sozzetti}, Alessandro and {Su{\'a}rez Mascare{\~n}o}, Alejandro and {Udry}, St{\'e}phane and {Zapatero Osorio}, Maria-Rosa and {Zerbi}, Filippo},
        title = "{Fundamental physics with ESPRESSO: Towards an accurate wavelength calibration for a precision test of the fine-structure constant}",
      journal = {\aap},
         year = 2021,
        month = feb,
       volume = {646},
          eid = {A144},
        pages = {A144},
          doi = {10.1051/0004-6361/202039345},
archivePrefix = {arXiv},
       eprint = {2011.13963},
 primaryClass = {astro-ph.IM},
       adsurl = {https://ui.adsabs.harvard.edu/abs/2021A&A...646A.144S}
}

@ARTICLE{Schmidt2025,
       author = {{Schmidt}, Tobias M. and {Reiners}, Ansgar and {Murphy}, Michael T. and {Lo Curto}, Gaspare and {Martins}, Carlos J.~A.~P. and {Huke}, Philipp},
        title = "{Validation of the ESPRESSO wavelength calibration using iodine absorption cell spectra}",
      journal = {\mnras},
         year = 2025,
        month = jun,
       volume = {539},
       number = {4},
        pages = {3301-3318},
          doi = {10.1093/mnras/staf588},
archivePrefix = {arXiv},
       eprint = {2504.18485},
 primaryClass = {astro-ph.IM},
       adsurl = {https://ui.adsabs.harvard.edu/abs/2025MNRAS.539.3301S}
}

@INPROCEEDINGS{Pepe2024,
       author = {{Pepe}, F. and {Bugatti}, M. and {Baur}, A. and {Berney}, J. and {Bozzo}, E. and {Broeg}, C. and {Crazzolara}, B. and {Deep}, A. and {Droz}, F. and {Figueiredo}, J. and {Haile}, H. and {Jolissaint}, L. and {Lecomte}, S. and {Moerschell}, J. and {Mordasini}, Ch. and {Obrzud}, E. and {Praplan}, C. and {Sarajlic}, M. and {Soja}, B. and {Van den Broeck}, M.},
        title = "{{\ensuremath{\nu}}Ancestor: an artificial satellite-borne star for accurate frequency calibration of ground-based EPRV spectrographs}",
    booktitle = {Advances in Optical and Mechanical Technologies for Telescopes and Instrumentation VI},
         year = 2024,
       editor = {{Navarro}, Ram{\'o}n and {Jedamzik}, Ralf},
       series = {Society of Photo-Optical Instrumentation Engineers (SPIE) Conference Series},
       volume = {13100},
        month = aug,
          eid = {131005W},
        pages = {131005W},
          doi = {10.1117/12.3016547},
       adsurl = {https://ui.adsabs.harvard.edu/abs/2024SPIE13100E..5WP}
}

@ARTICLE{Suarez-Mascareno2020,
       author = {{Su{\'a}rez Mascare{\~n}o}, A. and {Faria}, J.~P. and {Figueira}, P. and {Lovis}, C. and {Damasso}, M. and {Gonz{\'a}lez Hern{\'a}ndez}, J.~I. and {Rebolo}, R. and {Cristiani}, S. and {Pepe}, F. and {Santos}, N.~C. and {Zapatero Osorio}, M.~R. and {Adibekyan}, V. and {Hojjatpanah}, S. and {Sozzetti}, A. and {Murgas}, F. and {Abreu}, M. and {Affolter}, M. and {Alibert}, Y. and {Aliverti}, M. and {Allart}, R. and {Allende Prieto}, C. and {Alves}, D. and {Amate}, M. and {Avila}, G. and {Baldini}, V. and {Bandi}, T. and {Barros}, S.~C.~C. and {Bianco}, A. and {Benz}, W. and {Bouchy}, F. and {Broeng}, C. and {Cabral}, A. and {Calderone}, G. and {Cirami}, R. and {Coelho}, J. and {Conconi}, P. and {Coretti}, I. and {Cumani}, C. and {Cupani}, G. and {D'Odorico}, V. and {Deiries}, S. and {Delabre}, B. and {Di Marcantonio}, P. and {Dumusque}, X. and {Ehrenreich}, D. and {Fragoso}, A. and {Genolet}, L. and {Genoni}, M. and {G{\'e}nova Santos}, R. and {Hughes}, I. and {Iwert}, O. and {Kerber}, F. and {Knusdstrup}, J. and {Landoni}, M. and {Lavie}, B. and {Lillo-Box}, J. and {Lizon}, J. and {Lo Curto}, G. and {Maire}, C. and {Manescau}, A. and {Martins}, C.~J.~A.~P. and {M{\'e}gevand}, D. and {Mehner}, A. and {Micela}, G. and {Modigliani}, A. and {Molaro}, P. and {Monteiro}, M.~A. and {Monteiro}, M.~J.~P.~F.~G. and {Moschetti}, M. and {Mueller}, E. and {Nunes}, N.~J. and {Oggioni}, L. and {Oliveira}, A. and {Pall{\'e}}, E. and {Pariani}, G. and {Pasquini}, L. and {Poretti}, E. and {Rasilla}, J.~L. and {Redaelli}, E. and {Riva}, M. and {Santana Tschudi}, S. and {Santin}, P. and {Santos}, P. and {Segovia}, A. and {Sosnowska}, D. and {Sousa}, S. and {Span{\`o}}, P. and {Tenegi}, F. and {Udry}, S. and {Zanutta}, A. and {Zerbi}, F.},
        title = "{Revisiting Proxima with ESPRESSO}",
      journal = {\aap},
         year = 2020,
        month = jul,
       volume = {639},
          eid = {A77},
        pages = {A77},
          doi = {10.1051/0004-6361/202037745},
archivePrefix = {arXiv},
       eprint = {2005.12114},
 primaryClass = {astro-ph.EP},
       adsurl = {https://ui.adsabs.harvard.edu/abs/2020A&A...639A..77S}
}

@ARTICLE{Nari2025,
       author = {{Nari}, N. and {Dumusque}, X. and {Hara}, N.~C. and {Su{\'a}rez Mascare{\~n}o}, A. and {Cretignier}, M. and {Gonz{\'a}lez Hern{\'a}ndez}, J.~I. and {Stefanov}, A.~K. and {Passegger}, V.~M. and {Rebolo}, R. and {Pepe}, F. and {Santos}, N.~C. and {Cristiani}, S. and {Faria}, J.~P. and {Figueira}, P. and {Sozzetti}, A. and {Zapatero Osorio}, M.~R. and {Adibekyan}, V. and {Alibert}, Y. and {Allende Prieto}, C. and {Bouchy}, F. and {Benatti}, S. and {Castro-Gonz{\'a}lez}, A. and {D'Odorico}, V. and {Damasso}, M. and {Delisle}, J.~B. and {Di Marcantonio}, P. and {Ehrenreich}, D. and {G{\'e}nova-Santos}, R. and {Hobson}, M.~J. and {Lavie}, B. and {Lillo-Box}, J. and {Lo Curto}, G. and {Lovis}, C. and {Martins}, C.~J.~A.~P. and {Mehner}, A. and {Micela}, G. and {Molaro}, P. and {Mordasini}, C. and {Nunes}, N. and {Palle}, E. and {Quanz}, S.~P. and {S{\'e}gransan}, D. and {Silva}, A.~M. and {Sousa}, S.~G. and {Udry}, S. and {Unger}, N. and {Venturini}, J.},
        title = "{Revisiting the multi-planetary system of the nearby star HD 20794: Confirmation of a low-mass planet in the habitable zone of a nearby G-dwarf}",
      journal = {\aap},
         year = 2025,
        month = jan,
       volume = {693},
          eid = {A297},
        pages = {A297},
          doi = {10.1051/0004-6361/202451769},
archivePrefix = {arXiv},
       eprint = {2501.17092},
 primaryClass = {astro-ph.EP},
       adsurl = {https://ui.adsabs.harvard.edu/abs/2025A&A...693A.297N}
}

@ARTICLE{Gonzalez2024,
       author = {{Gonz{\'a}lez Hern{\'a}ndez}, J.~I. and {Su{\'a}rez Mascare{\~n}o}, A. and {Silva}, A.~M. and {Stefanov}, A.~K. and {Faria}, J.~P. and {Tabernero}, H.~M. and {Sozzetti}, A. and {Rebolo}, R. and {Pepe}, F. and {Santos}, N.~C. and {Cristiani}, S. and {Lovis}, C. and {Dumusque}, X. and {Figueira}, P. and {Lillo-Box}, J. and {Nari}, N. and {Benatti}, S. and {Hobson}, M.~J. and {Castro-Gonz{\'a}lez}, A. and {Allart}, R. and {Passegger}, V.~M. and {Zapatero Osorio}, M. -R. and {Adibekyan}, V. and {Alibert}, Y. and {Allende Prieto}, C. and {Bouchy}, F. and {Damasso}, M. and {D'Odorico}, V. and {Di Marcantonio}, P. and {Ehrenreich}, D. and {Lo Curto}, G. and {Santos}, R. G{\'e}nova and {Martins}, C.~J.~A.~P. and {Mehner}, A. and {Micela}, G. and {Molaro}, P. and {Nunes}, N. and {Palle}, E. and {Sousa}, S.~G. and {Udry}, S.},
        title = "{A sub-Earth-mass planet orbiting Barnard's star}",
      journal = {\aap},
         year = 2024,
        month = oct,
       volume = {690},
          eid = {A79},
        pages = {A79},
          doi = {10.1051/0004-6361/202451311},
archivePrefix = {arXiv},
       eprint = {2410.00569},
 primaryClass = {astro-ph.EP},
       adsurl = {https://ui.adsabs.harvard.edu/abs/2024A&A...690A..79G}
}

@ARTICLE{CostaSilva2024,
       author = {{Costa Silva}, A.~R. and {Demangeon}, O.~D.~S. and {Santos}, N.~C. and {Ehrenreich}, D. and {Lovis}, C. and {Chakraborty}, H. and {Lendl}, M. and {Pepe}, F. and {Cristiani}, S. and {Rebolo}, R. and {Zapatero-Osorio}, M.~R. and {Adibekyan}, V. and {Alibert}, Y. and {Allart}, R. and {Allende Prieto}, C. and {Azevedo Silva}, T. and {Borsa}, F. and {Bourrier}, V. and {Cristo}, E. and {Di Marcantonio}, P. and {Esparza-Borges}, E. and {Figueira}, P. and {Gonz{\'a}lez Hern{\'a}ndez}, J.~I. and {Herrero-Cisneros}, E. and {Lo Curto}, G. and {Martins}, C.~J.~A.~P. and {Mehner}, A. and {Nunes}, N.~J. and {Palle}, E. and {Pelletier}, S. and {Seidel}, J.~V. and {Silva}, A.~M. and {Sousa}, S.~G. and {Sozzetti}, A. and {Steiner}, M. and {Su{\'a}rez Mascare{\~n}o}, A. and {Udry}, S.},
        title = "{ESPRESSO reveals blueshifted neutral iron emission lines on the dayside of WASP-76 b}",
      journal = {\aap},
         year = 2024,
        month = sep,
       volume = {689},
          eid = {A8},
        pages = {A8},
          doi = {10.1051/0004-6361/202449935},
archivePrefix = {arXiv},
       eprint = {2409.13519},
 primaryClass = {astro-ph.EP},
       adsurl = {https://ui.adsabs.harvard.edu/abs/2024A&A...689A...8C}
}

@ARTICLE{Hobson2024,
       author = {{Hobson}, M.~J. and {Bouchy}, F. and {Lavie}, B. and {Lovis}, C. and {Adibekyan}, V. and {Allende Prieto}, C. and {Alibert}, Y. and {Barros}, S.~C.~C. and {Castro-Gonz{\'a}lez}, A. and {Cristiani}, S. and {D'Odorico}, V. and {Damasso}, M. and {Di Marcantonio}, P. and {Dumusque}, X. and {Ehrenreich}, D. and {Figueira}, P. and {G{\'e}nova Santos}, R. and {Gilbert}, E.~A. and {Gonz{\'a}lez Hern{\'a}ndez}, J.~I. and {Lillo-Box}, J. and {Lo Curto}, G. and {Martins}, C.~J.~A.~P. and {Mehner}, A. and {Micela}, G. and {Molaro}, P. and {Nunes}, N.~J. and {Palle}, E. and {Pepe}, F. and {Rebolo}, R. and {Rodrigues}, J. and {Santos}, N. and {Sousa}, S.~G. and {Sozzetti}, A. and {Su{\'a}rez Mascare{\~n}o}, A. and {Tabernero}, H.~M. and {Udry}, S. and {Zapatero Osorio}, M. -R. and {Armstrong}, D.~J. and {Ciardi}, D.~R. and {Collins}, K.~A. and {Collins}, K.~I. and {Everett}, M. and {Gandolfi}, D. and {Howell}, S.~B. and {Jenkins}, J.~M. and {Kielkopf}, J. and {Livingston}, J.~H. and {Lund}, M.~B. and {Mireles}, I. and {Ricker}, G.~R. and {Schwarz}, R.~P. and {Seager}, S. and {Tan}, T. -G. and {Ting}, E.~B. and {Winn}, J.~N.},
        title = "{Three super-Earths and a possible water world from TESS and ESPRESSO}",
      journal = {\aap},
         year = 2024,
        month = aug,
       volume = {688},
          eid = {A216},
        pages = {A216},
          doi = {10.1051/0004-6361/202450505},
archivePrefix = {arXiv},
       eprint = {2406.06278},
 primaryClass = {astro-ph.EP},
       adsurl = {https://ui.adsabs.harvard.edu/abs/2024A&A...688A.216H}
}

@ARTICLE{Suarez-Mascareno2024,
       author = {{Su{\'a}rez Mascare{\~n}o}, A. and {Passegger}, V.~M. and {Gonz{\'a}lez Hern{\'a}ndez}, J.~I. and {Armstrong}, D.~J. and {Nielsen}, L.~D. and {Lovis}, C. and {Lavie}, B. and {Sousa}, S.~G. and {Silva}, A.~M. and {Allart}, R. and {Rebolo}, R. and {Pepe}, F. and {Santos}, N.~C. and {Cristiani}, S. and {Sozzetti}, A. and {Zapatero Osorio}, M.~R. and {Tabernero}, H.~M. and {Dumusque}, X. and {Udry}, S. and {Adibekyan}, V. and {Allende Prieto}, C. and {Alibert}, Y. and {Barros}, S.~C.~C. and {Bouchy}, F. and {Castro-Gonz{\'a}lez}, A. and {Collins}, K.~A. and {Damasso}, M. and {D'Odorico}, V. and {Demangeon}, O.~D.~S. and {Di Marcantonio}, P. and {Ehrenreich}, D. and {Hadjigeorghiou}, A. and {Hara}, N. and {Hawthorn}, F. and {Jenkins}, J.~M. and {Lillo-Box}, J. and {Lo Curto}, G. and {Martins}, C.~J.~A.~P. and {Mehner}, A. and {Micela}, G. and {Molaro}, P. and {Nunes}, N. and {Nari}, N. and {Osborn}, A. and {Pall{\'e}}, E. and {Ricker}, G.~R. and {Rodrigues}, J. and {Rowden}, P. and {Seager}, S. and {Stefanov}, A.~K. and {Str{\o}m}, P.~A. and {Villase{\~n}or}, J.~N.~S. and {Watkins}, C.~N. and {Winn}, J. and {Wohler}, B. and {Zambelli}, R.},
        title = "{TESS and ESPRESSO discover a super-Earth and a mini-Neptune orbiting the K-dwarf TOI-238*}",
      journal = {\aap},
         year = 2024,
        month = may,
       volume = {685},
          eid = {A56},
        pages = {A56},
          doi = {10.1051/0004-6361/202348958},
archivePrefix = {arXiv},
       eprint = {2402.04113},
 primaryClass = {astro-ph.EP},
       adsurl = {https://ui.adsabs.harvard.edu/abs/2024A&A...685A..56S}
}

@ARTICLE{Passegger2024,
       author = {{Passegger}, V.~M. and {Su{\'a}rez Mascare{\~n}o}, A. and {Allart}, R. and {Gonz{\'a}lez Hern{\'a}ndez}, J.~I. and {Lovis}, C. and {Lavie}, B. and {Silva}, A.~M. and {M{\"u}ller}, H.~M. and {Tabernero}, H.~M. and {Cristiani}, S. and {Pepe}, F. and {Rebolo}, R. and {Santos}, N.~C. and {Adibekyan}, V. and {Alibert}, Y. and {Allende Prieto}, C. and {Barros}, S.~C.~C. and {Bouchy}, F. and {Castro-Gonz{\'a}lez}, A. and {D'Odorico}, V. and {Dumusque}, X. and {Di Marcantonio}, P. and {Ehrenreich}, D. and {Figueira}, P. and {G{\'e}nova Santos}, R. and {Lo Curto}, G. and {Martins}, C.~J.~A.~P. and {Mehner}, A. and {Micela}, G. and {Molaro}, P. and {Nari}, N. and {Nunes}, N.~J. and {Pall{\'e}}, E. and {Poretti}, E. and {Rodrigues}, J. and {Sousa}, S.~G. and {Sozzetti}, A. and {Udry}, S. and {Zapatero Osorio}, M.~R.},
        title = "{The compact multi-planet system GJ 9827 revisited with ESPRESSO}",
      journal = {\aap},
         year = 2024,
        month = apr,
       volume = {684},
          eid = {A22},
        pages = {A22},
          doi = {10.1051/0004-6361/202348592},
archivePrefix = {arXiv},
       eprint = {2401.06276},
 primaryClass = {astro-ph.EP},
       adsurl = {https://ui.adsabs.harvard.edu/abs/2024A&A...684A..22P}
}

@ARTICLE{Damasso2023,
       author = {{Damasso}, M. and {Rodrigues}, J. and {Castro-Gonz{\'a}lez}, A. and {Lavie}, B. and {Davoult}, J. and {Zapatero Osorio}, M.~R. and {Dou}, J. and {Sousa}, S.~G. and {Owen}, J.~E. and {Sossi}, P. and {Adibekyan}, V. and {Osborn}, H. and {Leinhardt}, Z. and {Alibert}, Y. and {Lovis}, C. and {Delgado Mena}, E. and {Sozzetti}, A. and {Barros}, S.~C.~C. and {Bossini}, D. and {Ziegler}, C. and {Ciardi}, D.~R. and {Matthews}, E.~C. and {Carter}, P.~J. and {Lillo-Box}, J. and {Su{\'a}rez Mascare{\~n}o}, A. and {Cristiani}, S. and {Pepe}, F. and {Rebolo}, R. and {Santos}, N.~C. and {Allende Prieto}, C. and {Benatti}, S. and {Bouchy}, F. and {Brice{\~n}o}, C. and {Di Marcantonio}, P. and {D'Odorico}, V. and {Dumusque}, X. and {Egger}, J.~A. and {Ehrenreich}, D. and {Faria}, J. and {Figueira}, P. and {G{\'e}nova Santos}, R. and {Gonzales}, E.~J. and {Gonz{\'a}lez Hern{\'a}ndez}, J.~I. and {Law}, N. and {Lo Curto}, G. and {Mann}, A.~W. and {Martins}, C.~J.~A.~P. and {Mehner}, A. and {Micela}, G. and {Molaro}, P. and {Nunes}, N.~J. and {Palle}, E. and {Poretti}, E. and {Schlieder}, J.~E. and {Udry}, S.},
        title = "{A compact multi-planet system transiting HIP 29442 (TOI-469) discovered by TESS and ESPRESSO. Radial velocities lead to the detection of transits with low signal-to-noise ratio}",
      journal = {\aap},
         year = 2023,
        month = nov,
       volume = {679},
          eid = {A33},
        pages = {A33},
          doi = {10.1051/0004-6361/202347240},
archivePrefix = {arXiv},
       eprint = {2308.13310},
 primaryClass = {astro-ph.EP},
       adsurl = {https://ui.adsabs.harvard.edu/abs/2023A&A...679A..33D}
}

@ARTICLE{Castro-Gonzalez2023,
       author = {{Castro-Gonz{\'a}lez}, A. and {Demangeon}, O.~D.~S. and {Lillo-Box}, J. and {Lovis}, C. and {Lavie}, B. and {Adibekyan}, V. and {Acu{\~n}a}, L. and {Deleuil}, M. and {Aguichine}, A. and {Zapatero Osorio}, M.~R. and {Tabernero}, H.~M. and {Davoult}, J. and {Alibert}, Y. and {Santos}, N. and {Sousa}, S.~G. and {Antoniadis-Karnavas}, A. and {Borsa}, F. and {Winn}, J.~N. and {Allende Prieto}, C. and {Figueira}, P. and {Jenkins}, J.~M. and {Sozzetti}, A. and {Damasso}, M. and {Silva}, A.~M. and {Astudillo-Defru}, N. and {Barros}, S.~C.~C. and {Bonfils}, X. and {Cristiani}, S. and {Di Marcantonio}, P. and {Gonz{\'a}lez Hern{\'a}ndez}, J.~I. and {Curto}, G. Lo and {Martins}, C.~J.~A.~P. and {Nunes}, N.~J. and {Palle}, E. and {Pepe}, F. and {Seager}, S. and {Su{\'a}rez Mascare{\~n}o}, A.},
        title = "{An unusually low-density super-Earth transiting the bright early-type M-dwarf GJ 1018 (TOI-244)}",
      journal = {\aap},
         year = 2023,
        month = jul,
       volume = {675},
          eid = {A52},
        pages = {A52},
          doi = {10.1051/0004-6361/202346550},
archivePrefix = {arXiv},
       eprint = {2305.04922},
 primaryClass = {astro-ph.EP},
       adsurl = {https://ui.adsabs.harvard.edu/abs/2023A&A...675A..52C}
}

@ARTICLE{Lavie2023,
       author = {{Lavie}, B. and {Bouchy}, F. and {Lovis}, C. and {Zapatero Osorio}, M. and {Deline}, A. and {Barros}, S. and {Figueira}, P. and {Sozzetti}, A. and {Gonz{\'a}lez Hern{\'a}ndez}, J.~I. and {Lillo-Box}, J. and {Rodrigues}, J. and {Mehner}, A. and {Damasso}, M. and {Adibekyan}, V. and {Alibert}, Y. and {Allende Prieto}, C. and {Cristiani}, S. and {D'Odorico}, V. and {Di Marcantonio}, P. and {Ehrenreich}, D. and {G{\'e}nova Santos}, R. and {Lo Curto}, G. and {Martins}, C.~J.~A.~P. and {Micela}, G. and {Molaro}, P. and {Nunes}, N. and {Palle}, E. and {Pepe}, F. and {Poretti}, E. and {Rebolo}, R. and {Santos}, N. and {Sousa}, S. and {Su{\'a}rez Mascare{\~n}o}, A. and {Tabrenero}, H. and {Udry}, S.},
        title = "{Planetary system around LTT 1445A unveiled by ESPRESSO: Multiple planets in a triple M-dwarf system}",
      journal = {\aap},
         year = 2023,
        month = may,
       volume = {673},
          eid = {A69},
        pages = {A69},
          doi = {10.1051/0004-6361/202143007},
archivePrefix = {arXiv},
       eprint = {2210.09713},
 primaryClass = {astro-ph.EP},
       adsurl = {https://ui.adsabs.harvard.edu/abs/2023A&A...673A..69L}
}

@ARTICLE{Suarez-Mascareno2023,
       author = {{Su{\'a}rez Mascare{\~n}o}, A. and {Gonz{\'a}lez-{\'A}lvarez}, E. and {Zapatero Osorio}, M.~R. and {Lillo-Box}, J. and {Faria}, J.~P. and {Passegger}, V.~M. and {Gonz{\'a}lez Hern{\'a}ndez}, J.~I. and {Figueira}, P. and {Sozzetti}, A. and {Rebolo}, R. and {Pepe}, F. and {Santos}, N.~C. and {Cristiani}, S. and {Lovis}, C. and {Silva}, A.~M. and {Ribas}, I. and {Amado}, P.~J. and {Caballero}, J.~A. and {Quirrenbach}, A. and {Reiners}, A. and {Zechmeister}, M. and {Adibekyan}, V. and {Alibert}, Y. and {B{\'e}jar}, V.~J.~S. and {Benatti}, S. and {D'Odorico}, V. and {Damasso}, M. and {Delisle}, J. -B. and {Di Marcantonio}, P. and {Dreizler}, S. and {Ehrenreich}, D. and {Hatzes}, A.~P. and {Hara}, N.~C. and {Henning}, Th. and {Kaminski}, A. and {L{\'o}pez-Gonz{\'a}lez}, M.~J. and {Martins}, C.~J.~A.~P. and {Micela}, G. and {Montes}, D. and {Pall{\'e}}, E. and {Pedraz}, S. and {Rodr{\'\i}guez}, E. and {Rodr{\'\i}guez-L{\'o}pez}, C. and {Tal-Or}, L. and {Sousa}, S. and {Udry}, S.},
        title = "{Two temperate Earth-mass planets orbiting the nearby star GJ 1002}",
      journal = {\aap},
         year = 2023,
        month = feb,
       volume = {670},
          eid = {A5},
        pages = {A5},
          doi = {10.1051/0004-6361/202244991},
archivePrefix = {arXiv},
       eprint = {2212.07332},
 primaryClass = {astro-ph.EP},
       adsurl = {https://ui.adsabs.harvard.edu/abs/2023A&A...670A...5S}
}

@ARTICLE{AzevedoSilva2022,
       author = {{Azevedo Silva}, T. and {Demangeon}, O.~D.~S. and {Santos}, N.~C. and {Allart}, R. and {Borsa}, F. and {Cristo}, E. and {Esparza-Borges}, E. and {Seidel}, J.~V. and {Palle}, E. and {Sousa}, S.~G. and {Tabernero}, H.~M. and {Zapatero Osorio}, M.~R. and {Cristiani}, S. and {Pepe}, F. and {Rebolo}, R. and {Adibekyan}, V. and {Alibert}, Y. and {Barros}, S.~C.~C. and {Bouchy}, F. and {Bourrier}, V. and {Lo Curto}, G. and {Di Marcantonio}, P. and {D'Odorico}, V. and {Ehrenreich}, D. and {Figueira}, P. and {Gonz{\'a}lez Hern{\'a}ndez}, J.~I. and {Lovis}, C. and {Martins}, C.~J.~A.~P. and {Mehner}, A. and {Micela}, G. and {Molaro}, P. and {Mounzer}, D. and {Nunes}, N.~J. and {Sozzetti}, A. and {Su{\'a}rez Mascare{\~n}o}, A. and {Udry}, S.},
        title = "{Detection of barium in the atmospheres of the ultra-hot gas giants WASP-76b and WASP-121b. Together with new detections of Co and Sr+ on WASP-121b}",
      journal = {\aap},
         year = 2022,
        month = oct,
       volume = {666},
          eid = {L10},
        pages = {L10},
          doi = {10.1051/0004-6361/202244489},
archivePrefix = {arXiv},
       eprint = {2210.06892},
 primaryClass = {astro-ph.EP},
       adsurl = {https://ui.adsabs.harvard.edu/abs/2022A&A...666L..10A}
}

@ARTICLE{Barros2022,
       author = {{Barros}, S.~C.~C. and {Demangeon}, O.~D.~S. and {Alibert}, Y. and {Leleu}, A. and {Adibekyan}, V. and {Lovis}, C. and {Bossini}, D. and {Sousa}, S.~G. and {Hara}, N. and {Bouchy}, F. and {Lavie}, B. and {Rodrigues}, J. and {Gomes da Silva}, J. and {Lillo-Box}, J. and {Pepe}, F.~A. and {Tabernero}, H.~M. and {Zapatero Osorio}, M.~R. and {Sozzetti}, A. and {Su{\'a}rez Mascare{\~n}o}, A. and {Micela}, G. and {Allende Prieto}, C. and {Cristiani}, S. and {Damasso}, M. and {Di Marcantonio}, P. and {Ehrenreich}, D. and {Faria}, J. and {Figueira}, P. and {Gonz{\'a}lez Hern{\'a}ndez}, J.~I. and {Jenkins}, J. and {Lo Curto}, G. and {Martins}, C.~J.~A.~P. and {Micela}, G. and {Nunes}, N.~J. and {Pall{\'e}}, E. and {Santos}, N.~C. and {Rebolo}, R. and {Seager}, S. and {Twicken}, J.~D. and {Udry}, S. and {Vanderspek}, R. and {Winn}, J.~N.},
        title = "{HD 23472: a multi-planetary system with three super-Earths and two potential super-Mercuries}",
      journal = {\aap},
         year = 2022,
        month = sep,
       volume = {665},
          eid = {A154},
        pages = {A154},
          doi = {10.1051/0004-6361/202244293},
archivePrefix = {arXiv},
       eprint = {2209.13345},
 primaryClass = {astro-ph.EP},
       adsurl = {https://ui.adsabs.harvard.edu/abs/2022A&A...665A.154B}
}

@article{Dumusque-2015b,
	adsurl = {http://adsabs.harvard.edu/abs/2015ApJ...814L..21D},
	archiveprefix = {arXiv},
	arxivurl = {http://arXiv.org/abs/1511.02267},
	author = {{Dumusque}, X. and {Glenday}, A. and {Phillips}, D.~F. and {Buchschacher}, N. and {Collier Cameron}, A. and {Cecconi}, M. and {Charbonneau}, D. and {Cosentino}, R. and {Ghedina}, A. and {Latham}, D.~W. and {Li}, C.-H. and {Lodi}, M. and {Lovis}, C. and {Molinari}, E. and {Pepe}, F. and {Udry}, S. and {Sasselov}, D. and {Szentgyorgyi}, A. and {Walsworth}, R.},
	doi = {10.1088/2041-8205/814/2/L21},
	eid = {L21},
	eprint = {1511.02267},
	journal = {\apjl},
	month = dec,
	pages = {L21},
	primaryclass = {astro-ph.EP},
	title = {{HARPS-N Observes the Sun as a Star}},
	volume = 814,
	year = 2015}

\end{document}